\documentclass[longauth]{aa} % for the long lists of affiliations 
\usepackage{natbib}
\usepackage[varg]{txfonts}
\usepackage{graphicx}
\usepackage{multirow}
\usepackage{relsize}
\DeclareUnicodeCharacter{0300}{ }

\begin{document}
%\begin{frontmatter}
	
\vspace{-2cm}

\title{Multiband variability studies and novel broadband SED modeling of Mrk~501 in 2009}
\titlerunning{Multiband variability studies and novel broadband SED
  modeling of Mrk~501 in 2009} %temporal evolution of the broad-band emission of Mrk~501}
%\include{authors_last_ApJ}

% authors 15.02.2016  Format AA
%
\author{
%\small
%\footnotesize
\tiny
M.~L.~Ahnen\inst{1} \and
S.~Ansoldi\inst{2} \and
L.~A.~Antonelli\inst{3} \and
P.~Antoranz\inst{4} \and
A.~Babic\inst{5} \and
B.~Banerjee\inst{6} \and
P.~Bangale\inst{7} \and
U.~Barres de Almeida\inst{7,}\inst{24} \and
J.~A.~Barrio\inst{8} \and
J.~Becerra Gonz\'alez\inst{9,}\inst{25}
% ...
\and \\ 
(The MAGIC collaboration) \\
A.~U.~Abeysekara\inst{35} \and
S.~Archambault\inst{36} \and
A.~Archer\inst{37} \and
W.~Benbow\inst{38} \and
R.~Bird\inst{39} \and
M.~Buchovecky\inst{40} \and
J.~H.~Buckley\inst{37} \and
V.~Bugaev\inst{37} \and
M.~P.~Connolly\inst{41} \and
W.~Cui\inst{42, 43} \and
H.~J.~Dickinson\inst{44} \and
A.~Falcone\inst{45} \and
Q.~Feng\inst{42} \and
J.~P.~Finley\inst{42} \and
H.~Fleischhack\inst{11} \and
A.~Flinders\inst{35} \and
L.~Fortson\inst{46} \and
G.~H.~Gillanders\inst{41} \and
S.~Griffin\inst{36} \and
J.~Grube\inst{47} \and
M.~H\"utten\inst{11} \and
D.~Hanna\inst{36} \and
J.~Holder\inst{48} \and
T.~B.~Humensky\inst{49} \and
P.~Kaaret\inst{50} \and
P.~Kar\inst{35} \and
N.~Kelley-Hoskins\inst{11} \and
M.~Kertzman\inst{51} \and
D.~Kieda\inst{35} \and
M.~Krause\inst{11} \and
F.~Krennrich\inst{44} \and
M.~J.~Lang\inst{41} \and
G.~Maier\inst{11} \and
A.~McCann\inst{36} \and
P.~Moriarty\inst{41} \and
R.~Mukherjee\inst{52} \and
D.~Nieto\inst{49} \and
S.~O'Brien\inst{39} \and
R.~A.~Ong\inst{40} \and
N.~Otte\inst{57} \and
N.~Park\inst{53} \and
J.~Perkins\inst{25} \and
A.~Pichel\inst{54} \and
M.~Pohl\inst{55, 25} \and
A.~Popkow\inst{40} \and
E.~Pueschel\inst{39} \and
J.~Quinn\inst{39} \and
K.~Ragan\inst{36} \and
P.~T.~Reynolds\inst{56} \and
G.~T.~Richards\inst{57} \and
E.~Roache\inst{38} \and
A.~C.~Rovero\inst{54} \and
C.~Rulten\inst{46} \and
I.~Sadeh\inst{11} \and
M.~Santander\inst{52} \and
G.~H.~Sembroski\inst{42} \and
K.~Shahinyan\inst{46} \and
I.~Telezhinsky\inst{55, 25} \and
J.~V.~Tucci\inst{42} \and
J.~Tyler\inst{36} \and
S.~P.~Wakely\inst{53} \and
A.~Weinstein\inst{44} \and
P.~Wilcox\inst{50} \and
A.~Wilhelm\inst{55, 25} \and
D.~A.~Williams\inst{58} \and
B.~Zitzer\inst{59} \and \\
(the VERITAS Collaboration),  \\ 
S.~Razzaque\inst{60}  \and
M.~Villata\inst{61} \and
C.~M.~Raiteri\inst{61} \and
H.~D.~Aller\inst{62} \and
M.~F.~Aller\inst{62} \and
V.~M.~Larionov\inst{63,64} \and
A.~A.~Arkharov\inst{64} \and
D.~A.~Blinov\inst{63,65,66} \and
N.~V.~Efimova\inst{64} \and
T.~S.~Grishina\inst{63} \and
V.~A.~Hagen-Thorn\inst{63} \and
E. N. Kopatskaya\inst{63} \and
L. V. Larionova\inst{63} \and
E. G. Larionova\inst{63} \and
D. A. Morozova\inst{63} \and
I. S. Troitsky\inst{63} \and
R.~Ligustri\inst{67} \and
P.~Calcidese\inst{68} \and
A.~Berdyugin\inst{17} \and
O.~M.~Kurtanidze\inst{69,70,71} \and
M.~G.~Nikolashvili\inst{69} \and
G. N. Kimeridze\inst{69} \and 
L. A. Sigua\inst{69} \and
S. O. Kurtanidze\inst{69} \and
R.A. Chigladze\inst{69} \and
W. P. Chen\inst{72} \and
E. Koptelova\inst{72} \and 
T.~Sakamoto\inst{73} \and
A.~C.~Sadun\inst{74} \and
J.~W.~Moody\inst{75} \and
C.~Pace\inst{75} \and
R.~Pearson~III\inst{75} \and
Y.~Yatsu\inst{76} \and
Y.~Mori\inst{76} \and
A.~Carraminyana\inst{77}  \and 
L.~Carrasco\inst{77}  \and 
E.~de~la~Fuente\inst{78}  \and 
J.P.~Norris\inst{79}  \and 
P.~S.~Smith\inst{80}  \and
A.~Wehrle\inst{81}  \and  
M. A. Gurwell\inst{82}  \and  
Alma Zook\inst{83}  \and  
C.~Pagani\inst{84}  \and  
M.~Perri\inst{85,86}  \and  
M. Capalbi\inst{85}  \and   
A. Cesarini\inst{87} \and 
H. A. Krimm\inst{25,88,89} \and
Y. Y. Kovalev\inst{90,91} \and
Yu. A. Kovalev\inst{90} \and
E. Ros\inst{91,92,93} \and
A.B. Pushkarev\inst{90,94} \and
M.L. Lister\inst{42} \and
K.V. Sokolovsky\inst{90,95,96} \and 
M. Kadler\inst{13} \and
G. Piner\inst{97} \and
A. L\"ahteenm\"aki\inst{98} \and 
M. Tornikoski\inst{98} \and  
E. Angelakis\inst{91} \and  
T. P. Krichbaum\inst{91} \and   
I. Nestoras\inst{91} \and  
L Fuhrmann\inst{91} \and  
J. A. Zensus\inst{91} \and   
P. Cassaro\inst{99} \and
A. Orlati\inst{100} \and 
G. Maccaferri\inst{100} \and  
P. Leto\inst{101} \and
M. Giroletti\inst{100} \and
J. L. Richards\inst{42} \and
W. Max-Moerbeck\inst{91} \and   
A. C. S. Readhead\inst{102}
 }
 \cleardoublepage
\institute {ETH Zurich, CH-8093 Zurich, Switzerland
\and Universit\`a di Udine, and INFN Trieste, I-33100 Udine, Italy
\and INAF National Institute for Astrophysics, I-00136 Rome, Italy
\and Universit\`a  di Siena, and INFN Pisa, I-53100 Siena, Italy
\and Croatian MAGIC Consortium, Rudjer Boskovic Institute, University of Rijeka, University of Split and University of Zagreb, Croatia
\and Saha Institute of Nuclear Physics, 1/AF Bidhannagar, Salt Lake, Sector-1, Kolkata 700064, India
\and Max-Planck-Institut f\"ur Physik, D-80805 M\"unchen, Germany
\and Universidad Complutense, E-28040 Madrid, Spain
\and Inst. de Astrof\'isica de Canarias, E-38200 La Laguna, Tenerife, Spain; Universidad de La Laguna, Dpto. Astrof\'isica, E-38206 La Laguna, Tenerife, Spain
\and University of \L\'od\'z, PL-90236 Lodz, Poland
\and Deutsches Elektronen-Synchrotron (DESY), D-15738 Zeuthen, Germany
\and Institut de Fisica d'Altes Energies (IFAE), The Barcelona Institute of Science and Technology, Campus UAB, 08193 Bellaterra (Barcelona), Spain
\and Universit\"at W\"urzburg, D-97074 W\"urzburg, Germany
\and Universit\`a di Padova and INFN, I-35131 Padova, Italy
\and Institute for Space Sciences (CSIC/IEEC), E-08193 Barcelona, Spain
\and Technische Universit\"at Dortmund, D-44221 Dortmund, Germany
\and Finnish MAGIC Consortium, Tuorla Observatory, University of Turku and Astronomy Division, University of Oulu, Finland
\and Unitat de F\'isica de les Radiacions, Departament de F\'isica, and CERES-IEEC, Universitat Aut\`onoma de Barcelona, E-08193 Bellaterra, Spain
\and Universitat de Barcelona, ICC, IEEC-UB, E-08028 Barcelona, Spain
\and Japanese MAGIC Consortium, ICRR, The University of Tokyo, Department of Physics and Hakubi Center, Kyoto University, Tokai University, The University of Tokushima, KEK, Japan
\and Inst. for Nucl. Research and Nucl. Energy, BG-1784 Sofia, Bulgaria
\and Universit\`a di Pisa, and INFN Pisa, I-56126 Pisa, Italy
\and ICREA and Institute for Space Sciences (CSIC/IEEC), E-08193 Barcelona, Spain
\and now at Centro Brasileiro de Pesquisas F\'isicas (CBPF/MCTI), R. Dr. Xavier Sigaud, 150 - Urca, Rio de Janeiro - RJ, 22290-180, Brazil
\and NASA Goddard Space Flight Center, Greenbelt, MD 20771, USA 
\and Humboldt University of Berlin, Institut f\"ur Physik Newtonstr. 15, 12489 Berlin Germany
\and also at University of Trieste
\and now at Ecole polytechnique f\'ed\'erale de Lausanne (EPFL), Lausanne, Switzerland
\and now at Max-Planck-Institut fur Kernphysik, P.O. Box 103980, D 69029 Heidelberg, Germany
\and also at Japanese MAGIC Consortium
\and now at Finnish Centre for Astronomy with ESO (FINCA), Turku, Finland
\and also at INAF-Trieste and Dept. of Physics \& Astronomy, University of Bologna
\and also at ISDC - Science Data Center for Astrophysics, 1290, Versoix (Geneva)
\and also at Department of Physics and Department of Astronomy, University of Maryland, College Park, MD 20742, USA
\and Department of Physics and Astronomy, University of Utah, Salt Lake City, UT 84112, USA
\and Physics Department, McGill University, Montreal, QC H3A 2T8, Canada
\and Department of Physics, Washington University, St. Louis, MO 63130, USA
\and Fred Lawrence Whipple Observatory, Harvard-Smithsonian Center for Astrophysics, Amado, AZ 85645, USA
\and School of Physics, University College Dublin, Belfield, Dublin 4, Ireland
\and Department of Physics and Astronomy, University of California, Los Angeles, CA 90095, USA
\clearpage
\and School of Physics, National University of Ireland Galway, University Road, Galway, Ireland
\and Department of Physics and Astronomy, Purdue University, West Lafayette, IN 47907, USA
\and Department of Physics and Center for Astrophysics, Tsinghua University, Beijing 100084, China.
\and Department of Physics and Astronomy, Iowa State University, Ames, IA 50011, USA
\and Department of Astronomy and Astrophysics, 525 Davey Lab, Pennsylvania State University, University Park, PA 16802, USA
\and School of Physics and Astronomy, University of Minnesota, Minneapolis, MN 55455, USA
\and Astronomy Department, Adler Planetarium and Astronomy Museum, Chicago, IL 60605, USA
\and Department of Physics and Astronomy and the Bartol Research Institute, University of Delaware, Newark, DE 19716, USA
\and Physics Department, Columbia University, New York, NY 10027, USA
\and Department of Physics and Astronomy, University of Iowa, Van Allen Hall, Iowa City, IA 52242, USA
\and Department of Physics and Astronomy, DePauw University, Greencastle, IN 46135-0037, USA
\and Department of Physics and Astronomy, Barnard College, Columbia University, NY 10027, USA
\and Enrico Fermi Institute, University of Chicago, Chicago, IL 60637, USA
\and Instituto de Astronomia y Fisica del Espacio, Casilla de Correo 67 - Sucursal 28, (C1428ZAA) Ciudad Automa de Buenos Aires, Argentina
\and Institute of Physics and Astronomy, University of Potsdam, 14476 Potsdam-Golm, Germany
\and Department of Physical Sciences, Cork Institute of Technology, Bishopstown, Cork, Ireland
\and School of Physics and Center for Relativistic Astrophysics, Georgia Institute of Technology, 837 State Street NW, Atlanta, GA 30332-0430
\and Santa Cruz Institute for Particle Physics and Department of Physics, University of California, Santa Cruz, CA 95064, USA
\and Argonne National Laboratory, 9700 S. Cass Avenue, Argonne, IL 60439, USA
\and Department of Physics, University of Johannesburg, P.O. Box 524, Auckland Park 2006, South Africa
\and INAF—Osservatorio Astrofisico di Torino, I-10025 Pino Torinese (TO), Italy
\and Department of Astronomy, University of Michigan, Ann Arbor, MI 48109-1042, USA
\and Astronomical Institute, St. Petersburg State University, Universitetskij Pr. 28, Petrodvorets, 198504 St. Petersburg, Russia
\and Pulkovo Observatory, St.-Petersburg, Russia
\and Department of Physics and Institute for Plasma Physics, University of Crete, 71003, Heraklion, Greece
\and Foundation for Research and Technology—Hellas, IESL, Voutes, 71110 Heraklion, Greece
\and Circolo Astrofili Talmassons, I-33030 Campoformido (UD), Italy
\and Osservatorio Astrofisico della Regione Autonoma Valle d'Aosta, Italy 
\and Abastumani Observatory, Mt. Kanobili, 0301 Abastumani, Georgia
\and Engelhard Astronomical Observatory, Kazan Federal University, Tatarstan, Russia
\and Center for Astrophysics, Guangzhou University, Guangzhou 510006, China
\and Graduate Institute of Astronomy, National Central University, 300 Zhongda Road, Zhongli 32001, Taiwan
\and Department of Physics and Mathematics, College of Science and 952 Engineering, Aoyama Gakuin University, 5-10-1 Fuchinobe, Chuoku, Sagamihara-shi Kanagawa 252-5258, Japan
\and Department of Physics, University of Colorado Denver, Denver, Colorado, CO 80217-3364, USA
\and Department of Physics and Astronomy, Brigham Young University, Provo, Utah 84602, USA
\clearpage
\and Department of Physics, Tokyo Institute of Technology, Meguro City, Tokyo 152-8551, Japan
\and Instituto Nacional de Astrof\'{i}sica, \'{O}ptica y Electr\'{o}nica, Tonantzintla, Puebla 72840, Mexico
\and Instututo de Astronomia y Meteorologia, Dpto. de Fisica, CUCEI, Universidad de Guadalajara
\and Department of Physics and Astronomy, University of Denver, Denver, CO 80208, USA
\and Steward Observatory, University of Arizona, Tucson, AZ 85721 USA
\and Space Science Institute, Boulder, CO 80301, USA
\and Harvard-Smithsonian Center for Astrophysics, Cambridge, MA 02138, USA
\and Department of Physics and Astronomy, Pomona College, Claremont CA 91711-6312, USA
\and Department of Physics and Astronomy, University of Leicester, Leicester, LE1 7RH, UK
\and ASI Science Data Center, Via del Politecnico snc I-00133, Roma, Italy
\and INAF—Osservatorio Astronomico di Roma, via di Frascati 33, I-00040 Monteporzio, Italy
\and Department of Physics, University of Trento, I38050, Povo, Trento, Italy
\and Center for Research and Exploration in Space Science and Technology (CRESST)
\and Universities Space Research Association (USRA), Columbia, MD 21044, USA
\and Astro Space Center of Lebedev Physical Institute, Profsoyuznaya 84/32, 117997 Moscow, Russia
\and Max-Planck-Institut f\"{u}r Radioastronomie, Auf dem H\''{u}gel 69, D-53121 Bonn, Germany
\and Observatori Astronòmic, Universitat de Val\`{e}ncia, Parc Cient\'{i}fic, C. Catedr\'{a}tico Jos\'{e} Beltr\'{a}n 2, E-46980 Paterna, Val\`{e}ncia, Spain
\and Departament d’Astronomia i Astrof\'{i}sica, Universitat de Val\`{e}ncia, C. Dr. Moliner 50, E-46100 Burjassot, Val\`{e}ncia, Spain
\and  Crimean Astrophysical Observatory, 98409 Nauchny, Crimea, Ukraine
\and Sternberg Astronomical Institute, M.V. Lomonosov Moscow State University, Universiteskij prosp. 13, Moscow 119991, Russia
\and Institute of Astronomy, Astrophysics, Space Applications and Remote Sensing, National Observatory of Athens, Vas. Pavlou \& I. Metaxa, 15 236 Penteli, Greece
\and Department of Physics and Astronomy, Whittier College, Whittier, CA, USA
\and Aalto University Mets\"ahovi Radio Observatory, FIN-02540 Kylm\"al\"a, Finland 
\and INAF Istituto di Radioastronomia, Sezione di Noto, Contrada Renna Bassa, 96017 Noto (SR), Italy
\and INAF Istituto di Radioastronomia, Stazione Radioastronomica di Medicina, I-40059 Medicina (Bologna), Italy
\and INAF Osservatorio Astrofisico di Catania, via S. Sofia 78, 95123 Catania, Italy
\and Cahill Center for Astronomy and Astrophysics, California Institute of Technology, Pasadena, CA 91125, USA
\and {*} Corresponding authors: Marlene Doert (marlene.doert@tu-dortmund.de) and David Paneque (dpaneque@mppmu.mpg.de)
}

%\authorrunning{MAGIC, VERITAS, Fermi-LAT}

%\begin{abstract}

\abstract {} 
{%Text of aims
\tiny We present an extensive study of the BL Lac object Mrk~501 based on a data set collected during the multi-instrument campaign spanning from 2009 March 15 to 2009 August 1, 
which includes, among other instruments, MAGIC, VERITAS, Whipple 10\,m, and {\it Fermi}-LAT  to cover the $\gamma$-ray range from 0.1~GeV to 20~TeV, {\it RXTE} and {\it Swift} to cover wavelengths from UV to hard X-rays, and GASP-WEBT that provides coverage of radio and optical wavelengths. Optical polarization measurements were provided for a fraction of the campaign by the Steward and St.Petersburg observatories. We evaluate the variability of the source and interband correlations, the $\gamma$-ray flaring activity occurring in May 2009, and interpret the results within two synchrotron self-Compton (SSC) scenarios. 
} {%Text of methods
\tiny The multiband variability observed during the full campaign is addressed in terms of the \emph{fractional variability}, and the possible correlations are studied by calculating the \emph{discrete correlation function} for each pair of energy bands, where the significance was evaluated with dedicated Monte Carlo simulations. The space of SSC model parameters is probed following a dedicated grid-scan strategy, allowing for a wide range of models to be tested and offering a study of the degeneracy of model-to-data agreement in the individual model parameters, hence providing a less biased interpretation than the ``single-curve SSC model adjustment'' typically reported in the literature.
} {%Text of results
\tiny We find an increase in the fractional variability with energy, while no significant interband correlations of flux changes are found on the basis of the acquired data set. 
The SSC model grid-scan shows that the flaring activity around May 22 cannot be modeled adequately with a one-zone SSC scenario (using an electron energy distribution with two breaks), while it can be suitably described within a two-independent-zone SSC scenario. Here, one zone is responsible for the quiescent emission from the averaged 4.5-month observing period, while the other one, which is spatially separated from the first, dominates the flaring emission occurring at X-rays and very high energy ($>$ 100\,GeV, VHE) $\gamma$-rays. 
The flaring activity from May 1, which coincides with a rotation of
the electric vector polarization angle (EVPA), cannot be
satisfactorily reproduced by either a one-zone or a two-independent-zone SSC model, yet this is partially affected by the lack of strictly
simultaneous observations and the presence of large flux changes on
sub-hour timescales (detected at VHE $\gamma$-rays). 
} {% Conclusions
\tiny The higher variability in the VHE emission and lack of
correlation with the X-ray emission indicate that, at least during the
4.5-month long observing campaign in 2009, the highest-energy (and
most variable) electrons that are responsible for the VHE
$\gamma$-rays do not make a dominant contribution to the $\sim$1~keV
emission. Alternatively, there could be a very variable component
contributing to the VHE $\gamma$-ray emission in addition to that
coming from the SSC scenario. The studies with our dedicated SSC
grid-scan show that there is some degeneracy in both the one-zone and
the two-zone SSC scenarios probed, with several combinations of model
parameters yielding a similar model-to-data agreement, and some
parameters better constrained than others. The observed $\gamma$-ray
flaring activity, with the EVPA rotation coincident with the first
$\gamma$-ray flare, resembles those reported previously for low
frequency peaked blazars, hence suggesting that there are many similarities in the flaring mechanisms of blazars with different jet properties.}

%\end{abstract}

%Active Galactic Nuclei, blazars, Markarian~501, multi-wavelength, multi-instrument}

%\end{frontmatter}
\maketitle

%\linenumbers
%\input{introduction}

\section{Introduction}
\noindent
The BL Lac type object Markarian (Mrk)~501 is among the most prominent members of the class of blazars. Due to its brightness, almost the entire broadband spectral energy distribution (SED) of Mrk~501 can be measured accurately with current instrumentation. It is also known as one of the most active blazars, showing very strong and fast variability on timescales as short as a few minutes \citep{Albert:2007bt}. Moreover, because of its low redshift of $z=0.034$, even the multi-TeV $\gamma$-rays are influenced only weakly by the absorption on the extragalactic background light (EBL). Altogether, this makes Mrk~501 an excellent candidate source to study flux and spectral variability in the broadband emission of blazars.

%Together with its proximity at a redshift of $z=0.034$, this makes Mrk~501 the ideal candidate source to study the intrinsic properties of blazars: different flux states can be studied quite easily and measurements at the highest energies are influenced only weakly by absorption on the extragalactic background light (EBL).\\
Being the second extragalactic object to be detected in very high energy ($>$100~GeV, hereafter VHE) $\gamma$-rays \citep[][]{Quinn:1996jf,1997A&A...320L...5B}, Mrk~501 has been the subject of extensive studies in the different accessible energy bands within the last two decades. %The source's behaviour has been studied throughout this time, covering longer quiescient periods as well as strong outburst. 
Based on its SED, it has been classified as a high-frequency peaked BL Lac type source (HBL) according to \cite{1995ApJ...444..567P}, or 
high-synchrotron peaked BL Lac (HSP) if following the classification given in \cite{2010ApJ...716...30A}.

In 1997 %, shortly after its first detection at VHE $\gamma$-rays,
Mrk~501 was found to be in an exceptionally high state, with the emission at VHE energies being up to 10 times the flux of the Crab Nebula \citep{1997astro.ph.10118P,1999A&A...350...17D}. During this large flare, the synchrotron bump appeared to peak at or above 100\,keV, indicating a shift of the peak position compared to the quiescent state by at least two orders of magnitude \citep[][]{Catanese:1997iv,1998ApJ...492L..17P,1999A&A...347...30V,Tavecchio:2001hk}. During the following years, the source was intensively monitored at X-rays and VHE $\gamma$-rays \citep[e.g.][]{1999APh....11..149K,1999ApJ...518..693Q,Sambruna:2000dg,2001ApJ...546..898A,2004A&A...422..103M}, and additional studies were done with the collected data a posteriori \citep[e.g.][]{Gliozzi:2006it}. The observations could be well reproduced in the scope of one-zone synchrotron self-Compton (SSC) models. In 2005, the source showed another strong flaring event, for which flux-doubling times down to two minutes were measured at VHE \citep{Albert:2007bt}. This fast variability is a strong argument for a comparatively small emission region (with R$\approx10^{15}$\,cm), while the typical activity of the source could still be accommodated in models assuming a radius of the emission region which is larger by one to two orders of magnitude \citep[e.g.][]{2011ApJ...727..129A}. Throughout the observations, the SED at the highest energies appeared to be harder in higher flux states \citep[e.g.][]{Albert:2007bt}. Together with the observed shift of the synchrotron peak during the 1997 event, this suggests a change in the electron energy distribution as the cause for flaring events \citep[][]{1998ApJ...492L..17P}, but long-term changes in the Doppler factor or the size of the emission region are also a possibility being discussed \citep[][]{2012ApJ...753..154M}.

High-resolution radio images revealed a comparatively slow moving jet which features a limb brightening structure \citep[][]{2008ApJ...678...64P,2008A&A...488..905G,2009ApJ...690L..31P}. The radio core position of Mrk~501 has been found to be stationary within 2 parsec (pc), using  observations from the observing campaign in 2011 with the VLBI Exploration of Radio Astrometry \citep[VERA,][]{2015PASJ...67...67K}, although one cannot exclude variations in its location on year timescales. High resolution Global mm-VLBI Array (GMVA) observations at 86 GHz during the observing campaign in 2012 detected a new feature in the jet of Mrk~501, located 0.75 milliarcseconds (mas)  southeast of the radio core (which corresponds to $\sim$0.5 pc de-projected distance), and one order of magnitude dimmer  than the core \citep[][]{2015arXiv150504433K}. This radio feature is consistent with the one reported in \citet{2008A&A...488..905G} using GMVA data from 2005. This confirms that there are several distinct regions in the jet of Mrk~501, possibly stationary on year timescales, with the presence of high-energy electrons which could potentially produce optical, X-ray and $\gamma$-ray emission, in addition to the emission detected with these high-resolution radio instruments.
%Both of these observations, together with the fast variability seen in $\gamma$-rays, suggest models which identify different regions in the jet to be responsible for the emission in radio and $\gamma$-rays, such as the spine-layer-model \citep[][]{2005A&A...432..401G}.

Despite the fact that Mrk~501 has been studied over a comparatively long duration, clear constraints on the properties of the highest activity regions, as well as the particle populations involved, are still to be set. In this paper we present an extensive multi-instrument campaign on Mrk~501, which was conducted in 2009 in order to shed light on some of these open questions. This paper is a sequel to \cite{2011ApJ...727..129A}, where, among other things, the averaged broadband SED from the campaign was studied in detail. A study focussed on the flaring activity of May 1 (MJD~54952), 
 which includes very fast variability detected with the Whipple 10\,m telescope, VERITAS light curves and spectra, and  some measurements of the optical polarization performed by the Steward Observatory are reported in  \cite{Pichel:2011we} and \cite{PichelMrk501MW2009}.  In the work presented here, we address the variability seen during the full campaign, possible interband correlation of flux changes, and the characterization of the measured SED during two states of increased activity. While \cite{PichelMrk501MW2009} looks at the average X-ray spectrum for a low-state covering three weeks and a high state covering three days of the first VHE enhancement, we do a detailed investigation characterizing the X-ray spectra for each pointing available for the campaign, hence providing a better quantification of the X-ray spectral variability.  Furthermore, we consider an expanded data set, which also includes radio observations performed with the Very Long Baseline Array (VLBA), measuring the radio flux coming from the entire source and the radio flux from the compact core region only, and additional measurements of the optical polarization performed by the Steward and St. Petersburg observatories before and after the flaring activity of May 1. 

This paper is structured as follows: in Sect.~\ref{sec:campaign} an overview of the multi-instrument campaign is given, and updates with respect to the information provided in \cite{2011ApJ...727..129A} are discussed. In Sect.~\ref{sec:variability} the collected light curves and spectra are assessed for variability and interband correlation. The discussion of the broadband spectral energy distributions and a quantification of these measured spectra within synchrotron self-Compton scenarios by means of a novel technique based on a scan over the full parameter range is reported in Sect.~\ref{sec:sed}. Finally, the results are discussed in Sect.~\ref{sec:discussion} and a short summary and concluding remarks are given in Sect.~\ref{sec:summary}.

%\input{mwlcampaign}   
%\newpage

\section{Multi-instrument observing campaign performed in 2009}
\label{sec:campaign}
\noindent
The presented multiwavelength (MWL) campaign was conducted over 4.5 months in 2009. The aim of this campaign was to sample the SED over all wavelengths every $\sim$5 days. This way, the intrinsic flux variability of the source could be probed during non-flaring activity, hence reducing the observational bias towards states of high activity, which are the main focus of Target of Opportunity (ToO) campaigns. The covered frequency range spans from radio to VHE $\gamma$-rays, including data from $\sim$30 different instruments. The campaign took place from 2009 March 15 (MJD~54905) to 2009 August 1 (MJD~55044). Good coverage was achieved, while the sampling density varies among the different wavelengths because of different duty cycles and observational constraints of the participating instruments. The individual datasets and the data reduction are described in detail in Tab.~1 and Sect.~5 of  \cite{2011ApJ...727..129A}, and hence will not be reported again in this paper. In this section we only briefly mention the various observations performed, and report on the updates of some data analyses, as well as about extended datasets.

In the radio band, several single-dish instruments took part in the measurements, namely the Effelsberg 100-m radio telescope, the 32-m Medicina radio telescope, the 14-m Mets\"{a}hovi radio telescope, the 32-m Noto radio telescope, the Owens Valley Radio Observatory (OVRO) 40-m telescope, the 26-m University of Michigan Radio Astronomy Observatory (UMRAO) and the 600\,m ring radio telescope RATAN-600. The mm-interferometer Submillimeter Array (SMA) and the Very Long Baseline Array (VLBA) also participated in the campaign. These single-dishes and the SMA monitored the total flux of Mrk~501 as a point-like unresolved source at frequencies between 2.6~GHz and 225\,GHz. The VLBA took data ranging from 5\,GHz to 43\,GHz through various programs (BP143, BK150 and MOJAVE). Due to the better angular resolution of the latter, in addition to the total flux of the source, measurements of the flux from the compact ($\sim$10$^{-3}$pc) core region of the jet could be obtained through 2D Gaussian fits to the observed data.

Observations in optical frequency ranges have been performed by numerous instruments distributed all over the globe. In the R band, the Abastumani, Lulin, Roque de los Muchachos (Kungliga Vetenskaplika Academy, KVA), St.~Petersburg, Talmassons, and Valle d’Aosta observatories performed observations as part of GASP-WEBT, the GLAST-AGILE Support Program of the Whole Earth Blazar Telescope  \citep[e.g.][]{2008A&A...481L..79V,2009A&A...504L...9V}. 
Additional data with several optical filters were provided by the Goddard Robotic Telescope (GRT), the Remote Observatory for Variable Object Research (ROVOR) and the Multicolor Imaging Telescopes for Survey and Monstrous Explosions (MITSuME). At near-infrared wavelengths, measurements performed by the Guillermo Haro Observatory (OAGH) have been included in the data set. %The 2.1\,m telescope at the Guillermo Haro Observatory (OAGH) and the University of Wyoming 2.3-meter Telescope (WIRO) performed measurements at near-infrared wavelengths. 
Also within the GASP-WEBT program, the Campo Imperatore took measurements in near-infrared frequencies (JHK bands). The data obtained in the optical and near-infrared regime used the calibration stars reported in \cite{1998A&AS..130..305V}, and have been corrected for Galactic extinction following \cite{1998ApJ...500..525S}.

Through various observing proposals related to this extensive MWL campaign, 29 pointing observations were performed with the {\it Rossi-X-ray Timing Explorer} ({\it RXTE}), and 44 pointing observations performed with the \textit{Swift} satellite\footnote{Several \textit{Swift}  observations took place thanks to a ToO proposal which concentrates on the states of increased activity of the source.}. These observations provided coverage in the  ultraviolet frequencies with  the \textit{Swift} Ultraviolet/Optical Telescope (UVOT), and in the X-ray regime with the {\it RXTE} Proportional Counter Array (PCA) and the \textit{Swift}  X-ray Telescope (XRT).  \textit{Swift}/XRT performed 41 snapshot observations in Windowed Timing (WT) mode throughout the whole campaign,  and three observations in Photon Counting (PC) mode around MJD 54952. The PC observations had not been used in \cite{2011ApJ...727..129A}.  For PC mode data, events for the spectral analysis were selected within a circle of 20 pixel ($\sim$46\,arcsec) radius, which encloses about 80\% of the point spread function (PSF),
centered on the source position. The source count rate was
above $\sim$5\,counts $s^{-1}$ and data were significantly affected by pile-up in the inner part of the PSF. After comparing the observed PSF profile with the analytical model derived by \cite{2005SPIE.5898..360M}, pile-up effects were removed by excluding events within a 4 pixel radius circle centred on the source position, and an outer radius of 30 pixels was used.  Occasionally, during the first $\sim$100 seconds of a WT mode observation, Swift-XRT data will display a deviation in the light curve that is not due to the source variability, but is instead due to the settling of the spacecraft pointing causing a hot column to come in and out of either the source or background region. We inspected these data for any such deviations that could significantly impact our analysis, and none were found.

While Mrk~501 can be significantly detected with XRT and PCA for each single observation ($\sim$0.3 hours), integration times of $\sim$30 days are required in order to obtain significant detections with the \textit{RXTE} All-Sky Monitor (ASM) and the \textit{Swift} Burst Alert Telescope (BAT). The advantage of ``all-sky instruments'' like \textit{RXTE}/ASM and \textit{Swift}/BAT is that they can observe Mrk~501 without specifically pointing to the source, and hence provide a more uniform and continuous coverage than pointed instruments like \textit{Swift}/XRT and \textit{RXTE}/PCA. Details on the analysis of the \textit{RXTE}/ASM and \textit{Swift}/BAT data were given in \cite{2011ApJ...727..129A}.

The range of high-energy $\gamma$-rays was covered with the \textit{Fermi} Large Area Telescope (\textit{Fermi}-LAT). As is the case with \textit{RXTE}/ASM and \textit{Swift}/BAT, the sensitivity of \textit{Fermi}-LAT to detect Mrk~501 is quite moderate and one typically needs to integrate over $\sim$15--30 days in order to have significant detections, but provide a more uniform temporal coverage than the pointing instruments.  Besides the observations from the coordinated MWL campaign, here we also report on the X-ray/$\gamma$-ray activity of Mrk~501 measured with \textit{RXTE}/ASM, \textit{Swift}/BAT and \textit{Fermi}-LAT for a time interval spanning from MJD~54800 to MJD~55100, 
%MJD~54683 to MJD~55162
which exceeds the time span of the campaign.

The  \textit{Fermi}-LAT  data were re-analyzed using the Pass~8 SOURCE class events, and the ScienceTools software\footnote{http://fermi.gsfc.nasa.gov/ssc/data/analysis/software/} package version v10r1p1. 
We used all events (from MJD~54800 to MJD~55100) with energies from 200 MeV to 300 GeV and within a 10$^{\circ}$ Region of Interest (RoI) centred at the position of Mrk~501.  In order to avoid contamination from the Earth limb {$\gamma$-rays}, only events with zenith angle below 100$^{\circ}$ were  used. We used the \emph{P8R2\_SOURCE\_V6} instrument response functions, and the \emph{gll\_iem\_v06} and \emph{iso\_P8R2\_SOURCE\_V6\_v06} models to parameterize the Galactic and extragalactic diffuse emission \citep{2016ApJS..223...26A}\footnote{{\scriptsize \url{http://fermi.gsfc.nasa.gov/ssc/data/access/lat/BackgroundModels.html}}}. 
Given the fact that Mrk~501 is a relatively hard source, we only use events above 300 MeV  for the spectral analysis, as was done in \cite{2011ApJ...727..129A}.
%reported in section \ref{sec:SpectralVariability}, 
All point sources in the third \textit{Fermi}-LAT source catalog \citep[3FGL,][]{2015ApJS..218...23A} located in
the 10$^{\circ}$ RoI and an additional surrounding 5$^{\circ}$ wide annulus (called ``source region'')
were modeled in the fits, with the spectral parameters set to the values from the 3FGL, and the normalization parameters kept free only for the nine sources identified as variable (in the 3FGL) and located within 10$^{\circ}$ of Mrk~501. The normalization parameters for the two diffuse components were also kept free.  The spectral analysis performed on 15 and 30-day time intervals from MJD~54800 to MJD~55100 led to spectra successfully described by a power-law (PL) function with an index compatible\footnote{The power-law index light curve can be fitted with a constant, yielding average power-law indices of 1.75$\pm$0.03 and 1.76$\pm$0.03, respectively for the 15 and 30-day time intervals.}  with $\Gamma=1.75$.  For the determination of the light curves in the two energy bands 0.2--2\,GeV and $>$2\,GeV that are reported in Sect.~\ref{sec:LCs}, we decided to fix the value of the PL index to $\Gamma=1.75$.

%% ONLY COMMENTED OUT FOR TESTING
%In the VHE $\gamma$-ray band, three instruments observed Mrk~501 as part of the MWL campaign: the Major Atmospheric Gamma-ray Imaging Cherenkov \citep[MAGIC,][]{2004NewAR..48..339L} telescopes, the Very Energetic Radiation Imaging Telescope Array System \citep[VERITAS,][]{2008AIPC.1085..657H} and the Whipple 10\,m telescope \citep[][]{2007APh....28..182K}. Due to the requirement of good weather and night sky background conditions together with a comparatively small number of instruments, the coverage in VHE $\gamma$-rays is naturally more sparse than for the other energy bands.

MAGIC observations were carried out with a single telescope, as the second telescope was under construction during the campaign period. Due to a scheduled upgrade, no data were taken with MAGIC between MJD~54948 and MJD~54960. All observations were carried out in ``wobble'' mode \citep[][]{wobble}. For the work presented here, the data underwent a revised quality check and were re-analyzed with an improved analysis pipeline, with respect to the one presented in \cite{2011ApJ...727..129A}. Compared to the analysis presented in the first publication, the data set has been expanded by several nights (MJD~54937, 54941, 54944, 54945, 54973, 54975, 55035, 55038). Three nights were rejected because of revised quality criteria (MJD~54919, 54977, 55026). After all data selection and analysis cuts, the effective observation time covered by the data comprises 17.4 hours, while the first analysis yielded 16.2 hours of selected data.

VERITAS observed Mrk~501 with different telescope configurations over the duration of this campaign. The data presented here amounts to 9.7 hours of effective time, and are identical to those presented in \cite{2011ApJ...727..129A}. However the work in this paper presents the VERITAS light curve for the first time. 
%
%
%VERITAS observed Mrk~501 with different setups over the duration of this campaign \citep[for details see][]{2011ApJ...727..129A}. %Most of the data were taken with all four telescopes or with only three telescopes due to an ongoing upgrade of the array. In two nights, operations were conducted with only two telescopes because of technical problems. 
%All data were taken in ``wobble'' mode% \citep[][]{wobble}
%. The data which remain after quality selection and analysis cuts amount to 9.7 hours.

The Whipple 10\,m telescope observed Mrk~501 for 120 hours throughout the campaign, separately from the VERITAS array. The data taken with the Whipple 10\,m have not been used in the first publication which focused on the average state of the source throughout the campaign \citep[][]{2011ApJ...727..129A}.  However, the Whipple 10\,m data over a flaring period around May 1 have been recently reported in a separate paper \citep{PichelMrk501MW2009}.
For better comparison to the other VHE instruments, Whipple 10\,m fluxes, originally computed as flux in Crab Units (C.U.) above 400 GeV, were converted into fluxes above 300 GeV using the Crab flux above 300 GeV of  F$_{>300\,\mathrm{GeV}}=1.2$ x $10^{-10}$ cm$^{-2}$ s$^{-1}$ \citep{2012APh....35..435A}.

For more details on the observation strategy, list of instruments and analysis procedures performed for the different instruments, the reader is referred to \cite{2011ApJ...727..129A} and references therein.

In addition to the MWL observations conducted as part of the campaign,
the data set was expanded with measurements of the optical
polarization performed by the Steward (Bok telescope) and
St. Petersburg (LX-200) observatories from February to September 2009.
The LX-200 polarization measurements were obtained from R-band imaging
polarimetry, while the measurements from the Steward Observatory were
derived from spectropolarimetry  between 4000 and 7500$\r{A}$ with a
resolution of $\sim$15$\r{A}$, and the reported values are constructed
from the median Q/I and U/I in the 5000–7000 $\r{A}$ band. 
The effective wavelength of this bandpass is similar to the Kron-Cousins R band, and the wavelength dependence in the polarization of Mrk~501 seen in the spectro-polarimetry is small and does not significantly affect the results. The details related to the observations and analysis of the polarization data is reported in \cite{2008A&A...492..389L} and \cite{2009arXiv0912.3621S}.  The Steward observations are part of the public Steward Observatory program to monitor $\gamma$-ray-bright blazars during the \textit{Fermi}-LAT mission\footnote{
\url{http://james.as.arizona.edu/~psmith/Fermi}}, and a fraction of these polarisation observations have been recently reported in \cite{PichelMrk501MW2009}.

%\input{variability}   
%\newpage

\vspace{-0.5cm}

\section{Multi-instrument flux and spectral variability }
\label{sec:variability}
\noindent
During the 4.5 month long MWL campaign, Mrk~501 was observed by numerous instruments covering the entire broadband SED. In the following section, we report the measured multiband flux and spectral variability, as well as multiband correlations. 

\vspace{-0.5cm}

\subsection{Multi-instrument light curves}
\label{sec:LCs}

\vspace{-0.2cm}

\begin{figure*}[!ht]
\centering
\includegraphics[width=1.0\textwidth]{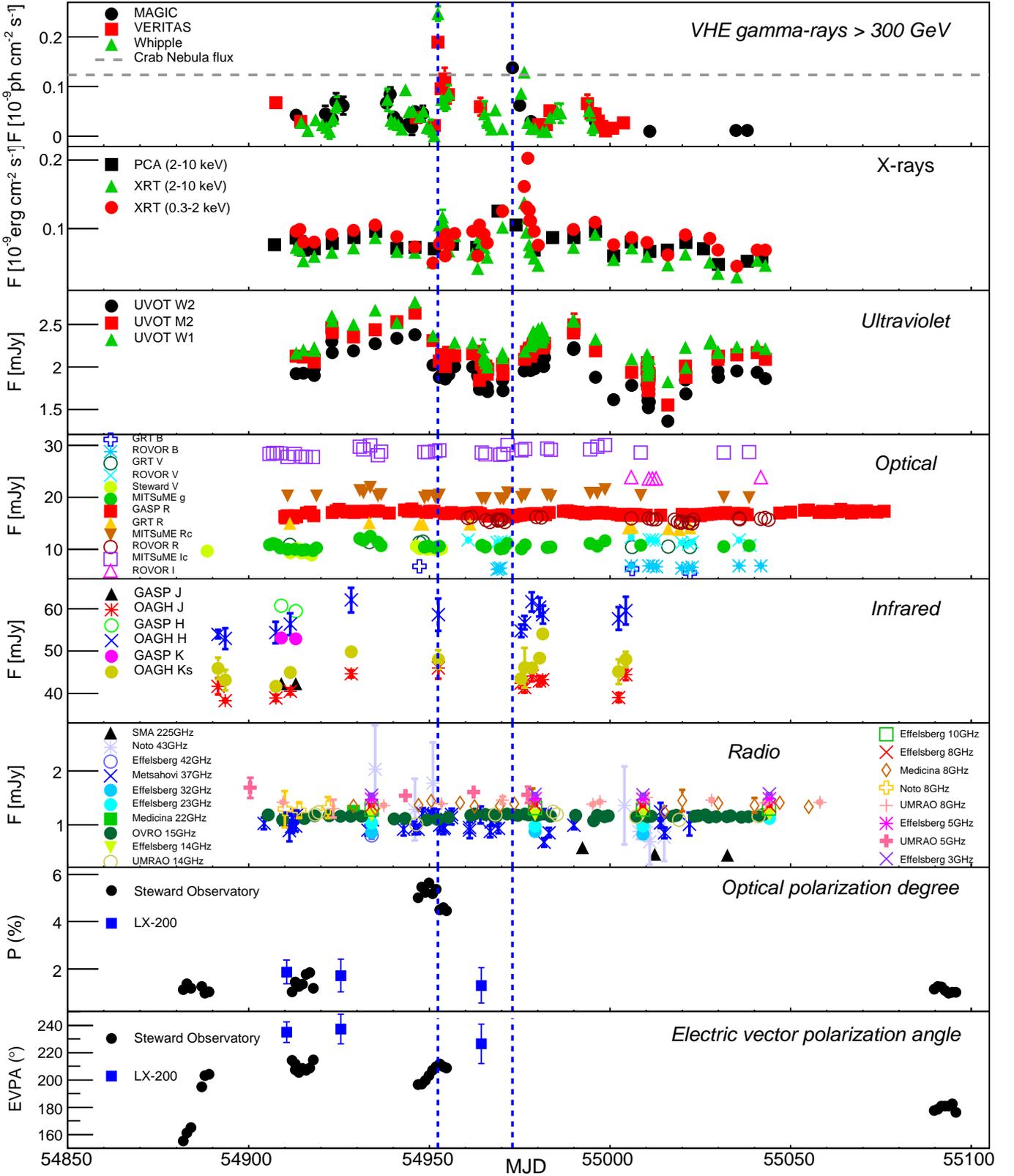}
\caption{Light curves compiled based on pointing observations in various energy bands. The lowest two panels show measurements of the optical polarization. The two vertical blue lines indicate the location of the two VHE $\gamma$-ray flares at MJD~54952 and MJD~54973 that are discussed in Sect.~\ref{sec:sed1} and \ref{sec:sed2}. \label{fig:mwllcs}}
\end{figure*}

\begin{figure*}
	\centering
		\includegraphics[width=17cm]{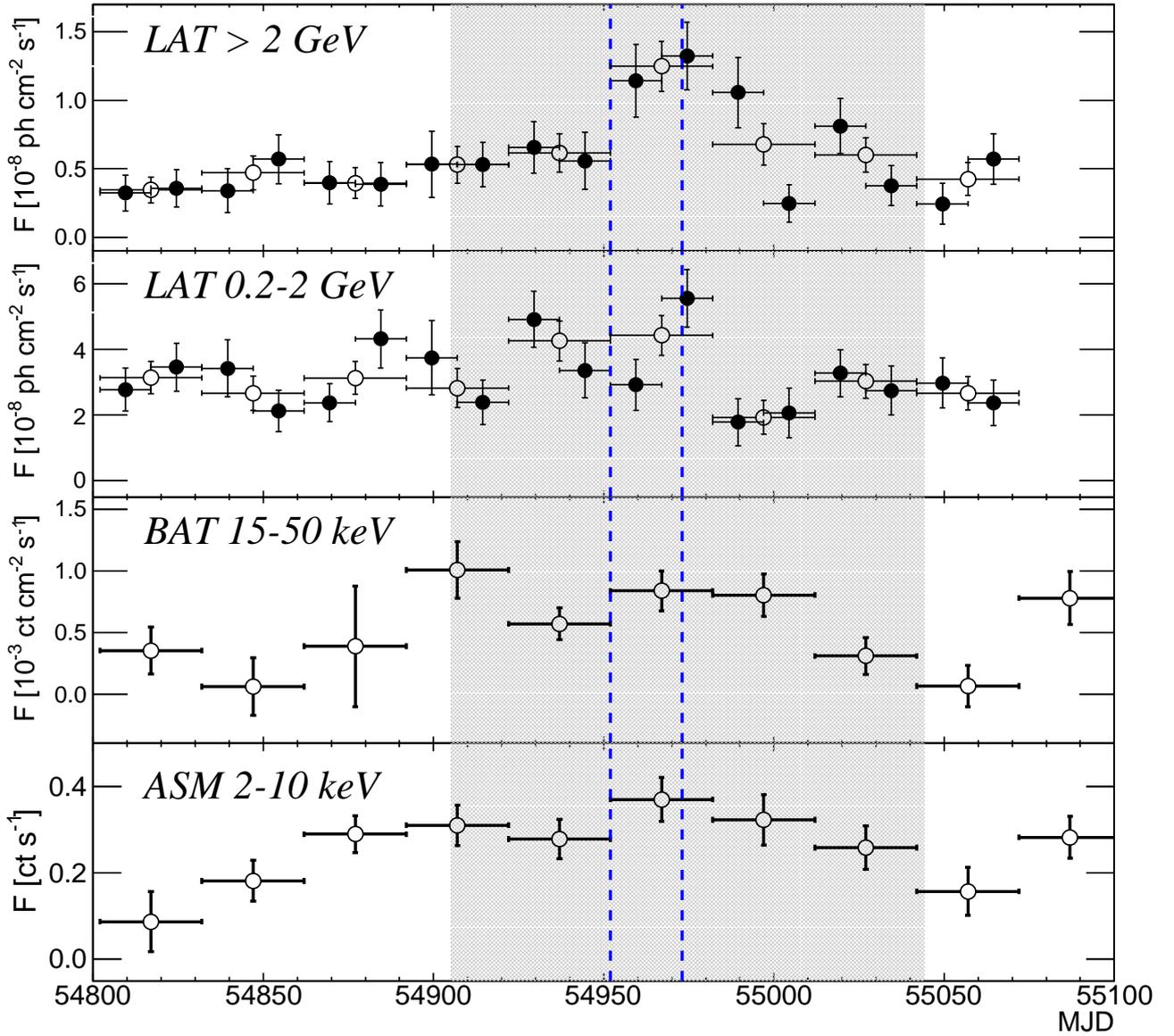}
		\caption{Light curves of instruments with longer
                  integration times. From top to bottom: \textit{Fermi}-LAT  above 2\,GeV, \textit{Fermi}-LAT 0.2-2\,GeV, \textit{Swift}/BAT and \textit{RXTE}/ASM. Flux points with integration times of 30 days are shown as open markers, while for \textit{Fermi}-LAT also flux points integrated over 15 days have been derived and are added with filled markers. The grey shaded area depicts the time interval related to the multi-instrument campaign. The two vertical blue lines indicate the location of the two VHE $\gamma$-ray flares at MJD~54952 and MJD~54973  that are discussed in Sect. \ref{sec:sed1} and \ref{sec:sed2}. }
	  \label{fig:30daylcs}
\end{figure*}

The light curves which were derived from pointed observations in the different energy bands, spanning from radio to VHE $\gamma$-rays, are shown in Fig.~\ref{fig:mwllcs}. 
Fig. \ref{fig:30daylcs} reports on the X-ray and $\gamma$-ray activity as measured with the all-sky instruments \textit{RXTE}/ASM, \textit{Swift}/BAT and \textit{Fermi}-LAT. 

The light curves obtained during pointing observations in the radio regime exhibit a nearly constant flux at a level of $\sim$ 1.2\,Jy. The well-sampled light curve taken with the OVRO telescope shows constant emission of 1.158$\pm$0.003 Jy. 

\begin{figure*}
	\centering
		\includegraphics[width=17cm]{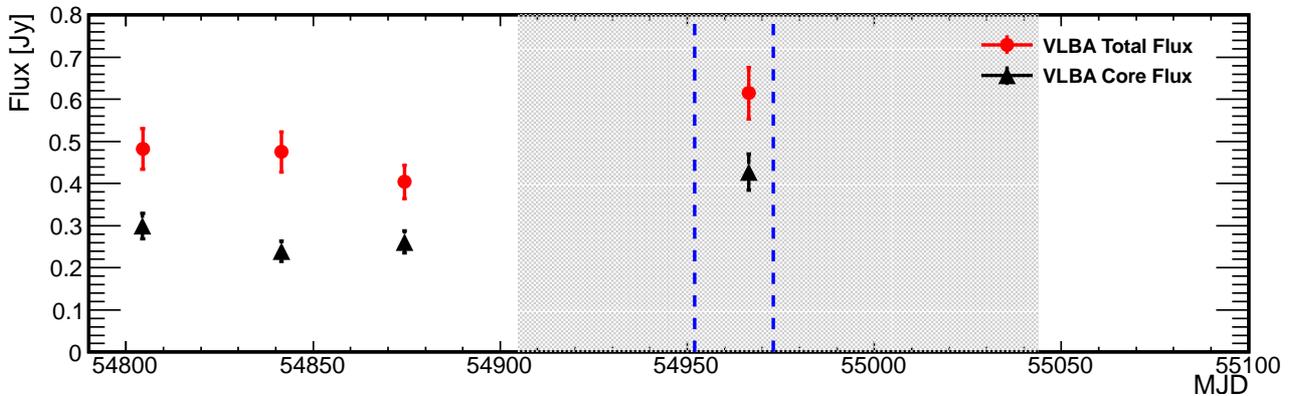}
		\caption{Light curves obtained with the VLBA at 43\,GHz. The total flux and the flux from the core region are shown. The grey shaded area depicts the time interval related to the multi-instrument campaign. The two vertical blue lines indicate the location of the two VHE $\gamma$-ray flares at MJD~54952 and MJD~54973 that are discussed in Sect. \ref{sec:sed1} and \ref{sec:sed2}. }
	  \label{fig:vlbalc}
\end{figure*}

The measurements performed with the VLBA at a frequency of 43\,GHz are presented in Fig.~\ref{fig:vlbalc}. A constant fit delivers a reduced $\chi^2$ of 8.4/3 for the total flux and 15.6/3 for the core flux, yielding a probability for the data points to be well described by a constant fit of 3.8\% and 0.14\%, respectively. Despite being marginally significant, this suggests an increase in the radio flux in May 2009 (dominated by the core emission), in comparison to that measured during the other months.

For the near-infrared observations in Fig.~\ref{fig:mwllcs}, flux levels of $\sim$40-50\,mJy (J and K band) and $\sim$50-60\,mJy (H band) have been measured. Only small variations can be seen, even though the sampling  is less dense and the uncertainties of the measurements are comparatively large. 
For the extensive data sample in the optical regime, a nearly constant flux was measured, at flux levels of $\sim6$\,mJy (B band), 11\,mJy (V), $\sim$16\,mJy (R) and 24 - 29\,mJy (I / Ic). 
No correction for emission by the host galaxy has been applied. 
At ultraviolet frequencies, a flux level of $\sim$2\,mJy with flux variations of about 25\% over timescales of about 25 to 40 days can be seen.

The average \textit{Swift}/XRT measured fluxes during the entire
campaign are $\mathrm{F}_{0.3-2\,\mathrm{keV}}
=(9.2\pm0.3)\cdot10^{-11} $\,ergs cm$^{-2}$s$^{-1}$ in the energy
range between 0.3 and 2\,keV and  $\mathrm{F}_{2-10\,\mathrm{keV}} =
(7.2\pm0.3)\cdot10^{-11}$\,ergs cm$^{-2}$s$^{-1}$ in the range
2-10\,keV, while \textit{RXTE}/PCA, due to a slightly different
temporal coverage, measured an average \mbox{2--10\,keV} flux of
$\mathrm{F}_{2-10\,\mathrm{keV}}= (7.8\pm0.2)\cdot10^{-11}$\,ergs
cm$^{-2}$s$^{-1}$.

The \textit{Fermi}-LAT measured a variable flux in the two probed $\gamma$-ray bands, with an average flux of $\mathrm{F}_{0.2-2\,\mathrm{GeV}}=(2.75\pm0.14)\cdot10^{-8}$\,ph cm$^{-2}$\,s$^{-1}$ between 200\,MeV and 2\,GeV and  $\mathrm{F}_{>2\,\mathrm{GeV}}=(5.3\pm0.4)\cdot10^{-9}$\,ph cm$^{-2}$\,s$^{-1}$ at energies above 2\,GeV (shown in Fig.~\ref{fig:30daylcs}). The highest emission is seen in the 15-day time interval between MJD~54967 and MJD~54982.

The VHE $\gamma$-ray light curves are shown in the upper panel of Fig.~\ref{fig:mwllcs}. The average flux above 300~GeV of Mrk~501 during the campaign, including the flaring activities, is about $5\cdot10^{-11}$\,ph cm$^{-2}$\,s$^{-1}$ ($\sim$0.4 C.U.)\footnote{The average fluxes measured with MAGIC, VERITAS and Whipple during the observing campaign are somewhat different because of the distinct temporal coverage of these instruments. The average VHE flux with MAGIC is $\mathrm{F}_{>300\,\mathrm{GeV}}=(4.6\pm0.4)\cdot10^{-11}$\,ph cm$^{-2}$\,s$^{-1}$, with VERITAS is $\mathrm{F}_{>300\,\mathrm{GeV}}=(5.3\pm0.7)\cdot10^{-11}$\,ph cm$^{-2}$\,s$^{-1}$ and the one with Whipple is $\mathrm{F}_{>300\,\mathrm{GeV}}=(4.4\pm0.5)\cdot10^{-11}$\,ph cm$^{-2}$\,s$^{-1}$. }.  Flux variability is evident throughout the VHE light curve, in addition to two few-day long flaring episodes occurring in MJD~54952 (2009 May 1) and MJD~54973 (2009 May 22). 

In the following paragraphs we review the first VHE flare in a MWL context, and include additional details specifically on the X-ray data. We then provide details on the second VHE flare.

\subsubsection{VHE $\gamma$-ray flaring event starting at MJD~54952}
\label{subsec:firstflarevariability}
On 2009 May 1, the Whipple 10\,m telescope observed Mrk~501 for 2.3 hours and, in the first 0.5 hours (from MJD~54952.35 to MJD~54952.37), detected a VHE flux ($>$300~GeV) increase from $\sim$1.0--1.5 C.U. to $\sim$4.5 C.U., which implies a flux increase by about one order of magnitude with respect to the average VHE flux level recorded during the full campaign.  Following the alert by the Whipple 10\,m, VERITAS started to observe Mrk~501 after 1.4 hours (at MJD~54952.41) and detected the source  at a VHE flux of 1.5\,C.U. without statistically significant flux variations during the full observation (from MJD~54952.41 to MJD~54952.48). This VHE flux level was also measured by the Whipple 10\,m telescope during approximately the same time window (from MJD~54952.41 to MJD~54952.47), and corresponds to a VHE flux $\sim$4 times larger than the typical flux level of 0.4 C.U. measured during the full campaign.  The peak of the flare (which occurred at MJD~54952.37) was caught only by the Whipple 10\,m. Still, 
the Mrk~501 VHE $\gamma$-ray flux remained high for the rest of the night and the following 2 nights (until MJD~54955), which was measured by VERITAS and the Whipple 10\,m with very good agreement. Further details about the VERITAS and Whipple 10\,m intra-night variability measured on 2009 May 1, as well as the enhanced activity during the first days of May, can be found in \citet{Pichel:2011we} and \citet{PichelMrk501MW2009}.

During the period of the considered VHE $\gamma$-ray flare, no substantial increase in the X-ray regime can be claimed based on the \textit{Swift}/XRT observations:  the 0.3--2~keV and the 2--10~keV fluxes during this flaring episode are about $\sim 8\cdot10^{-11}$\,ergs cm$^{-2}$s$^{-1}$ and $\sim 1\cdot10^{-10}$\,ergs cm$^{-2}$s$^{-1}$, which are about $\sim$10\% lower and $\sim$30\% higher than the average X-ray flux values reported above.  However, the \textit{Swift}/XRT observations started seven hours after the Whipple 10\,m and VERITAS observations of this very high VHE state on MJD~54952. The reason for this is that the XRT observations were taken within a ToO activated by the enhanced VHE activity measured by the Whipple 10\,m and VERITAS, unlike most of the X-ray pre-planned observations from the MWL campaign which were coordinated with the VHE observations.

In the two lowest panels in Fig.~\ref{fig:mwllcs}, the evolution of the optical polarization degree and orientation are shown. The degree of polarization during the few days around the first VHE flaring activity is measured at 5\% compared to a 1\% measurement during several observations before and after this flaring activity. Additionally, there is also a rotation of the EVPA by 15 degrees, which comes to a halt at the time of the VHE outburst, when the degree of polarization drops from 5.4\% to 4.5\% \citep[see further details in][]{Pichel:2011we,PichelMrk501MW2009}.

\subsubsection{VHE $\gamma$-ray flaring event starting at MJD~54973} 
\label{subsec:secondflarevariability}
The MAGIC telescope observed Mrk~501 for 1.7 hours on 2009 May 22 (MJD~54973) and measured a flux of 1.2\,C.U., which corresponds to $\sim$ 3 times the low flux level. At the next observation on May 24 (MJD~54975.00 to MJD~54975.12), the flux had already decreased to a level of $\sim$0.5\,C.U. The Whipple 10\,m observed Mrk~501 later on the same date (from MJD~54975.25) and measured a flux of $\sim$0.7\,C.U., while the following day (from MJD~54976.23) an measured a flux increase to 1.1\,C.U. No VERITAS observations of Mrk~501 took place at this time due to scheduled telescope maintenance.

The MAGIC data of the flaring night were probed for variations on timescales down to minutes, but no significant intra-night variability was found. Moreover, tests for spectral variability within the night in terms of hardness ratios vs.~time in different energy bands showed no significant variations either.

Unfortunately, there are no X-ray observations which are strictly simultaneous with the MAGIC observations on MJD~54973. The closest \textit{RXTE}/PCA observations took place on MJD~56969 and MJD~54974, and the closest \textit{Swift}/XRT observations are from   MJD~54970 and MJD~54976, all of them showing a flux increase (up to a factor of $\sim$2) with respect to the average X-ray flux measured during the campaign.

Under the assumption that no unobserved intra-day variability occurred in the X-ray band, it can be inferred that Mrk~501 was in a state of increased X-ray and VHE activity over a period of up to 5 days. During this period there were no flux changes observed at optical or radio frequencies.

\subsection{Spectral variability in individual energy bands}
\label{sec:SpectralVariability}
In this section we report on the spectral variability observed during the two few-day long VHE flaring episodes around the peaks of the two SED bumps, namely at X-ray and $\gamma$-rays, where most of the energy is being emitted and where the flux variability is highest. 

\subsubsection{VHE $\gamma$-rays}
\label{subsec:vhespectra}
The VHE spectra measured with MAGIC and \mbox{VERITAS}, averaged over the entire campaign between 2009 March 15 ((MJD~54905) and 2009 August 1 ((MJD 55044), were reported in \cite{2011ApJ...727..129A}. Only the time span MJD~54952-54955, where VERITAS recorded VHE flaring activity, had been excluded for the average spectrum and was presented as a separate high-state spectrum \citep[see Fig.~8 of ][]{2011ApJ...727..129A}. The resulting average spectra relate to a VHE flux of about 0.3 C.U., which is the typical non-flaring VHE flux level of Mrk~501. 
%From the excluded time window, a high state VERITAS spectrum was calculated, which was also reported in \cite{2011ApJ...727..129A}. 
Additionally, two spectra have been obtained with the Whipple 10\,m for that night: a very-high state spectrum, spanning MJD~54952.35-54952.41, which seems to cover the peak of the flare, and a high-state spectrum, derived from the time interval MJD 54952.41 - 54955.00, which is simultaneous with the observations performed with VERITAS. These spectra were reported in \cite{Pichel:2011we}, following the general Whipple analysis technique described in \citet{2007ApJ...655..396H}, and further details from these spectra are reported in \citet{PichelMrk501MW2009}.

\begin{figure*}
\centering
		\includegraphics[width=15%17
cm]{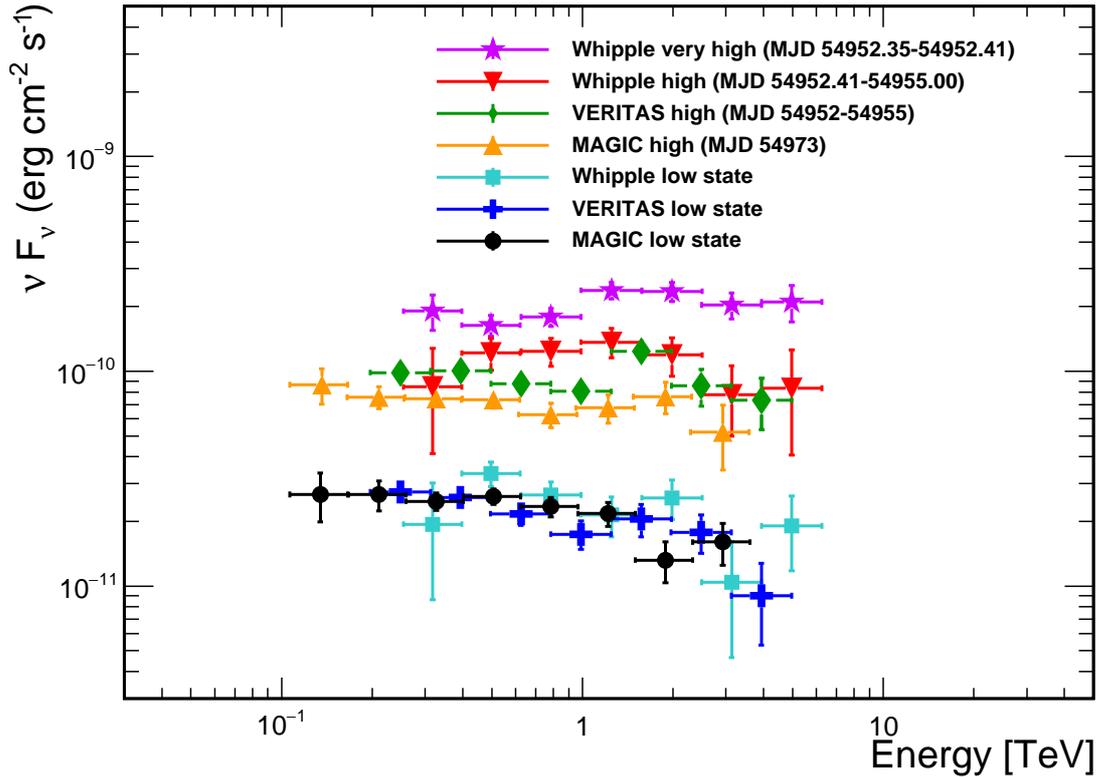}
		\caption{Spectral energy distributions measured by MAGIC, VERITAS and the Whipple 10\,m during the low state of the source and two states of increased VHE flux. The spectra have been corrected for EBL absorption using the model of \cite{Franceschini:2008vt}.}
	  \label{fig:vhespectra}
\end{figure*}

\begin{table*} [htbp]
\begin{center}
\caption{Fit parameters and goodness of fit describing the power-law function for the measured VHE $\gamma$-ray spectra. %In case of MAGIC spectra, the correlation between the individual spectral points, which is introduced in the unfolding of the energy spectrum, is taken into account during the fitting procedure. 
For low state spectra, the stated flaring time intervals have been excluded from the data. Spectral fits for the Whipple 10\,m and VERITAS are listed as presented in \cite{Pichel:2011we}.}
\label{tab:spectralfits}
\vspace{5mm}
	\begin{tabular}{ccc|ccc}
instrument & flux state & MJD& $F_0$ $[10^{-7}$ ph m$^{-2}$ s$^{-1}$ TeV$^{-1}]$ & $\Gamma$ & $\chi^2$ / ndf \\
\hline
Whipple &very-high &54952.35 - 54952.41& $16.1 \pm 0.4$ & $2.10 \pm 0.05$ & 13.5/8 \\
Whipple &high &54952.41 - 54955& $5.6 \pm 0.4$ & $2.31 \pm 0.11$ & 3.1/8 \\
VERITAS &high &54952.41 - 54955& $4.17 \pm 0.24$ & $2.26 \pm 0.06 $& 6.3/5 \\
\hline
MAGIC &high &54973&$3.1 \pm 0.2$&$2.28 \pm 0.06$&1.9/6\\
\hline
Whipple &low&54936 - 54951 & $1.16 \pm 0.09$ & $2.61 \pm 0.11$ & 3.4/8 \\
VERITAS &low &54907 - 55004& $0.88 \pm 0.06$ & $2.48 \pm 0.07$ & 3.8/5 \\
MAGIC &low &54913 - 55038&$0.93 \pm 0.04$&$2.40 \pm 0.05$& 8.4/6 \\
\end{tabular}
\end{center}
\end{table*}

The re-analysis of the MAGIC data (see Sect.~\ref{sec:campaign}), which contains some additional data compared to the analysis presented in \cite{2011ApJ...727..129A},  revealed a flaring state on MJD~54973, for which a dedicated spectrum was computed. An averaged spectrum was derived based on the remaining data set. The energy distribution of the differential photon flux can be well described by a power-law (PL) function of the form:
\begin{equation}
\label{eq:spectrum}
\frac{dN}{dE}=F_0 \times \left(E / 1 \mathrm{TeV}\right)^{-\Gamma},% m^{-2} s^{-1} TeV^{-1}.
\end{equation}
yielding $F_0=\left(9.3 \pm 0.4\right) \times 10^{-8} \mathrm{ph}$ $ \mathrm{m}^{-2}$ $ \mathrm{s}^{-1}$ $ \mathrm{TeV}^{-1}$ and $\Gamma=2.40 \pm 0.05$. This new MAGIC averaged spectrum was found to be in agreement with the previously presented one, where a power-law fit gave $F_0 = \left( 9.0 \pm 0.5\right) \times 10^{-8} \mathrm{ph}$ $ \mathrm{m}^{-2}$ $ \mathrm{s}^{-1}$ $ \mathrm{TeV}^{-1}$ and $\Gamma = 2.51 \pm 0.05$ \citep{2011ApJ...727..129A}. Here we only quote statistical uncertainties of the measurements. The systematic errors affecting data taken by the MAGIC telescope at the time of the presented campaign are discussed in \cite{2008ApJ...674.1037A} and are valid for both analyses. They are estimated as an energy scale error of 16\%, a systematic error on the flux normalization of 11\% and an error on the obtained spectral slope of $\pm 0.2$.
In the following, the more recent analysis result will be used.

All the VHE $\gamma$-ray spectra described above are presented in Fig.~\ref{fig:vhespectra}. The spectra displayed in the figure were corrected for absorption by the EBL using the model from \cite{Franceschini:2008vt}. Given the proximity of Mrk~501, the impact of the EBL on the spectrum is relatively weak: the attenuation of the flux reaches 50\% at an energy of 5\,TeV. Many other EBL models
\citep[e.g.][]{2010ApJ...712..238F,2011MNRAS.410.2556D}  provide compatible results at energies below 5 TeV, hence the results do not depend significantly on the EBL model used.
The power-law fit parameters (see Eq.~\ref{eq:spectrum}) of the measured spectra (i.e.  non-corrected-for-EBL spectra)  can be found in Table~\ref{tab:spectralfits}. For spectra measured with MAGIC, the presented fits also take into account the correlation between the individual spectral points which is introduced by the unfolding of the spectrum, while no explicit unfolding has been applied for the other instruments. %Please also note the differences in the chosen units compared to the representation in Fig.~\ref{fig:vhespectra}. 
The average state spectra measured by the three instruments (after subtracting the time intervals with strong flaring activity in the VHE) agree very well, despite the somewhat different observing periods. This suggests that these VHE spectra are a good representation of the typical VHE spectrum of Mrk~501 during this MWL campaign.  The high-state spectra show a spectral slope which is harder compared to the one from the non-flaring state, hence indicating a ``harder when brighter'' behaviour, as has been reported previously \citep[e.g.][]{Albert:2007bt}.

\subsubsection{GeV $\gamma$-rays}

The two few-day long VHE flaring episodes discussed in this paper occurred within the time interval MJD~54952-54982, which is the 30-day time interval with the highest flux and hardest GeV $\gamma$-ray spectrum reported in \cite{2011ApJ...727..129A}.  The flux above 300 MeV F$_{>300\mathrm{MeV}}$ and photon index $\Gamma$ for this 30-day time interval computed using the ScienceTools software package version v9r15p6 and the \emph{P6\_V3\_DIFFUSE} instrument response functions, are \mbox{F$_{>300\mathrm{MeV}}=(3.6\pm$0.5)$\times$10$^{-8}$ph cm$^{-2}$s$^{-1}$} and \mbox{$\Gamma=1.64\pm$0.09}, while values for the {\it Fermi}-LAT spectrum averaged for the entire MWL campaign are \mbox{F$_{>300\mathrm{MeV}}=(2.8\pm$0.2)$\times$10$^{-8}$ph cm$^{-2}$s$^{-1}$} and \mbox{$\Gamma=1.74\pm$0.05} \citep[for further details, see ][]{2011ApJ...727..129A}. Performing the analysis with the ScienceTools software package version v10r1p1 and the Pass 8 data (which implies somewhat different photon candidate events), as described in Sect.~\ref{sec:campaign}, led to a photon flux (above 300 MeV) of \mbox{F$_{>300\mathrm{MeV}}=(4.2\pm$0.5)$\times$10$^{-8}$ph cm$^{-2}$s$^{-1}$} and a PL index of \mbox{$\Gamma=1.68\pm$0.07} for the  time interval MJD~54952-54982, and a flux (above 300 MeV) of \mbox{F$_{>300\mathrm{MeV}}=(3.0\pm$0.2)$\times$10$^{-8}$ph cm$^{-2}$s$^{-1}$} and a PL index of \mbox{$\Gamma=1.75\pm$0.04} for the entire campaign. 
%Again here, only statistical uncertainties are given. The systematic uncertainties for measurements with the \textit{Fermi}-LAT are energy-dependent, but can be estimated to be around $5-10$\,\%. 
The spectral results derived with Pass 6 and Pass 8 are compatible, and show a marginal increase in the flux and the hardness of the spectra during the time interval MJD~54952-54982 with respect to the full campaign period.

The Pass 8 {\em Fermi}-LAT data analysis is more sensitive than the Pass 6 data analysis, and allows us to detect Mrk~501 significantly (TS$>$25)\footnote{``TS'' stands for test statistic from the maximum likelihood fit. A TS value of 25  corresponds to an estimated $\sim$4.6$\sigma$ \citep{1996ApJ...461..396M}.} and to determine the spectra around these two flares in time intervals as short as 2 days centered at MJD 54952 and 54973, for the two flares respectively. Additionally, for comparison purposes, we also computed the spectra for 7-day time intervals centered at MJD 54952 and 54973\footnote{A one week period is a natural time interval that, for instance, is also used in the LAT public light curves \url{http://fermi.gsfc.nasa.gov/ssc/data/access/lat/msl_lc/}. The spectral results would not change if we had used a 5-day or 10-day time interval. }. The \textit{Fermi}-LAT spectral results for the various time intervals in May 2009 are reported in Table~\ref{tab:LATspectralfits}. For the first flare, for both the 2-day and 7-day time intervals, the LAT analysis yields a signal with TS$\sim$40. This shows that increasing the time interval from 2 days to 7 days did not increase the $\gamma$-ray signal, and hence indicates that the 2-day time interval centred at MJD~54952 dominates the $\gamma$-ray signal from the 7-day time interval.   The spectrum is marginally harder than the average spectrum from the time interval MJD~54952-54982. For the second flare, the 7-day time interval yields a signal significance ($\sim\sqrt{TS}$) 2.6 times larger than that of the 2-day time interval, showing that, contrary to the first flare, increasing the time interval from 2 days to 7 days enhanced the $\gamma$-ray signal considerably.  The {\em Fermi}-LAT spectrum around the second flare is very similar to the average spectrum obtained for the 30-day time interval MJD~54952-54982.

\begin{table*} [htbp]
\begin{center}
\caption{Spectral parameters describing the measured power-law spectra with {\it Fermi}-LAT during several temporal intervals in May 2009.}
\label{tab:LATspectralfits}
\vspace{5mm}
	\begin{tabular}{cc|ccc}
Temporal interval & MJD range & $F_{>300~{\rm MeV}}$ $[10^{-8}$ ph m$^{-2}$ s$^{-1}]$ & $\Gamma$ & TS \\
\hline
May 2009, 30 days  &54952 - 54982& $4.2 \pm 0.5$ & $1.68 \pm 0.07$ & 595 \\
\hline
First Flare, 2 days &54951 - 54953& $2.5 \pm 1.3$ & $1.2 \pm 0.3$ & 43 \\
First Flare, 7 days &54948.5 - 54955.5& $1.7 \pm 0.8$ & $1.4 \pm 0.2$ & 41 \\
\hline
Second Flare, 2 days &54972 - 54974& $4.0 \pm 1.7$ & $1.8 \pm 0.3$ & 39 \\
Second Flare, 7 days &54969.5 - 54976.5& $5.3 \pm 1.0$ & $1.6 \pm 0.1$ & 263 \\
\end{tabular}
\end{center}
\end{table*}

For the MWL SEDs presented in Fig.~\ref{Fig:mrk501flareseds}, we show the {\em Fermi}-LAT spectral results for these two flares performed on three and five differential energy bins (starting from 300 MeV). Here, the shape of the spectrum was fixed to that obtained for the full range for each temporal bin. Upper limits at 95\% confidence level were computed whenever the TS value (for the $\gamma$-ray signal of the bin) was below six and/or the uncertainty was equal (or larger) than the energy flux value.

\subsubsection{X-rays}
\label{subsec:xrayspec}
\begin{figure*}
\begin{minipage}[b]{0.5\linewidth}
	\begin{center}
\includegraphics[width=\textwidth]{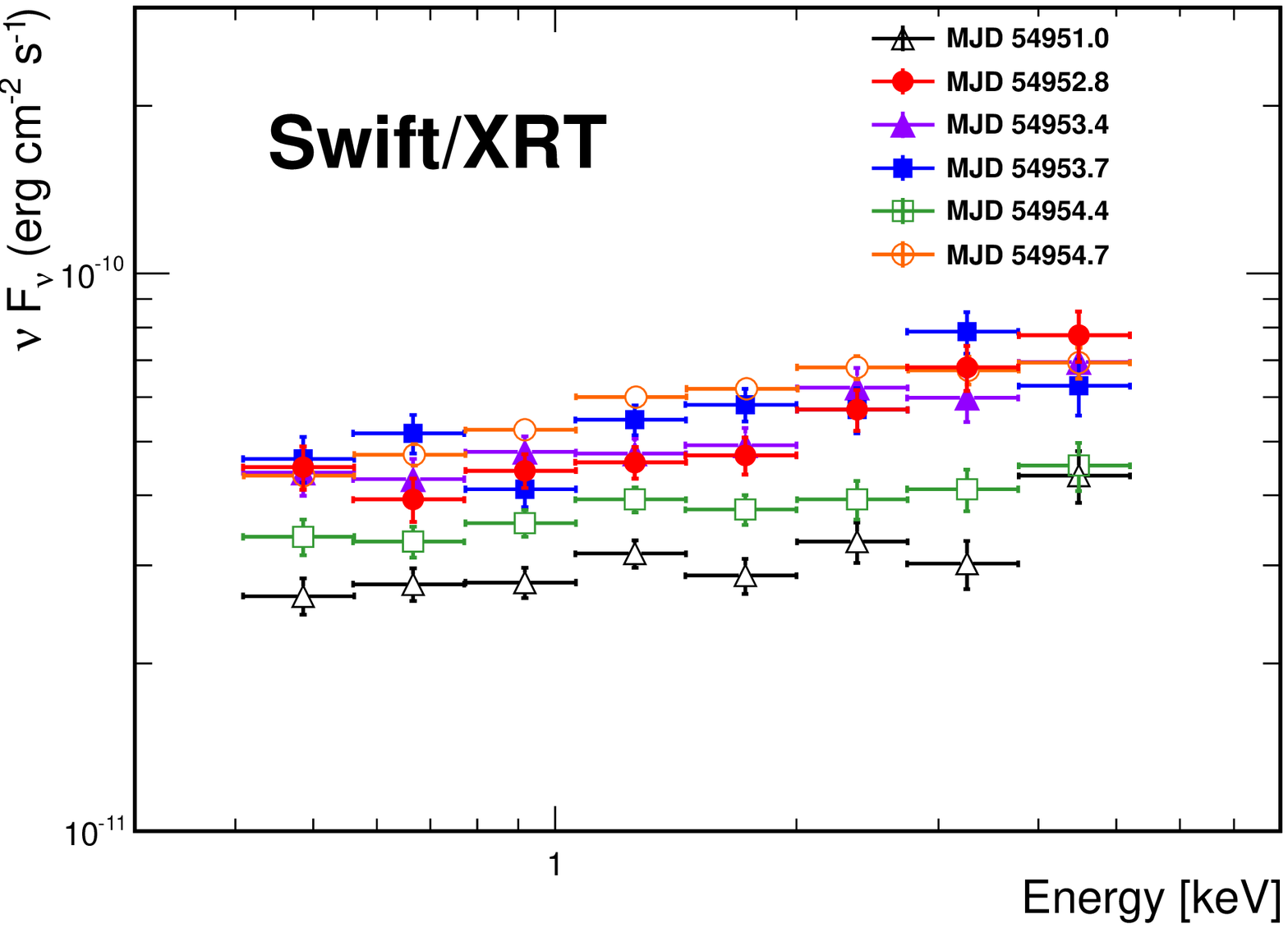}
\includegraphics[width=\textwidth]{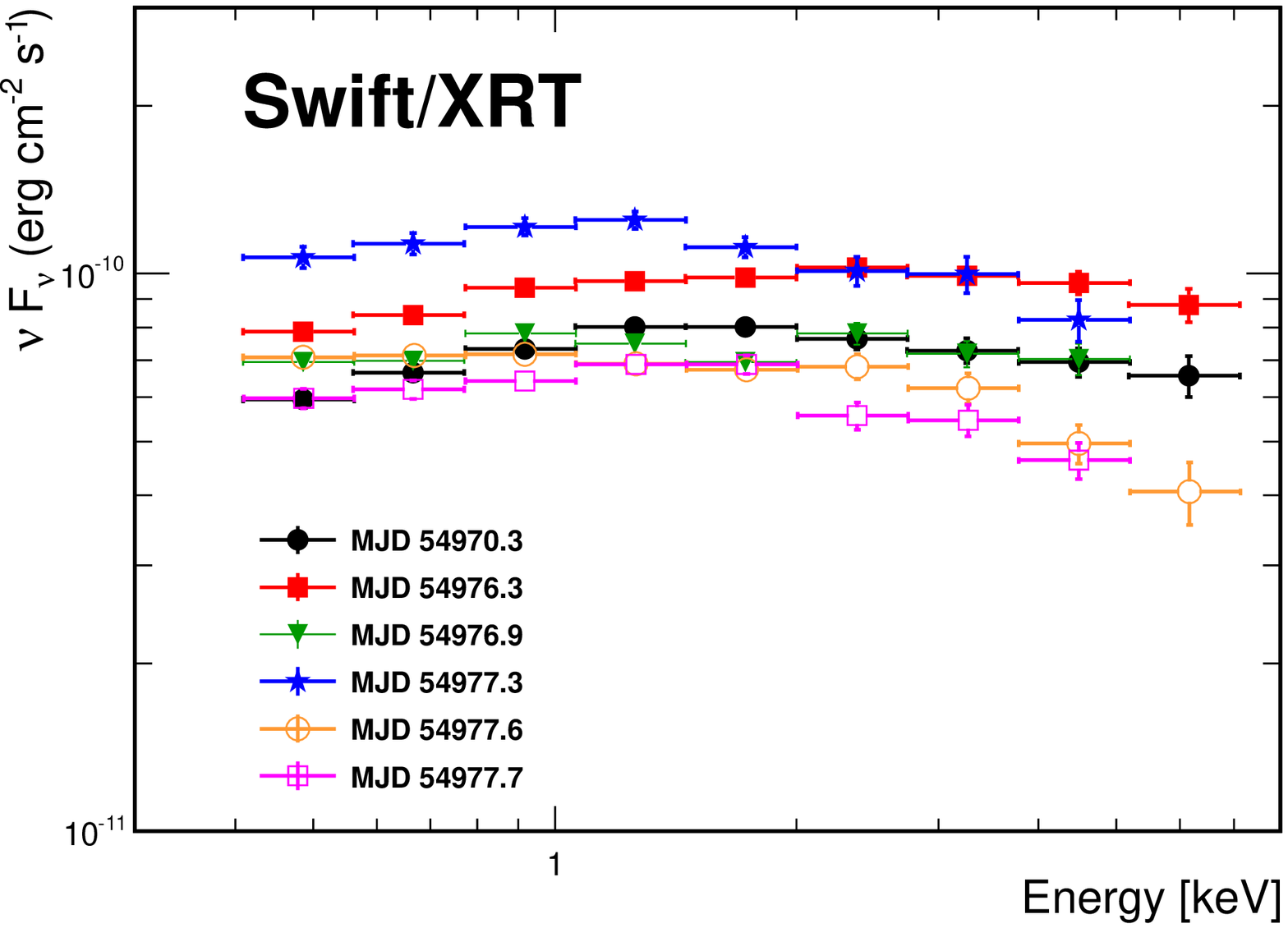}
\end{center}
\end{minipage}
\begin{minipage}[b]{0.5\linewidth}
	\begin{center}
\includegraphics[width=\textwidth]{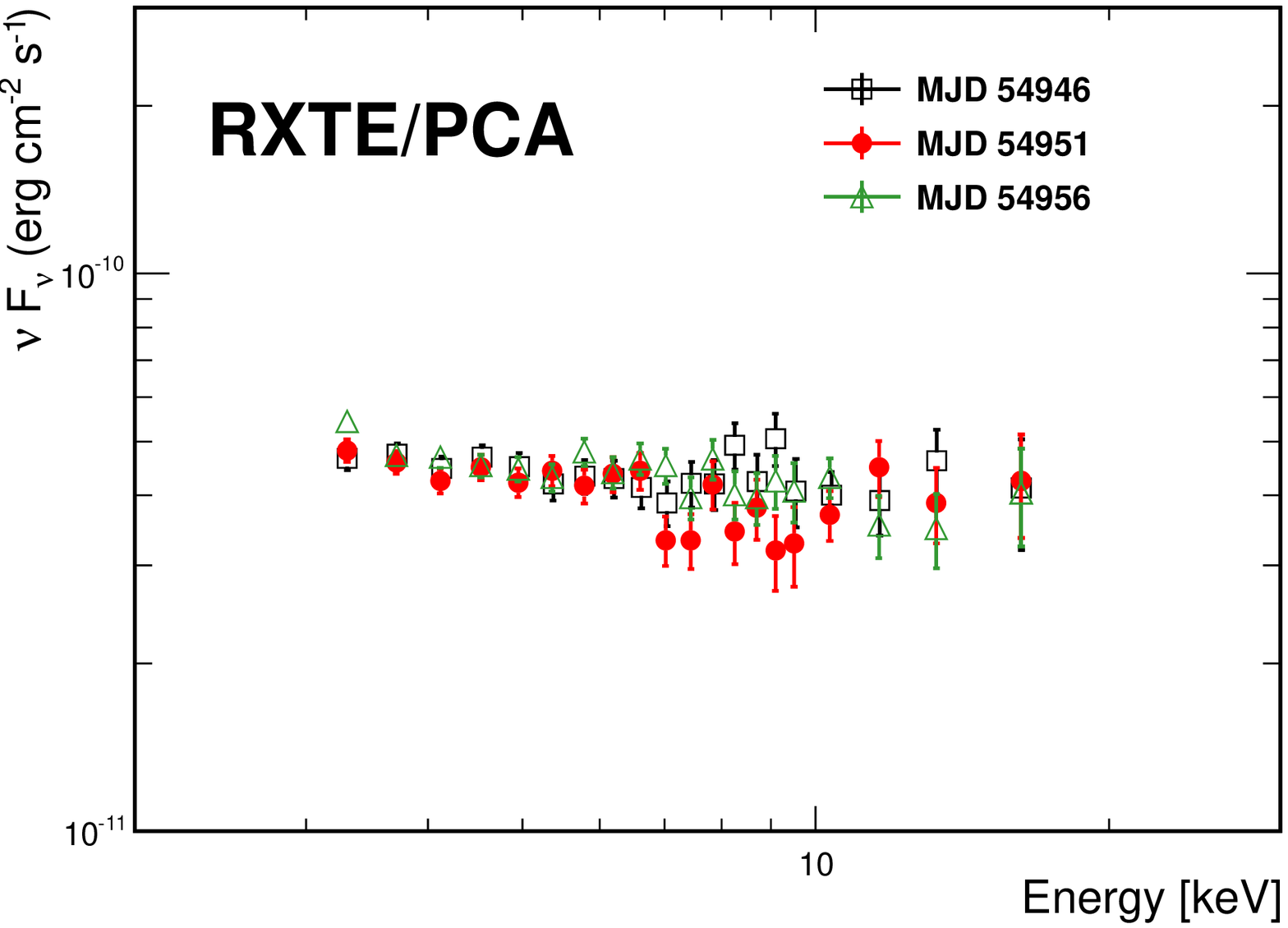}
\includegraphics[width=\textwidth]{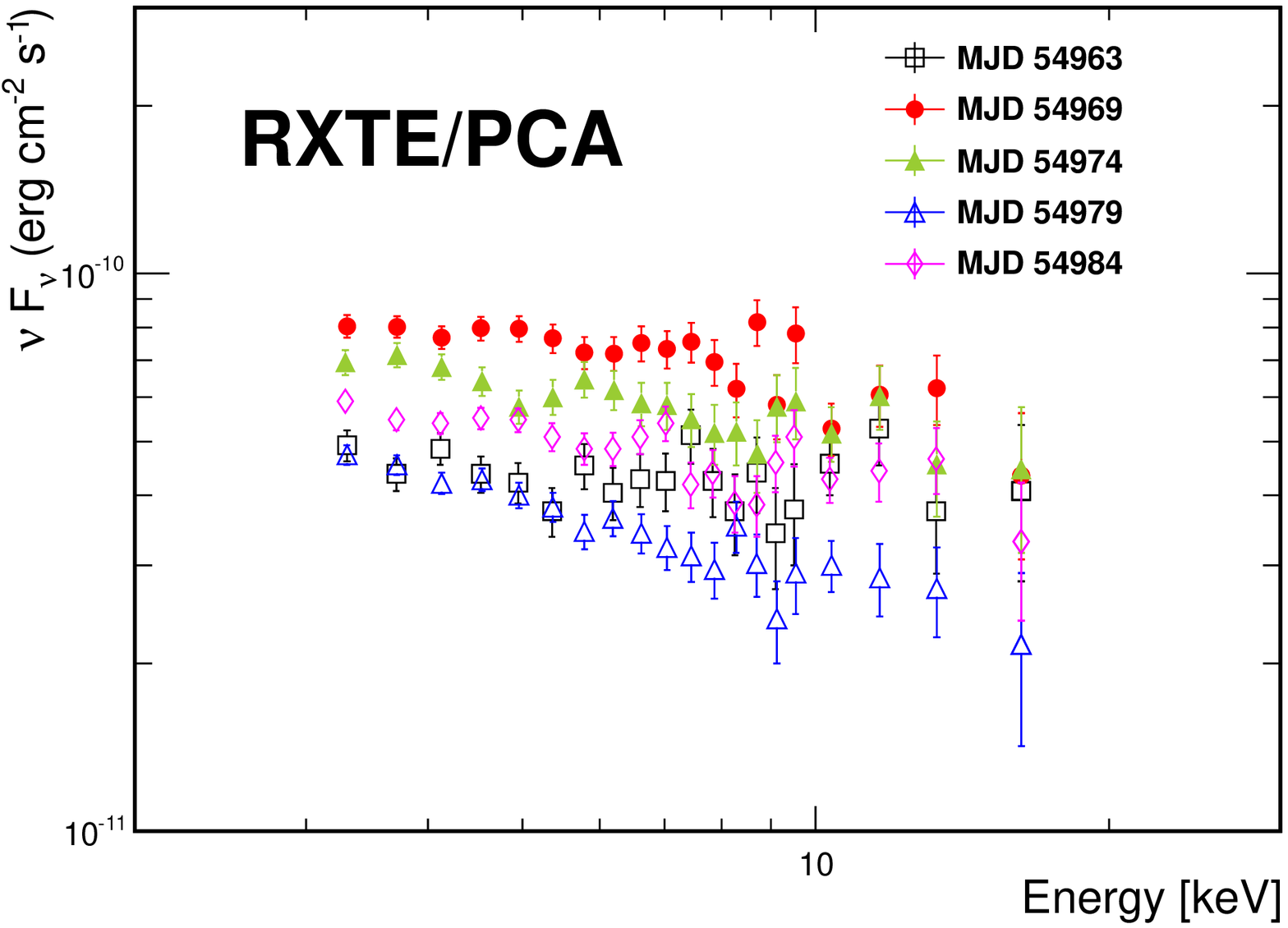}
	\end{center}
\end{minipage}
		\caption{X-ray spectra from single pointings. Left: \textit{Swift}/XRT. Right: \textit{RXTE}/PCA. Upper panels: spectra around the first flare (MJD~54952); lower panels: spectra around the second flare (MJD~54973).}
	  \label{fig:xrayspectra}
\end{figure*}

In the X-ray band, individual spectra could be derived for each pointing of the two instruments \textit{Swift}/XRT and \textit{RXTE}/PCA. Both indicated  significant variability in flux and spectral index during the course of the campaign. Fig.~\ref{fig:xrayspectra}  shows the XRT and PCA spectra around the times of the first and second flux increase in the VHE range. For the first flare, the variability in flux and spectral shape is larger for XRT than for PCA, but this is mostly due to the fact that many of the XRT observations were performed within a ToO program, and hence they provide a better characterisation of the enhanced activity (see Sect.~\ref{sec:campaign}).

\begin{figure*}
\sidecaption
\centering
\includegraphics[width=12cm]{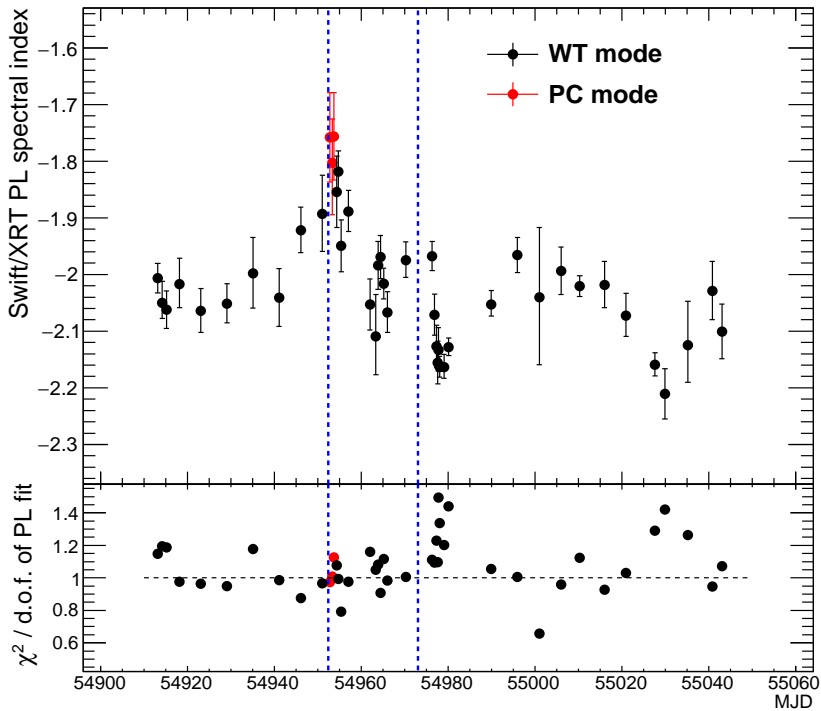}
\caption{Upper panel: Spectral index obtained from a power-law fit to the \textit{Swift}/XRT spectra vs. observation date. Lower panel: Reduced $\chi^2$ of the power-law fit to the X-ray spectra. The two vertical blue lines indicate the location of the two VHE $\gamma$-ray flares at MJD~54952 and MJD~54973 that are discussed in Sect. \ref{sec:sed1} and \ref{sec:sed2}. 
}
	  \label{fig:swiftindex}
\end{figure*}

\begin{table*} [htbp]
\begin{center}
\caption{Spectral results from the power-law (PL) fit to the measured \textit{Swift}/XRT spectra. For all spectra where the PL fit does not deliver a satisfactory result (fit probability $P<0.3 \% \ (3\,\sigma)$), additional results from a log-parabola fit are quoted in the following:\hspace{\textwidth} 
MJD~54977.7: $\alpha=2.01_{ -0.06}^{+0.05}$, $\beta=0.4_{-0.1}^{+0.1}$, $\chi^2$/d.o.f.$=214/163;\quad$
MJD~54978.0: $\alpha=2.05_{-0.03}^{+0.03}$, $\beta=0.32_{-0.06}^{+0.06}$, $\chi^2$/d.o.f.$=335/316;$\hspace{\textwidth} 
MJD~54980.1: $\alpha=2.03_{-0.02}^{+0.02}$, $\beta=0.312_{-0.05}^{+0.05}$, $\chi^2$/d.o.f.$=373/344;\quad$ 
MJD~55027.6: $\alpha=2.04_{-0.03}^{+0.03}$,$\beta=0.33_{-0.06}^{+0.06}$, $\chi^2$/d.o.f.$=293/291.$\hspace{\textwidth} 
}
\label{tab:spectralfitsxray}
\vspace{5mm}
\begin{tabular}{cccc}
MJD & Obs Mode & PL Index & $\chi^2$/\#d.o.f.\\
\hline
54913.1 & WT &	 $-2.01_{-0.03}^{+0.03}$ & 246/214\\
54914.2 & WT &	 $-2.05_{-0.03}^{+0.04}$ & 246/206\\
54915.2 & WT &	 $-2.06_{-0.03}^{+0.03}$ & 200/168\\
54918.2 & WT &	 $-2.02_{-0.04}^{+0.05}$ & 130/133\\
54923.1 & WT &	 $-2.06_{-0.04}^{+0.04}$ & 160/166\\
54929.1 & WT &	 $-2.05_{-0.03}^{+0.03}$ & 179/189\\
54935.0 & WT &	 $-2.00_{-0.06}^{+0.06}$ & 88/75\\
54941.1 & WT &	 $-2.04_{-0.05}^{+0.05}$ & 111/113\\
54946.1 & WT &	 $-1.92_{-0.04}^{+0.04}$ & 139/159\\
54951.0 & WT &	 $-1.89_{-0.07}^{+0.07}$ & 62/64\\
54952.8 & PC &	 $-1.76_{-0.08}^{+0.08}$ & 56/58\\
54953.4 & PC &	 $-1.80_{-0.09}^{+0.08}$ & 57/56\\
54953.7 & PC &	 $-1.76_{-0.08}^{+0.08}$ & 66/59\\
54954.4 & WT &	 $-1.85_{-0.06}^{+0.06}$ & 84/78\\
54954.7 & WT &	 $-1.82_{-0.04}^{+0.04}$ & 165/166\\
54955.4 & WT &	 $-1.95_{-0.05}^{+0.05}$ & 100/126\\
54957.1 & WT &	 $-1.89_{-0.03}^{+0.04}$ & 163/167\\
54962.0 & WT &	 $-2.05_{-0.04}^{+0.05}$ & 125/108\\
54963.4 & WT &	 $-2.11_{-0.07}^{+0.07}$ & 79/75\\
54963.9 & WT &	 $-1.98_{-0.04}^{+0.04}$ & 140/129\\
54964.4 & WT &	 $-1.97_{-0.04}^{+0.04}$ & 141/155\\
54965.1 & WT &	 $-2.02_{-0.03}^{+0.03}$ & 269/241\\
\end{tabular}
\hspace{10mm}
\begin{tabular}{cccc}
MJD & Obs Mode & PL Index & $\chi^2$/\#d.o.f.\\
\hline
54966.0 & WT &	 $-2.07_{-0.04}^{+0.04}$ & 180/183\\
54970.2 & WT &	 $-1.97_{-0.03}^{+0.03}$ & 200/199\\
54976.3 & WT &	 $-1.97_{-0.03}^{+0.03}$ & 271/244\\
54976.9 & WT &	 $-2.07_{-0.04}^{+0.04}$ & 201/184\\
54977.3 & WT &	 $-2.13_{-0.04}^{+0.04}$ & 171/139\\
54977.6 & WT &	 $-2.16_{-0.04}^{+0.04}$ & 178/162\\
54977.7 & WT &	 $-2.13_{-0.04}^{+0.04}$ & 245/164\\
54978.0 & WT &	 $-2.16_{-0.02}^{+0.02}$ & 424/317\\
54979.0 & WT &	 $-2.16_{-0.02}^{+0.02}$ & 359/298\\
54980.1 & WT &	 $-2.13_{-0.01}^{+0.02}$ & 497/345\\
54989.9 & WT &	 $-2.05_{-0.02}^{+0.02}$ & 303/287\\
54995.9 & WT &	 $-1.97_{-0.03}^{+0.03}$ & 197/196\\
55001.0 & WT &	 $-2.04_{-0.10}^{+0.10}$ & 14/21\\
55006.0 & WT &	 $-1.99_{-0.04}^{+0.04}$ & 152/159\\
55010.3 & WT &	 $-2.02_{-0.02}^{+0.02}$ & 331/295\\
55015.9 & WT &	 $-2.02_{-0.04}^{+0.04}$ & 147/158\\
55020.9 & WT &	 $-2.07_{-0.04}^{+0.04}$ & 180/175\\
55027.6 & WT &	 $-2.16_{-0.02}^{+0.02}$ & 377/292\\
55029.9 & WT &	 $-2.21_{-0.04}^{+0.04}$ & 144/101\\
55035.2 & WT &	 $-2.12_{-0.07}^{+0.08}$ & 89/70\\
55040.8 & WT &	 $-2.03_{-0.05}^{+0.05}$ & 101/107\\
55043.0 & WT &	 $-2.10_{-0.05}^{+0.05}$ & 109/102\\
\end{tabular}
\end{center}
\end{table*}

Around the first VHE flare, the XRT spectra tend to be much harder and appear to show an upward curvature towards higher energies. The hardening of the spectrum is confirmed by a spectral analysis performed using a power-law spectral model with the hydrogen density $N_H$ fixed to the Galactic value. Fig.~\ref{fig:swiftindex} shows the 
spectral index light curve  (see also Table~\ref{tab:spectralfitsxray}) and the reduced $\chi^2$ of the individual fits. Based on the reduced $\chi^2$ values, the representation by a simple power-law function is sufficient for most spectra. 
Around MJD~54952-54953, which roughly corresponds to the time of the first VHE flare, a peak in the hardness of the spectrum can be seen.

Around the second flux increase in the VHE $\gamma$-ray band, variability was seen by both \textit{Swift}/XRT and \textit{RXTE}/PCA, with flux changes by up to a factor of 2 with respect to the  flux average of  $\sim$(7--8)$\cdot10^{-11}$\,ergs cm$^{-2}$s$^{-1}$ in the 2--10 keV band  (see Fig.~\ref{fig:mwllcs}).
However, no particular hardening of the spectrum was found (see Fig.\ref{fig:swiftindex}), as observed for the first flare.

\subsection{Quantification of the multi-instrument variability}

As a quantitative study of the underlying variability seen at different wave-bands, the fractional variability $F_{\text{var}}$ has been determined for each instrument according to Eq.~10 in \cite{Vaughan:2003iu}:
\begin{equation}
F_{\text{var}}=\sqrt{\frac{S^2 - < \sigma^2_{\text{err}} > }{<F_{\gamma}>^2}},
\end{equation}
where $S^2$ represents the variance, $< \sigma^2_{\text{err}} >$ specifies the mean square error stemming from measurement uncertainties and ${<F_{\gamma}>}$ is the arithmetic mean of the measured flux. The term under the square root is also known as the normalised excess variance $\sigma^2_{\text{NXS}}$.

The uncertainty of $F_{\text{var}}$ is calculated following the prescription from \citet{2008MNRAS.389.1427P}, as described in \citet{2015A&A...576A.126A}, so that they are valid also in the case when $\Delta F_{\text{var}} \ll F_{\text{var}}$: 

\begin{equation}
\Delta F_{\text{var}}=\sqrt{F^2_{\text{var}}+err(\sigma^2_{\text{NXS}})}-F_{\text{var}},
\end{equation}
with the error of the normalised excess variance $err(\sigma^2_{\text{NXS}})$ as defined in Eq.~11 in \cite{Vaughan:2003iu}. 

\begin{figure*}
	\centering
\includegraphics[width=17cm]{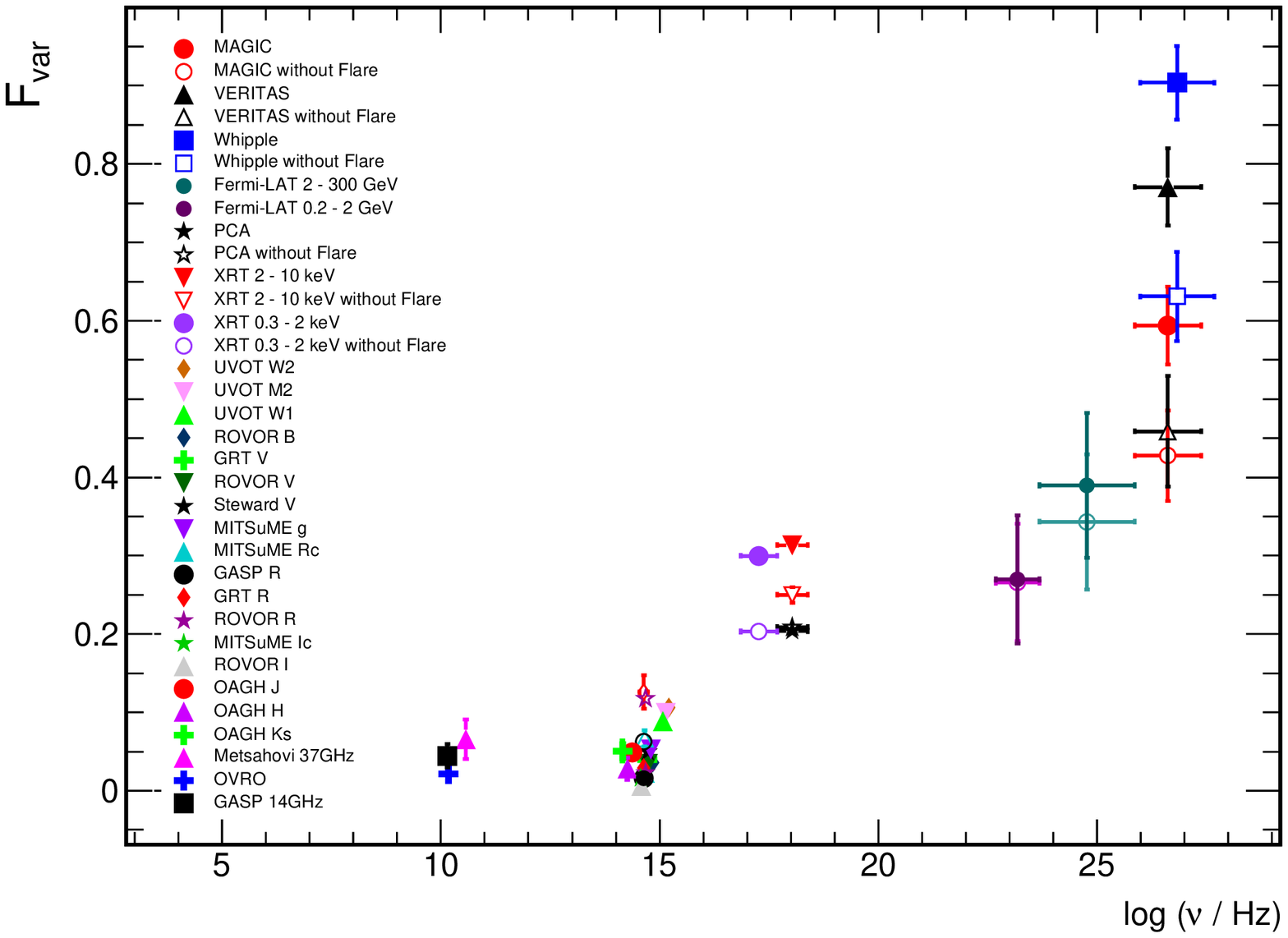}
		\caption{Fractional variability at different frequencies. All the $F_{\text{var}}$ values are computed with the single observations reported in Fig.~\ref{fig:mwllcs}, with the exception of the $F_{\text{var}}$ values related to \textit{Fermi}-LAT which were computed with 15-day and 30-day time intervals, and depicted with full circles and open light-coloured circles respectively. Open symbols for optical bands indicate the fractional variability after subtracting the host galaxy contribution, as determined in \cite{2007A&A...475..199N}. For the X-ray and the VHE $\gamma$-ray band, open markers depict the variability after removal of flaring episodes from the light curves as described in the text.}
	  \label{Fig:fvar}
\end{figure*}

This methodology to quantify the variability has the caveat that the resulting $F_{\text{var}}$ and related uncertainty  depend very much on instrument sensitivity and the observing sampling, which is different for the different energy bands.  In other words, a densely sampled light curve with small uncertainties in the flux measurements may allow us to see flux variations that are hidden otherwise, and hence may yield a larger $F_{\text{var}}$ and/or smaller uncertainties in the calculated values of $F_{\text{var}}$. Some practical issues in the application of this methodology in the context of multiwavelength campaigns are elaborated in \citet{2014A&A...572A.121A,2015A&A...573A..50A,2015A&A...576A.126A}.

For \textit{Swift}/XRT and \textit{RXTE}/PCA in the X-ray band, and MAGIC, VERITAS and the Whipple 10\,m  in the VHE regime, the fractional variability has been calculated for the full dataset and also after removal of the temporal intervals related to the two flaring episodes (MJD~54952-54955, MJD~54973-54978). The fractional variability specifically computed for the period around the first flaring episode has been recently reported in \citet{PichelMrk501MW2009}. For measurements in the optical R band, $F_{\text{var}}$ has additionally been calculated for optical fluxes corrected for the host galaxy emission as derived in \cite{2007A&A...475..199N}. For datasets containing fewer than five data points, no $F_{\text{var}}$ was calculated. The results are presented in Fig.~\ref{Fig:fvar}.

 A negative excess variance was obtained for datasets from the following instruments: UMRAO (at 5\,GHz and 8\,GHz), Noto (at 8\,GHz and 43\,GHz), Medicina (at 8\,GHz), Effelsberg (all bands) and the near-IR measurements within the GASP-WEBT program (all bands). Such a negative excess variance is interpreted as an absence of flux variability within the sensitivity range of the instrument. These datasets have not been included in Fig.~\ref{Fig:fvar}.

At low frequencies, from radio to optical, no substantial variability was detected, with $F_{\text{var}}$ ranging from $\approx 0.02-0.06$ %(formerly 5 \%$ 
in radio %($0.14$ for SMA, but based on only 3 data points) 
to $0.01-0.1$ %formerly $5-15\%$ 
in optical. %{\bf This was updated with the results after subtraction of the host galaxy. Valus are in the same range. DONE on 2014_08_27} 
In the X-ray band, we find $F_{\text{var}}\approx 0.3$, indicating
substantial variation in the flux during the probed time
interval. After removal of the flaring times, variabilities of
$F_{\text{var}}\approx 0.2-0.25$ are still seen. The fractional
variability in the $\gamma$-ray band covered by \textit{Fermi}-LAT is
of the order $F_{\text{var}}\approx 0.3-0.4$; yet the
\textit{Fermi}-LAT $F_{\text{var}}$ values are not directly comparable
to the other instruments, as GeV variability on day timescales, which
could be higher than that computed (separately) for the 15-day and the
30-day timescales, cannot be probed. 
Strong variability can be noted at VHE, with $F_{\text{var}}\geq 0.4$ for the datasets without the flares, and $F_{\text{var}}\geq 0.6$ (0.9 for Whipple 10\,m) for observations including the flaring episodes.

All in all, Mrk~501 showed a large increase in variability with increasing energy, ranging from an almost steady behaviour at the lowest frequencies to the highest variability observed in the VHE band.

\subsection{Multi-instrument correlations}

To study possible cross-correlations of flux changes between the different wavelengths, we determined the discrete correlation functions (DCF), following \cite{1988ApJ...333..646E}, based on the light curves obtained by the various instruments. The DCF allows a search for correlations with possible time lags, which could result e.g.~from a spatial separation of different emission regions. We probed time lags in steps of 5 days up to a maximum shift of 65 days. The step size corresponds to the overall sampling of the light curve and thus to the objective of the MWL campaign itself, which was to probe the source activity and spectral distribution every $\approx 5$ days. The maximum time span is governed by the duration of the campaign, as a good fraction of the light curve should be available for the calculation of cross-correlations. We chose a maximum of 65 days, which corresponds to roughly half the time span of the entire campaign. Because of the uneven sampling and varying exposure times, the significance of the correlations derived from the prescription given in \cite{1988ApJ...333..646E} might be overestimated \citep[][]{2003ApJ...584L..53U}. 
We derived an independent assessment of the significance of the correlation by means of dedicated Monte Carlo simulations as described in \citet{2009MNRAS3972004A}  and \citet{2015A&A...573A..50A, 2015A&A...576A.126A}.

In this study, possible cross-correlations between instruments of different wave-bands were examined. As already suggested by the low level of variability in the radio and optical band throughout the campaign, no correlations with any other wave-band were found for these instruments. A correlation with flux changes in the MeV-GeV range could not be probed on timescales of days due to the integration time of 15--30 days required by \textit{Fermi}-LAT for a significant detection. A similar situation occurs in the X-ray bands from \textit{Swift}/BAT and \textit{RXTE}/ASM, which also need integration times of the same order, and are thus also neglected for day-scale correlation studies.

\begin{figure*}[!ht]
\begin{minipage}[b]{0.5\linewidth}
	\begin{center}
		\includegraphics[width=3in]{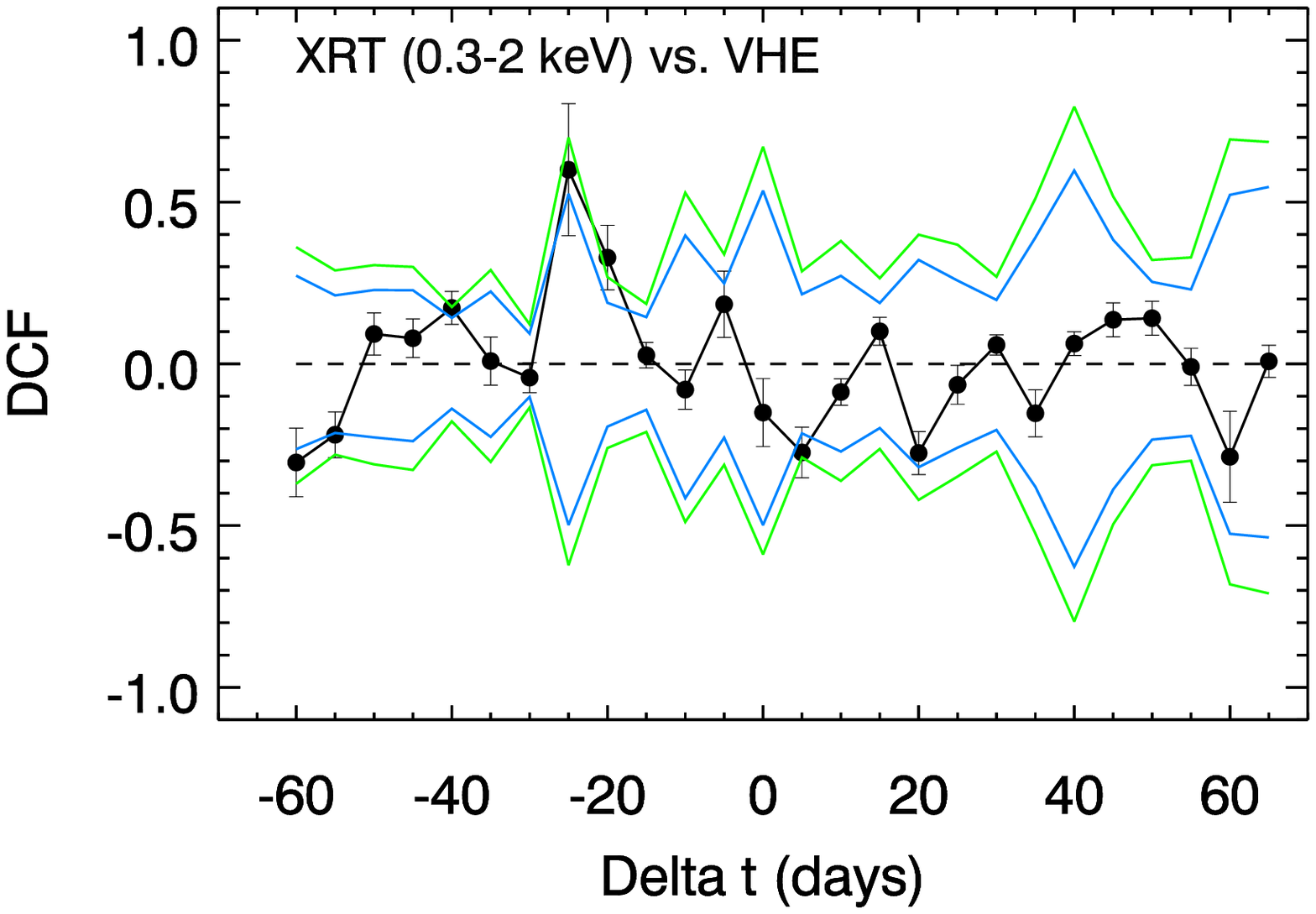}
\includegraphics[width=3.in]{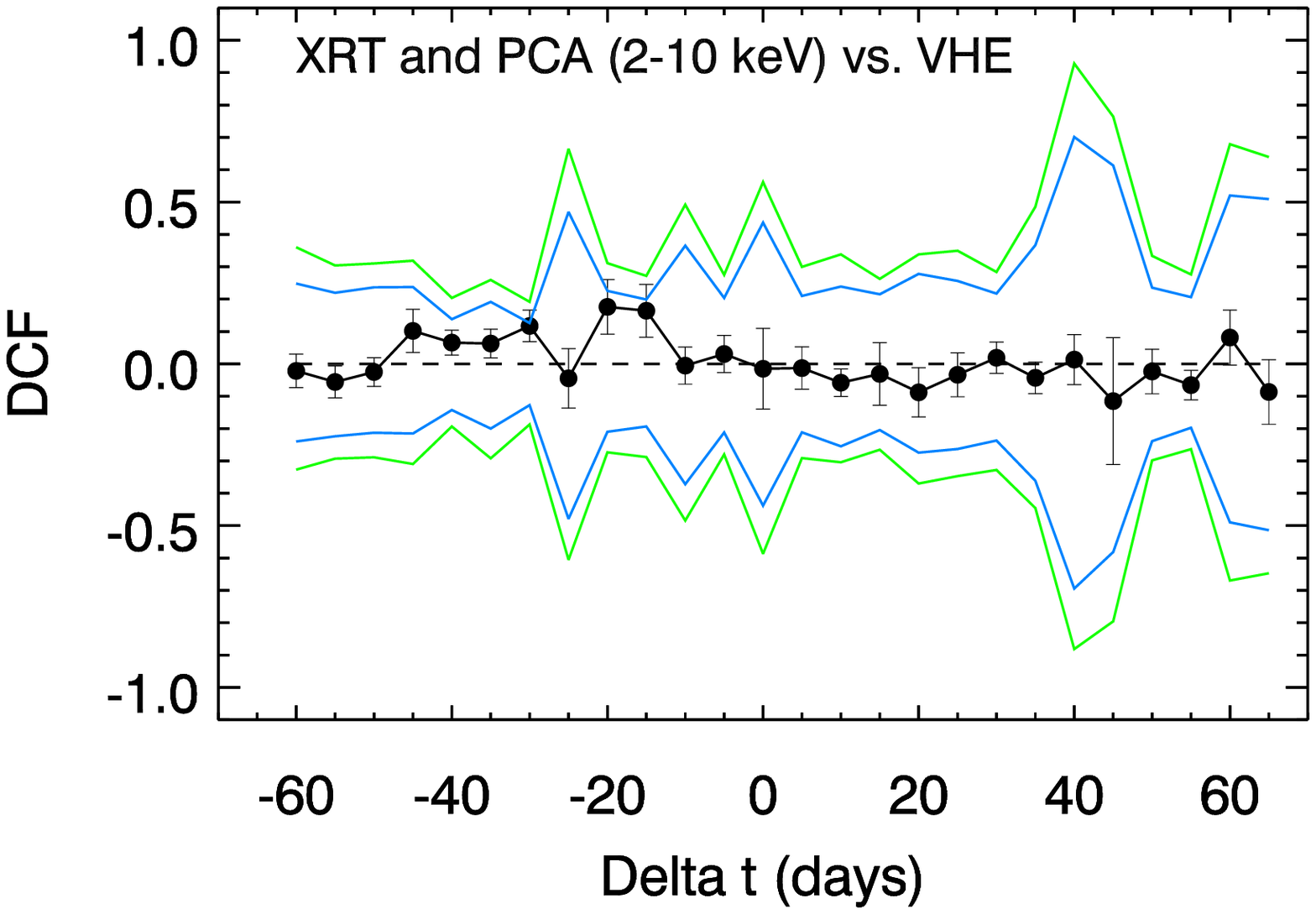}
	\end{center}
\end{minipage}
\hspace*{-5mm}
\begin{minipage}[b]{0.5\linewidth}
	\begin{center}
		\includegraphics[width=3in]{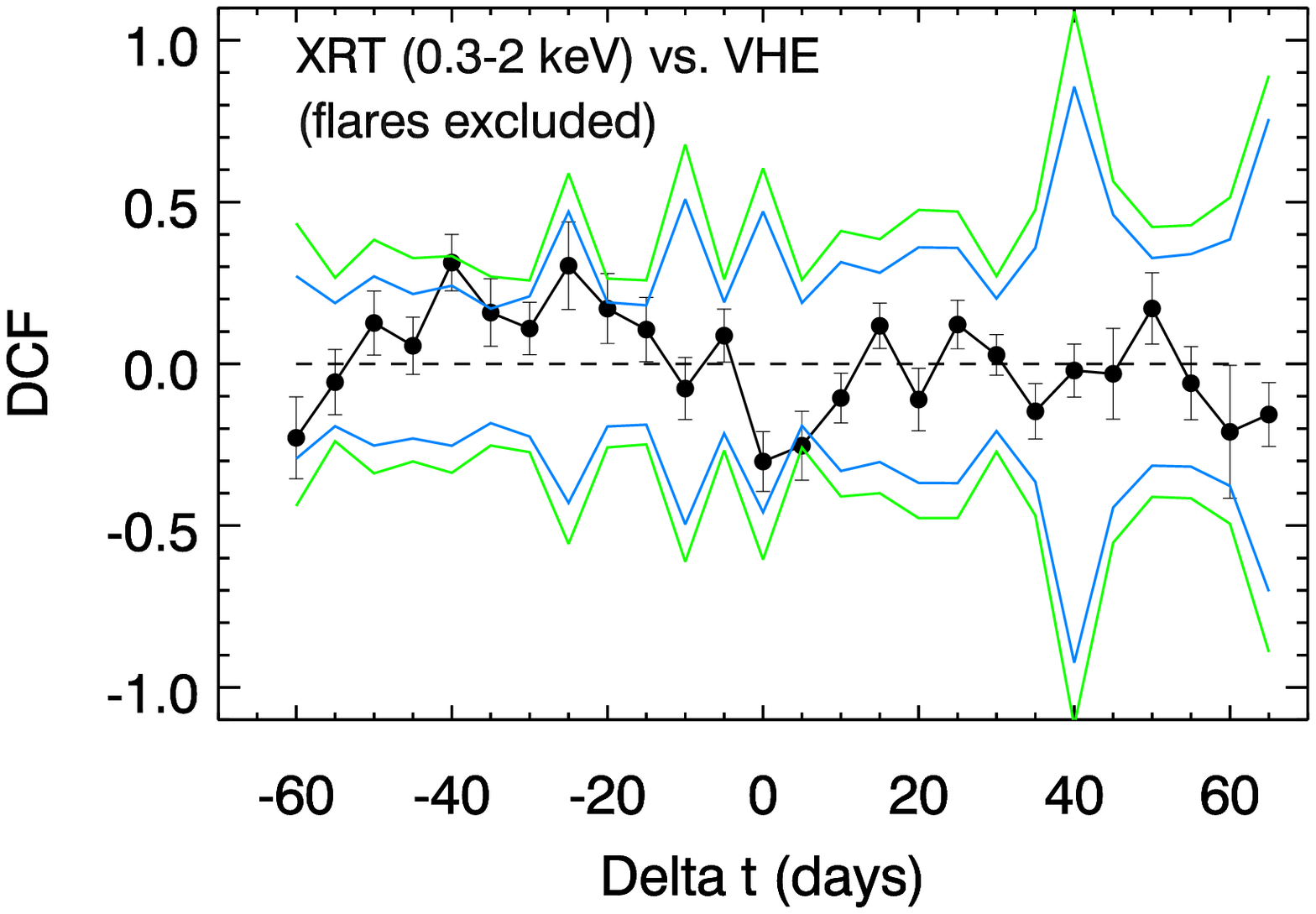}
		\includegraphics[width=3in]{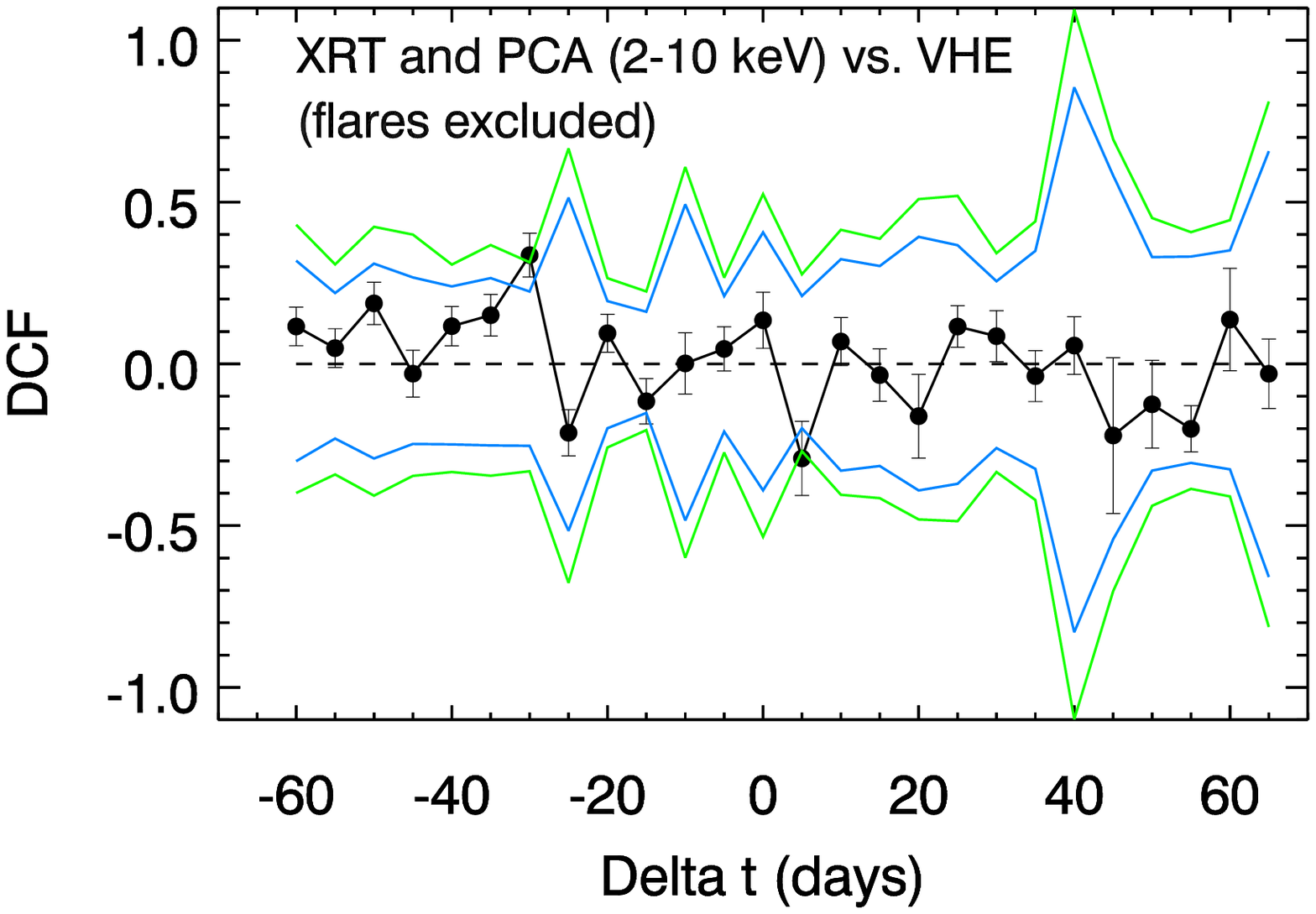}
	\end{center}
\end{minipage}
		\caption{DCF derived for VHE $\gamma$-rays (combined from MAGIC, VERITAS and Whipple measurements) and two X-ray bands (\textit{Swift}/XRT measurements within the 0.3 - 2 keV band; \textit{Swift}/XRT and \textit{RXTE}/PCA combined within the 2 - 10 keV band). The blue (green) lines depict the 95\% (99\%) 
confidence intervals derived from Monte Carlo generated light curves (see text for detailed explanation). 
Left: DCF of complete datasets. Right: DCF derived with the datasets after subtracting the two flaring periods (excluded time windows as explained in the text).  }
	  \label{fig:vhexray}
\end{figure*}

Therefore, the study focuses on the highly sensitive X-ray and VHE $\gamma$-ray observations, namely the ones performed with \textit{Swift}/XRT, \textit{RXTE}/PCA, MAGIC, VERITAS and Whipple 10\,m, which are also the ones that report the highest variability, as shown in Fig.~\ref{Fig:fvar}. In the VHE $\gamma$-ray band, the number of observations is relatively small (in comparison to the number of X-ray observations performed with \textit{Swift} and \textit{RXTE}), and hence we compile a single light curve with a dense temporal sampling of Mrk~501, including the measured flux points from all three participating VHE $\gamma$-ray telescopes. This procedure is straight-forward, as VERITAS and MAGIC both measured the flux above 300\,GeV and the Whipple 10\,m measurements have been scaled to report a flux in the same energy range (see Sect.~\ref{sec:campaign}). We also combined measurements by \textit{Swift}/XRT in the 2-10\,keV band and data points from \textit{RXTE}/PCA to a single light curve, as the same energy range is covered by the two instruments. The light curve in the 0.3-2 keV band consists of only measurements performed by \textit{Swift}/XRT. %We evaluate the X-ray/VHE correlations separately for \textit{Swift}/XRT (in two energy bands, 0.3-2 keV and 2-10 keV) and \textit{RXTE}/PCA. 
The DCF vs.~time-shift distributions for the two X-ray bands and the VHE $\gamma$-ray measurements are shown on the left hand side of Fig.~\ref{fig:vhexray}.

%For a zero time lag, only the XRT 2-10\,keV band shows a hint of correlation to the VHE $\gamma$-ray band. This effect is likely dominated by the second flare event seen during this campaign, together with the almost-simultaneity of the largest flux values in this band compared to the VHE $\gamma$-ray fluxes around this episode. 
At a time lag $\Delta$T$=\left(\text{T}_{\text{VHE}} - \text{T}_{\text{X-ray}}\right)$ on the order of -20 to -25 days, a hint of correlation at the level of 2 sigma between 
fluxes in the soft X-ray band and the VHE $\gamma$-ray band is seen in the top left panel of Fig.~\ref{fig:vhexray}. %In the case of PCA an actually significant correlation is seen. %both energy bands accessed by XRT. This correlation is also found for the PCA light curve, although somewhat less significant. 
This feature is dominated by the two flaring events% in the VHE and X-ray light curves
, as the dominant flare in VHE $\gamma$-rays occured around MJD~54952, while the largest flux increase in soft X-rays was seen around MJD~54977%from MJD~54969 to MJD~54977
, with a separation of $24-25$ days. 
The right hand side of Fig.~\ref{fig:vhexray}  reports the evaluation of the correlations after the flaring episodes have been excluded from the X-ray and VHE $\gamma$-ray light curves. The above-mentioned feature at 20-25 days is no longer present. %The removed observationswere MJD~54973.1 for MAGIC; MJD~54952.4 for VERITAS; MJD~54952.4, 54975.4 and 54976.3 for Whipple 10\,m; MJD~54952.8, 54953.4, 54953.7, 54976.3 and 54977.3 for XRT. 
%{\bf what bout the points around 54953 for XRT ? They will probably not make a big difference because the flux increase is relatively small, but formally we need to remove those too.} 
%In this case, no significant correlation between the fluxes in soft X-rays and VHE $\gamma$-rays is seen anymore. 

The large growth of the confidence intervals apparent at time shifts of $\Delta$T $\approx 40$ days are caused by sparsely populated regions in the VHE $\gamma$-ray light curve, mainly towards the end of the campaign. In case the light curves are shifted by $\approx 40$ days with respect to each other, these regions overlap with densely populated regions in the X-ray light curves, which results in a larger uncertainty of the determined DCF. 

Overall, no significant correlation between X-ray and VHE $\gamma$-ray fluxes is found for any of the combinations probed.% once the two flaring episodes are removed from the light curves. 

%\input{sed}
%\newpage

\section{Evolution of the spectral energy distribution}
\label{sec:sed}

\noindent
The time-averaged broadband SED measured during this MWL campaign (from MJD 54905 to MJD 55044) was reported and modeled satisfactorily in the context of a one-zone synchrotron self-Compton (SSC) scenario \citep{2011ApJ...727..129A}. In such a model, several properties of the emission region are defined, such as the size of the region $R$, the local magnetic field $B$ and the Doppler factor $\delta$, which describes the relativistic beaming of the emission towards the observer. 
Furthermore, the radiating electron population is described by a local particle density $n_e$ and the spectral shape. For the averaged data set of this campaign, the underlying spectrum of the electron population was parameterized with a power-law distribution from a minimum energy $\gamma_{\text{min}}$ to a maximum energy $\gamma_{\text{max}}$, with two spectral breaks $\gamma_{\text{break,1}}$ and  $\gamma_{\text{break,2}}$. The two breaks in the electron energy distribution (EED) were required in order to properly model the entire broadband SED.
Because of the relatively small multiband variability during the 4.5-month long observing campaign (once the first VHE flare is removed) and the large number of observations performed with all the instruments, the average SED could be regarded as a high-quality representation of the typical broadband emission of Mrk~501 during the time interval covered by the campaign, and hence the one-zone SSC model was constrained to describe all the data points (including 230 GHz SMA and interferometric 43 GHz VLBA observations).

\begin{figure*}[!ht]
	\begin{center}
\vspace{-5mm}
\includegraphics[width=5.1in]{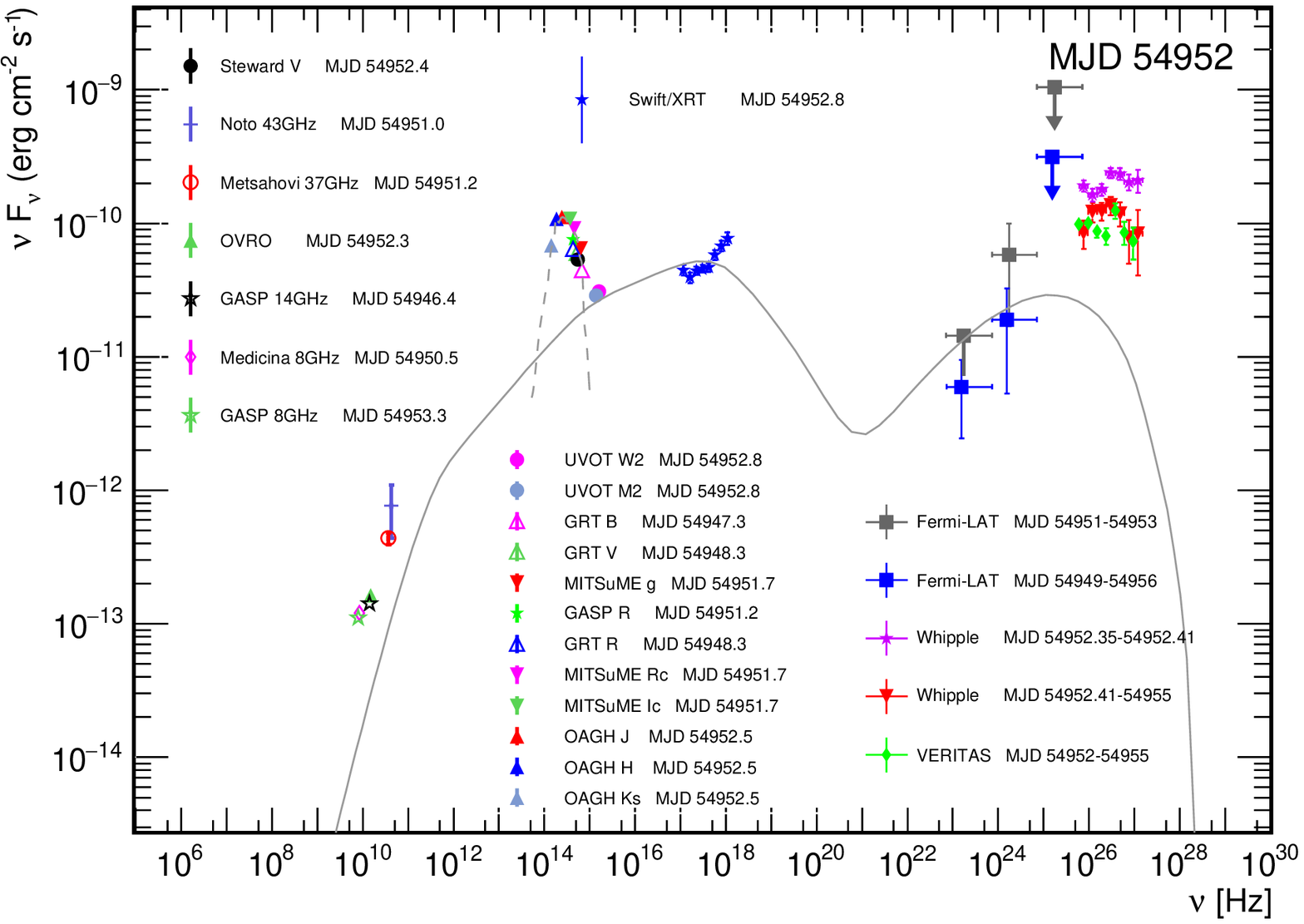}
\includegraphics[width=5.1in]{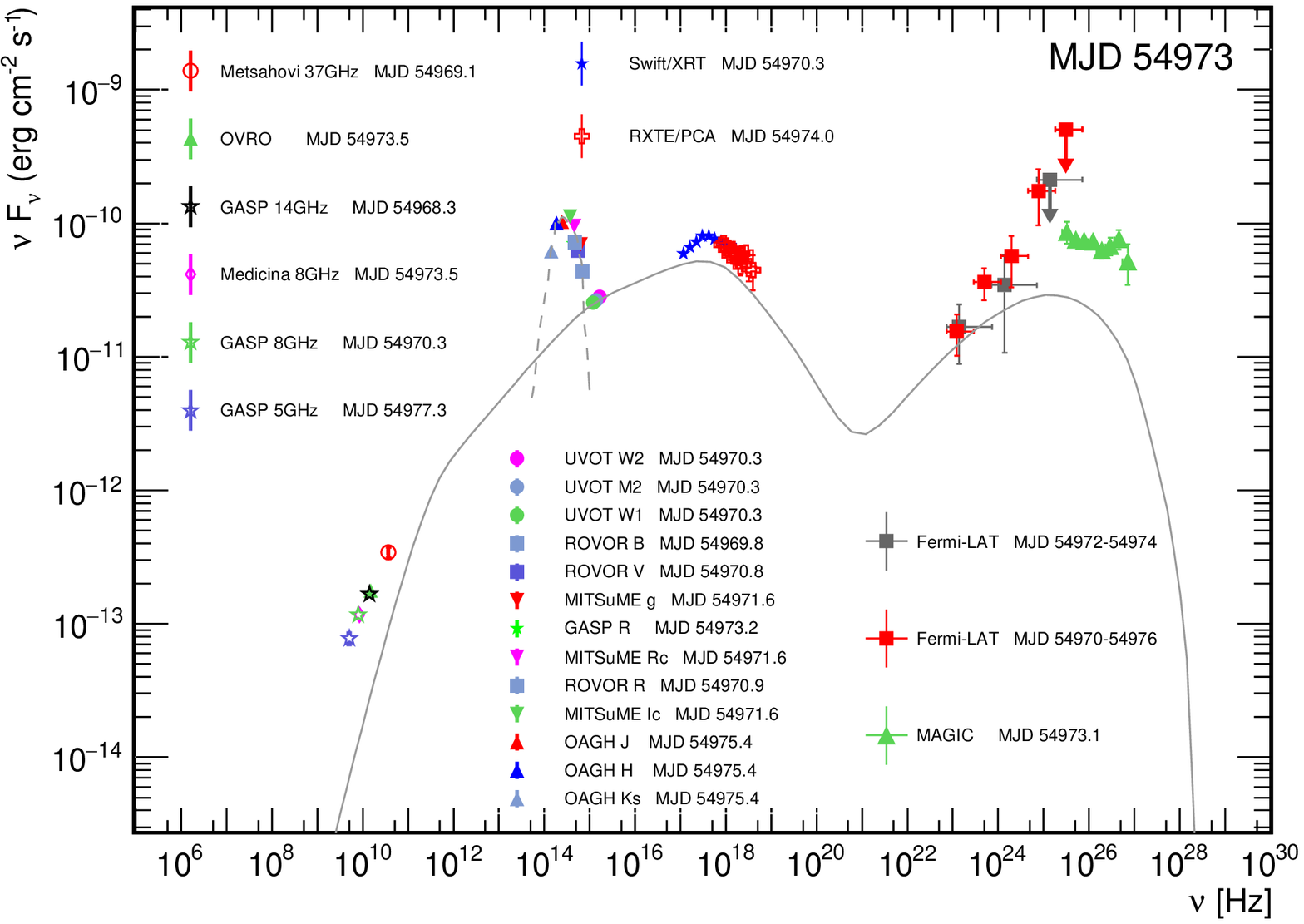}
\end{center}
\caption{
Broadband SED of Mrk~501 during the two VHE high states observed within the campaign (upper panel: MJD~54952, lower panel: MJD~54973). See text for details regarding the included spectral measurements. The data points have been corrected for EBL absorption according to the model by \cite{Franceschini:2008vt}. The emission of the host galaxy  parameterized according to \cite{1998ApJ...509..103S} is shown with a grey dashed line, while the one-zone SSC model describing the average broadband SED over the entire campaign \citep[see][]{2011ApJ...727..129A} is depicted with a grey solid line.
}
	  \label{Fig:mrk501flareseds}
\end{figure*}

In this work, we focus on the characterization of the broadband SED during the two flaring episodes occurring in May 2009. As reported in Sect.~\ref{sec:LCs}, these two flaring episodes start on MJD 54952 and MJD~54973, and last for approximately three and five days each, respectively. There is some flux and spectral variability throughout these two flaring episodes, but for the sake of simplicity, in this section we will attempt to model only the SEDs related to the VHE flares on MJD 54952 and 54973, which are the first days of these two flaring activities. We try to model these two SEDs with the simplest leptonic scenarios, namely a one-zone SSC and a two-independent-zone SSC model.  In the latter we assume that the quiescent or slowly changing emission is dominated by one region that is described by the SSC model parameters used for the average/typical broadband emission from the campaign \citep[see][]{2011ApJ...727..129A}, while the flaring emission (essentially only visible in the X-ray and $\gamma$-ray bands) is dominated by a second, independent and spatially separated region.

The assumption of a theoretical scenario consisting of one (or two) steady-state homogenous emission zone(s) could be an oversimplification of the real situation. The blazar emission may be produced in inhomogeneous regions, involving stratification of the emitting plasma both along and across a relativistic outflow, and the broad-band SED may be the superposition of the emission from all these different regions, characterized by different parameters and emission properties, as reported by various authors \citep[e.g.][]{2005A&A...432..401G,2008ApJ...689...68G,2009MNRAS.395L..29G,2011MNRAS.416.2368C,2014ApJ...789...66Z,2015MNRAS.447..530C}. In this paper we decided to continue using the same theoretical scenario used in \citet{2011ApJ...727..129A}, which we adopted as the reference paper for this dataset. We also kept the discussion of the model parameters at a basic level, and did not attempt to perform any profound study of the implications of those parameters.

In this work we used the SSC model code described in \cite{2011MNRAS.413.1845T}, which is qualitatively the same as the one used in \cite{2011ApJ...727..129A}, with the difference that the 
EED is parameterized as 
\begin{equation}
\label{eq:electrondistr}
\frac{dN}{d\gamma} =
\begin{cases} n_e \cdot \gamma^{-\alpha_1},         &     (\gamma_{\text{min}} < \gamma < \gamma_{\text{break},1})\\
                                        n_e \cdot \gamma_{\text{break},1}^{\alpha_2-\alpha_1}  \cdot \gamma^{-\alpha_2}, & ( \gamma_{\text{break},1}< \gamma < \gamma_{\text{break},2})\\
                                        n_e \cdot \gamma_{\text{break},1}^{\alpha_2-\alpha_1}  \cdot \gamma_{\text{break},2}^{\alpha_3-\alpha_2}  \cdot e^{\left(\frac{\gamma_{\text{break},2}}{\gamma_{\text{max}}}\right)}  \\
\;\;\;\;\; \cdot \gamma^{-\alpha_3}  \cdot e^{\left(-\frac{\gamma}{\gamma_{\text{max}}}\right)}& (\gamma_{\text{break},2}<\gamma),
\end{cases}
\end{equation}
where $n_e$  is the electron number density. For reasons of comparability, only this definition is applied in all the SED modelling results in this section, including that of the quiescent, averaged SED obtained over the full MWL campaign. The corresponding one-zone SSC model parameter values defining the averaged SED from the full 2009 multi-instrument campaign are listed in Table~\ref{tab:sscaveragestate}. The parameter values are identical to those from the ``Main SSC fit'' reported in Table~2 of \cite{2011ApJ...727..129A}, with the only difference being the usage of the electron number density $n_e$, instead of the equipartition parameter. The contribution of star light from the host galaxy can be approximately described with the template from \cite{1998ApJ...509..103S}, as was done in \cite{2011ApJ...727..129A}.
\begin{table*} [htbp]
\begin{center}
\caption{SSC model parameters which characterize the average emission over the entire MWL campaign. The parameters apply to a one-zone model defined by Eq.~\ref{eq:electrondistr} and are retrieved from the modelling presented in \cite{2011ApJ...727..129A}.}
\label{tab:sscaveragestate}
\vspace{5mm}
	\begin{tabular}{r|ccccccccccc}
			&$\gamma_{\text{min}}$ 	&$\gamma_{\text{max}}$ &	$\gamma_{\text{break},1}$&	$\gamma_{\text{break},2}$& 	$\alpha_{1}$ &	$\alpha_{2}$&	$\alpha_{3}$  &n$_{e}$ &	B/mG &	$\log(\frac{R}{cm}) $ &$\delta$ \\
\hline
av. state &600 & $1.5 \times 10^7$ & $4 \times 10^4$ & $9\times10^5$ & 2.2 & 2.7 & 3.65 & 635 & 15 & $17.11 $ & 12 \\
\end{tabular}
\end{center}
\end{table*}

For the characterization of the SEDs collected during the two flaring states, we allow for an EED with two spectral breaks in the case of one-zone SSC models. For the second zone in the two-zone SSC scenario, we keep the somewhat simpler description of the electron energy distribution as a power law with only one spectral break  (i.e.~$\alpha_2\equiv \alpha_1$ in Eq.~\ref{eq:electrondistr}). \\

\subsection{The grid-scan strategy for modeling the SED}
\label{TheGridScan}

In contrast to the commonly used method of adjusting the model curve to the measured SED data points \citep[e.g.][]{1998ApJ...509..608T,2001ApJ...554..725T, Albert:2007bt}, in this study we applied a novel variation on the {\em grid-scan} approach
 in the space of model parameters. Given a particular theoretical scenario (e.g. the one-zone or two-zone SSC model), we make a multi-dimensional grid with the $N$ model parameters that we want to sample. For each parameter, we define a range of allowed values and a step size for the variation within this range.  Theoretical (SSC) model curves are calculated for each point on the grid, i.e.~for each combination of the $N$ parameter values. Subsequently, the goodness of the resulting model curves in reconstructing the data points is quantified by means of the $\chi^2$ between data and model,  which takes into account the statistical uncertainties of the individual measurements.
 At the moment, systematic uncertainties are not considered for the evaluation of the agreement. This would require performing the entire procedure for various shifts in the flux and energy scale for each instrument, as well as for possible distortions in the individual spectra. The net impact of including systematic uncertainties in the single-instrument spectra would be a larger tolerance for the agreement between the experimental data and the theoretical model curves, which would yield a larger degeneracy in the parameter values that can model the data. While this will be investigated in the future, it is beyond the scope of this paper.
Therefore, the data-model agreements reported in this manuscript, which are based on the $\chi^2$  analysis using only the statistical uncertainties, provide a lower limit to the actual agreement between the presented experimental data and the theoretical model curves being tested, and we mostly use them to judge the relative agreement of the various theoretical model curves.

Depending on the complexity of the model itself, the model calculations for an entire grid can be very intensive in computing power. For instance, one of the simplest SSC scenarios, involving only one emission zone with an electron energy distribution with one spectral break, already leads to a grid spanning a nine-dimensional parameter space. With the ranges and grid spacings we are using in this work, the number of model curves to calculate and evaluate amounts to tens of million. For this reason, the access to cluster computing becomes essential for this grid-scan modelling approach. The model calculations in this work have been performed using the computing farms at SLAC\footnote{\url{https://www.slac.stanford.edu/comp/unix/unix-hpc.html}} and TU Dortmund.\footnote{\url{http://www.cs.tu-dortmund.de/nps/en/Home/}}

After the evaluation of all models regarding their level of agreement with the data, individual models can be chosen for the final set, according to the achieved probability of agreement (derived from the $\chi^2$ and the number of degrees of freedom). These sets of models can then be visualized both in the SED representation and in the space of parameter values defining the models, which could populate non-continuous regions in the parameter space.

One aim of the grid-scan strategy is to keep the range of model parameters as wide as possible. By sampling a large parameter phase space we can reduce the bias which is usually introduced into the modelling by adopting a set of assumptions or educated choices. 
Another advantage is the fact that, besides the obvious aim of finding parameter values which describe the data in the best way, the ``grid-scan'' approach also offers the possibility of investigating the degeneracy of the model-to-data agreement regarding each individual model parameter. In order to do this, sets of models within bands of achieved fit probabilities are compiled and their distributions in each of the model parameters are visualized. Based on such plots, interesting regions in the parameter space can be selected for a deeper search, which leads to models with an even better agreement with the data and to a more thorough study of the degeneracy of individual model parameters. Finally, the grid-scan method allows one to potentially find multiple clusters or regions in the model parameter phase space that could be related to different physical scenarios, which can be equally applicable to the data set at hand, but might be missed by statistical methods aiming at only "one best" solution.

A concept for ``grid-scan'' SED modelling has already been presented in \cite{2013A&A...558A..47C}, where model curves are computed for each point on the parameter space grid, but the assessment of the agreement between model and data is performed in a different way: the authors evaluate the agreement based on seven observables (i.e.~the frequency and luminosity of the synchrotron peak, the measured X-ray spectral slope and the GeV and TeV spectral slopes and flux normalizations), which are derived from the model curves and compared to the data. They also provide a family of solutions, involving any uncertainties in the observables. In the work presented here, the model-to-data agreement criterion, which is used to select a set of models, is derived directly from the $\chi^2$-distances between each data point and each model curve, without computing any secondary characteristics of the SED which may introduce additional uncertainties. \cite{2013A&A...558A..47C} also determine this distance for the models picked by their algorithm, but apply it only as a posteriori check of their result. Furthermore, the authors have reduced the dimensionality of the parameter space from nine to six, and used only five steps for each parameter, which implied the creation of a grid with $5^6$=15625 SED realizations. In the work presented here, the smallest grid-scan implied the creation of more than $40\times10^6$ SED realizations. Additionally, after selecting interesting regions in the various model parameters with the grid-scan, we went one step further and performed a second (dense) grid-scan focused only on those regions, and using a smaller step size.

The objective of finding uncertainty ranges of model parameters has also been addressed by \cite{2011ApJ...733...14M,2015arXiv150903319Z}. Here, a Markov-Chain Monte Carlo procedure is used to fit emission model curves (for a number of different emission models) to the observational results. While this approach delivers uncertainties or probability distribution functions for the particle distribution parameters, this is done only for one particular solution. Disjointed regions of equally good model configurations, i.e.~``holes'' in the probability distribution for the individual parameters, are not found following this method. 

A 3-dimensional parameter grid with 9504 (48$\times$22$\times$9) steps was used by \citet{2000ApJ...536..742P} to find the most suited model parameter set to describe weekly averaged SEDs of Mrk~501, where the "best" model was selected as the one with the smallest data-model difference, quantified with a $\chi^2$ approach. Despite the usage of a parameter grid, the goal and merits of that work differ from those of the methodology presented here. While \citet{2000ApJ...536..742P} used the 3-dimensional parameter grid to find the best model \citep[as in][with a $\chi2$ minimization procedure]{2011ApJ...733...14M}, in this work the 9-dimensional grid is used to find the family (or families) of parameter values that give a good representation of the broad-band SED, and to show the large degeneracy of the model parameters to describe the SED.

\begin{table*} [htbp]
\begin{center}
\caption{Grid of SSC model parameters which is probed for one-zone
  models within the coarse grid-scan. For each parameter the probed
  range is given by a minimum and maximum value, and the number of
  tested values is given by the number of steps between (and
  including) these limits. The number of SSC models required to
  realize this grid-scan amounts to 62 million. 
}
\label{tab:sscparameterspaceonezone}
\vspace{5mm}
	\begin{tabular}{r|ccccccccccc}
			coarse grid &&&&&&&&&&&\\
one-zone &$\gamma_{\text{min}}$ 	&$\gamma_{\text{max}}$ &	$\gamma_{\text{break},1}$ &	$\gamma_{\text{break},2}$& 	$\alpha_{1}$ &	$\alpha_{2}$  &	$\alpha_{3}$ &n$_{e}$ &	B/mG &	$\log(\frac{R}{cm}) $	&$\delta$ 	\\
\hline
min 	&	$1\times 10^{2}$ 	&	$1\times 10^{6}$ 	&	$1\times 10^{4}$	&	$1\times 10^{5}$			&	1.7 &	2.1 &	3.6 	&$1\times 10^{3}$ &	5 	&	14.0 &	1 		\\
max		&	$1\times 10^{4}$	&	$1\times 10^{8}$	&	$1 \times 10^{5}$	& $3.2 \times 10^{6}$		&		2.3 &	3.3 & 4.8	&		$1\times 10^{6}$	&250	&	16.0	&60		\\
$\#$ of steps	&	3		&	3	& 4 &		4		&		7& 7 	&	4		&7	&	9		&5		&7	\\
spacing	&	log		&	log		&	log			&	log &	lin	&	lin	&	lin &	log		&log	&	lin	&	log	\\
\end{tabular}
\end{center}
\end{table*}
\begin{table*} [htbp]
\begin{center}
\caption{Grid of SSC model parameters which is probed for two-zone models within the coarse grid-scan. In two-zone models only the second zone is defined by the parameters given here, while the first zone is given by the model derived in \cite{2011ApJ...727..129A}, and reported in Table \ref{tab:sscaveragestate}. The number of SSC models required to realize this grid-scan amounts to 40 million.
}
\label{tab:sscparameterspacetwozone}
\vspace{5mm}
	\begin{tabular}{r|ccccccccc}
coarse grid &&&&&&&&&\\
			two-zone&$\gamma_{\text{min}}$ 	&$\gamma_{\text{max}}$ &	$\gamma_{\text{break}}$& 	$\alpha_{1}$ &	$\alpha_{2}$ &n$_{e}$ &	B/mG &	$\log(\frac{R}{cm}) $	&$\delta$ 	\\
\hline
min 	&	$1\times 10^{2}$ 	&	$1\times 10^{5}$ 	&	$1\times 10^{4}$			&	1.7 &		2.0 	&100 &	5 	&	14.0&	1 		\\
max		&	$1\times 10^{6}$	&	$1\times 10^{8}$	& $1 \times 10^{7}$		&		2.3  &4.8	&		$1\times 10^{6}$	&250	&	18.0	&60		\\
$\#$ of steps	&	5		&	4	&		7		&		7 	&	8		&9	&	9		&9		&7	\\
spacing	&	log		&	log		&	log			 &	lin		&	lin &	log		&log	&	lin	&	log	\\
\end{tabular}
\end{center}
\end{table*}

For the theoretical SED modelling of the two flaring states of Mrk~501, following the grid-scan strategy outlined above, the parameter ranges given in Table~\ref{tab:sscparameterspaceonezone} and Table~\ref{tab:sscparameterspacetwozone} have been investigated for the one-zone and two-zone scenarios described at the beginning of this section. Given that we aim to sample a wide range of parameter values with a relatively coarse step (for each parameter), we denote these scans as ``coarse grid-scans''. The general orientation for the choice of parameter ranges is based on previous works on modelling of the SED of Mrk~501, e.g.~\cite{Albert:2007bt,2009ApJ...705.1624A,2011ApJ...727..129A,2012ApJ...753..154M}. Based on these values\footnote{Many of the previous works in the literature use $\gamma_{\text{min}}$=1 \citep[e.g.][]{2001ApJ...554..725T,Albert:2007bt,2012ApJ...753..154M}, but we decided to follow here the approach done in \cite{2011ApJ...727..129A}, where a $\gamma_{\text{min}} >>$1 had to be used in order to properly describe the simultaneous GeV data from {\em Fermi}-LAT and the high-frequency radio observations from SMA and VLBA, which did not exist in the previous publications parameterising the broadband SED of Mrk~501.}, one-zone SSC models have been built as well as second zones for the two-zone scenario. In the latter, the first zone is described by the model reproducing the average emission seen over the entire campaign \citep[see][]{2011ApJ...727..129A}, while only the second zone is varied as described by the model parameters from the grid presented here.  One could reduce the phase space of the grid-scan by imposing certain relations between the locations of the breaks  ($\gamma_{\text{break}}$) in the EED and the size $R$ and magnetic field $B$ values, as well as to constrain the index values before and after the breaks (e.g. $\Delta \alpha=1$). But cooling breaks with a spectral change twice larger than the canonical value of 1 were necessary to describe the broadband SED of Mrk\,421 within a SSC homogeneous model scenario \citep[see Sect.~7.1 of][]{2011ApJ...736..131A}, and the breaks needed by the SSC models are not always related to the cooling of the electrons, but instead could be related to the acceleration mechanism, as reported for Mrk\,501 in \citet{2011ApJ...727..129A}. Internal breaks (related to the electron acceleration) have been reported for various blazars \citep[e.g.][]{2009ApJ...699..817A,2010ApJ...710.1271A}. The origin of these internal breaks, as well as large spectral changes at the EED breaks, may be related to variations in the global field orientation, turbulence levels sampled by particles of different energy, or gradients in the physical quantities describing the system. These characteristics are not taken into account in the "relatively simple" homogenous SSC models, and argue for more sophisticated theoretical scenarios, such as the ones mentioned above. In order to keep the range of allowed model parameter values as broad as possible, in this exercise we did not impose constraints on the location of the EED breaks or in the index values before/after the breaks. The hardest index we use in this study is 1.7, which is harder than the canonical index values $>2$ derived from shock acceleration mechanisms and used very often to parameterize the broad-band SEDs of blazars. But this is actually not a problem, as various authors have shown that indices as hard as 1.5 can be produced through  stochastic acceleration \citep[e.g.][]{2005ApJ...621..313V}, or through diffusive acceleration in relativistic magnetohydrodynamic shocks, as reported in \citet{2007ApJ...667L..29S,2012ApJ...745...63S,2016MNRAS.tmp.1447B}. We also use $\gamma_{\text{min}}$ values extending up to $10^6$, which are substantially higher that those used in conventional SSC models (that go typically up to $\sim10^3$); but such high $\gamma_{\text{min}}$ values have also been used already by various authors \citep[e.g.][]{2006MNRAS.368L..52K,2009MNRAS.399L..59T,2011ApJ...743L..19L,2011ApJ...740...64L}.

In the evaluation of the models, we used two additional constraints, besides the requirement of presenting a good agreement with the SED data points. Equipartition arguments impose the condition that the energy densities held by the electron population ($u_e$) and the magnetic field ($u_B$) should be of comparable order. Typically, the parameterization of the broadband SED of Mrk501 (and all TeV blazars, in general) within SSC theoretical scenarios require $u_e \sim 10^{2-3} u_B$, which implies higher energy in the particles than in the magnetic field, at least locally, where the broadband blazar emission is produced \citep[see e.g. ][]{2001ApJ...554..725T,2011ApJ...727..129A,2015A&A...573A..50A}. There is no physical reason for any specific (somewhat arbitrary) cut value in the quantity $u_e/u_B$, but driven by previous works in the literature, in this study we only consider models fulfilling the requirement of $u_e/u_B < 10^3$.  Secondly, the observed variability timescales have to be taken into account. Following causality arguments, the observed variability should not happen on timescales which are shorter than the time needed to distribute information throughout the emitting region. Based on the given Doppler factor $\delta$ and the size of the emitting region $R$, the implied minimum variability timescale quantity for each model is derived according to
\begin{equation}
\label{eq:tvar}
t_{\text{var}_{\text{min}}} \simeq \frac{\left(1+z\right) R}{\delta c}.
\end{equation}
While for the first flare we observed large (up to factors of $\sim$4) flux changes within 0.5 hour \citep{Pichel:2011we,PichelMrk501MW2009}, the second flare shows substantial flux changes ($\sim$2) on timescales of several days. Consequently, we consider only models which yield a minimum variability timescale of $t_{\text{var}_{\text{min}}}\leq 0.25$ hours and $t_{\text{var}_{\text{min}}}\leq 1$ day for the 1st and 2nd VHE flare, respectively.

\subsection{1st VHE flare}
\label{sec:sed1}

All spectral points which were obtained at the time or close to the time of the VHE flare measured by VERITAS and the Whipple 10\,m telescope at MJD 54952 are shown in the top panel of Fig.~\ref{Fig:mrk501flareseds} (see Sect.~\ref{subsec:firstflarevariability} and \ref{subsec:firstflarevariability} for details on the individual observation times).

The attempt to apply the grid-scan to the broadband SED from this flaring episode is affected by the fact that some of the flux variations occurred on sub-hour timescales, and the observations performed were not strictly simultaneous as discussed in the previous sections. Therefore, the SED reported in this section is not necessarily a good representation of the true SED for this flaring episode, and hence any modelling results have to be regarded as inconclusive. 

In this SED modeling exercise, the data used are the measurements from \textit{Swift} (UV and X-rays), \textit{Fermi}-LAT (2-day spectrum) and Whipple 10\,m very-high state.  The optical and infrared, as well as the radio points, are not taken into account for the evaluation of the agreement of the SSC model curves with the data. The former two are strongly dominated by emission from the host galaxy, and the latter only serve as upper limits for the SSC flux, as the radio emission shows substantially lower variability timescales and is widely assumed to stem from a larger region than the emitting blob responsible for a few-day long flaring activity.

Exploiting the entire parameter grid space, neither the one-zone SSC model nor the two-zone SSC model can reconstruct the measured broadband SED, with the data-model agreement quantified with $\chi^2/d.o.f. > 300/20$, which would imply a probability of agreement $P$ (or p-value)\footnote{The conversion between $\chi^2/d.o.f$ and probability values assumes that the $\chi^2$ distribution (for the given degrees of freedom) is valid also for $\chi^2$ values that are very far away from the central value, which is not necessarily correct. In any case, when the model-to-data agreement is very bad (i.e. a very large $\chi^2$ value) the precise knowledge of the $P$ value is not relevant for the discussion, and hence the inaccuracy of the conversion between $\chi^2$ values and probabilities does not critically impact the results discussed in the paper.} between the SSC model curves and the data points of $P < 10^{-50}$.  When removing the tight constraint given by the cut in $t_{\text{var}_{\text{min}}}$, the best agreement obtained with the one-zone SSC scenario from the grid-scan defined by Table~\ref{tab:sscparameterspaceonezone} is  $\chi^2/d.o.f. = 180/20$  ($P \sim 10^{-27}$). The two-zone scenario with the quiescent emission characterized by the model parameters from the average SED reported in Table~\ref{tab:sscaveragestate} and the (spatially independent) region responsible for the flaring activity modeled based on the coarse grid parameter values reported in Table~\ref{tab:sscparameterspacetwozone} provides at best an agreement given by $\chi^2/d.o.f. =225/20$ ($P \sim 10^{-36}$).  Since the X-ray spectrum at low energies is already accounted for with the ``quiescent'' zone, the contribution from the ``flaring'' zone (which is needed to explain the increase in the flux at VHE) exceeds the measured flux at X-ray energies, and hence yields a bad agreement with the data points.

Besides trying with the grid-scan defined in Table~\ref{tab:sscparameterspaceonezone}, we also evaluated the model-to-data agreement when using a one-zone scenario with the grid-scan defined in Table~\ref{tab:sscparameterspacetwozone}, which provides a more simple theoretical scenario (only one break, instead of two, in the EED), but with somewhat extended ranges probed for the parameters  $\gamma_{\text{min}}$, $\gamma_{\text{max}}$, $\gamma_{\text{break}}$, n$_{e}$, and	$\log(\frac{R}{cm}) $. We found a few models with data-model agreement given by  $\chi^2/d.o.f. = 95/20$ ($P \sim 10^{-11}$). But as soon as the requirement for fast variability is applied, all these models (mostly featuring large emission regions with $R\ge 10^{16.5}$) are no longer applicable, and the agreement  between the SSC model curves and the data points become $\chi^2/d.o.f. > 300/20$.

One of the difficulties in modeling these data with a one-zone scenario is that it is difficult to describe the emission in the UV and the X-ray range with a synchrotron component. These UV flux points cannot be modeled only with the host galaxy template, and the one-zone models that could potentially describe well the shape of the X-ray spectrum would produce a flux that is many times below the measured UV flux, and hence would give a very bad data-model agreement.  Contrary to the mentioned caveat of a time offset between the X-ray and VHE $\gamma$-ray observations, the UV and the X-ray observations have been performed simultaneously and thus should be reconstructed consistently. The difficulty in modelling the UV and X-ray measurements in a consistent way suggests that a more complex scenario is needed to explain this emission. In \citet{PichelMrk501MW2009}, the host galaxy was modeled using a different template with respect to the one in \citet{2011ApJ...727..129A} that is used in this paper. The host galaxy template used in \citet{PichelMrk501MW2009} describes approximately the measured UV flux level from the 3-day broadband SED considered in \citet{PichelMrk501MW2009}, but it would not be consistent with the variability in the data set presented here. In Fig.~\ref{fig:mwllcs} one can see relative variations in the UV flux of more than 50\% (peak to peak), which cannot occur if this UV emission was dominated by the steady emission from the host galaxy.

\subsection{2nd VHE flare}
\label{sec:sed2}

\begin{figure*}[!th]
	\begin{center}
\includegraphics[width=5.1in]
{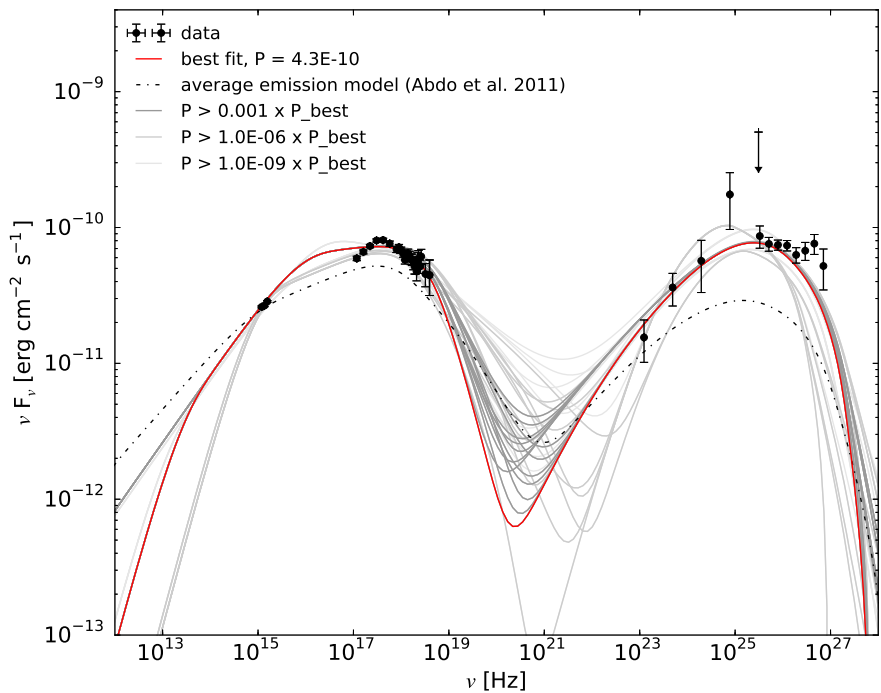}
\end{center}
\caption{SED grid-scan modelling results for the flaring episode around MJD 54973 in the scope of a one-zone SSC scenario. Shown are the model curve (red solid line) with the highest probability of agreement with the data as well as model curves within different probability bands. For comparison, also the SSC one-zone model found to describe the average state \citep{2011ApJ...727..129A} is given (black dash-dotted line). Data points have been corrected for EBL absorption according to the model by \cite{Franceschini:2008vt}.}
	  \label{fig:flare2onezonesedmodeling}
\end{figure*}

The SED of Mrk 501 built from spectra around the time of the second flux increase seen by MAGIC on May 22 (MJD 54973) is shown in the bottom panel of Fig.~\ref{Fig:mrk501flareseds}  (see Sect.~\ref{subsec:secondflarevariability} and \ref{subsec:firstflarevariability} for details on the individual observation times).
The data related to the second flare were not taken strictly simultaneously.  However, here the resulting caveat is not as strong as for the first flare. On one hand, the observed variability occurs on timescales of days, rather than tens of minutes, and the \textit{RXTE}/PCA measurements were performed within a day from the VHE observations. While this is not true for the \textit{Swift}/XRT measurements, the overall flux changes are relatively small,  and the derived \textit{Swift}/XRT spectrum is in very good agreement with the one derived from \textit{RXTE}/PCA, as can be seen in the bottom panel of Fig.~\ref{Fig:mrk501flareseds}.

\begin{figure*}[!ht]
\begin{minipage}[t]{0.33\linewidth}
	\begin{center}
\includegraphics[width=2.5in]{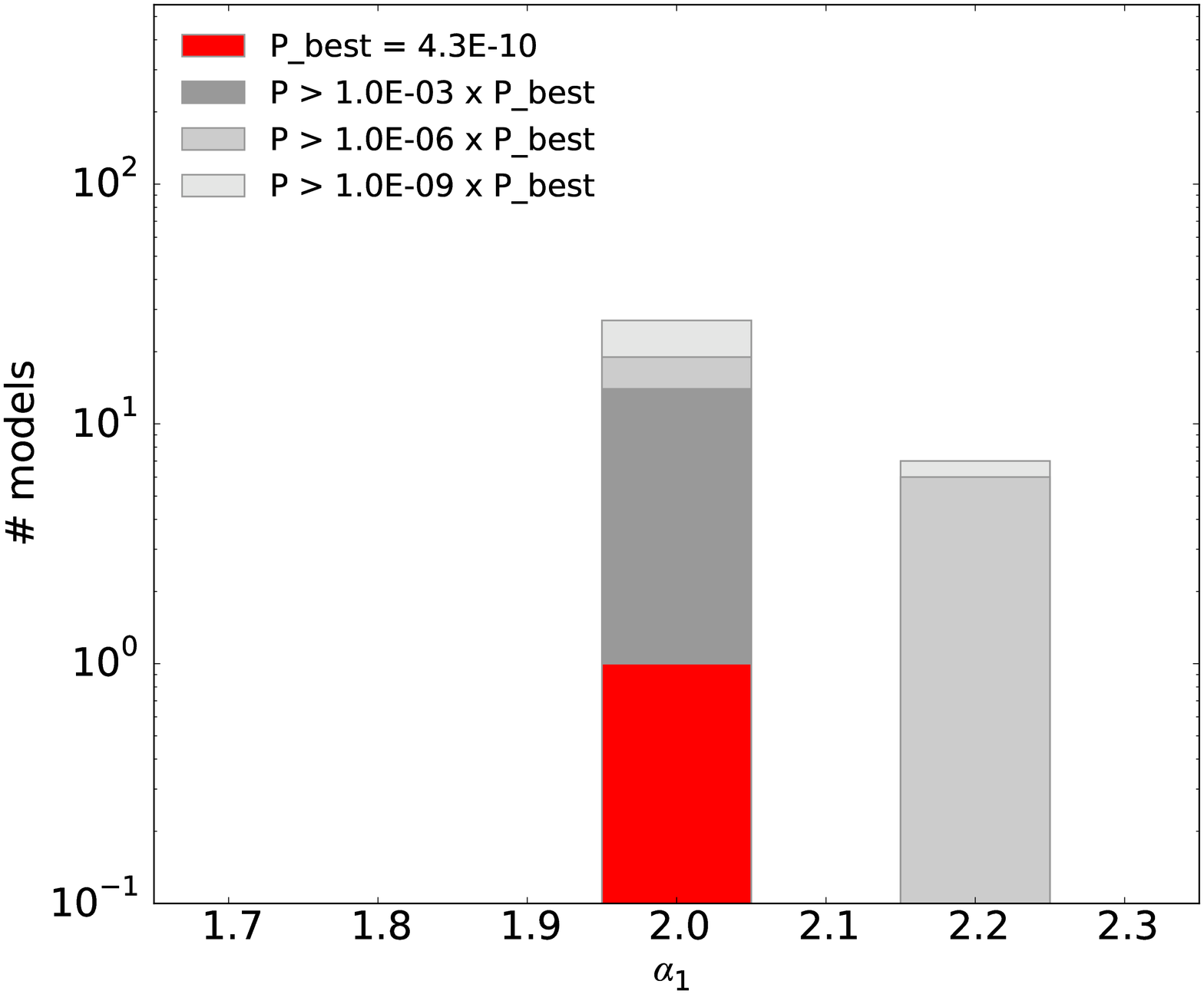}
\includegraphics[width=2.5in]{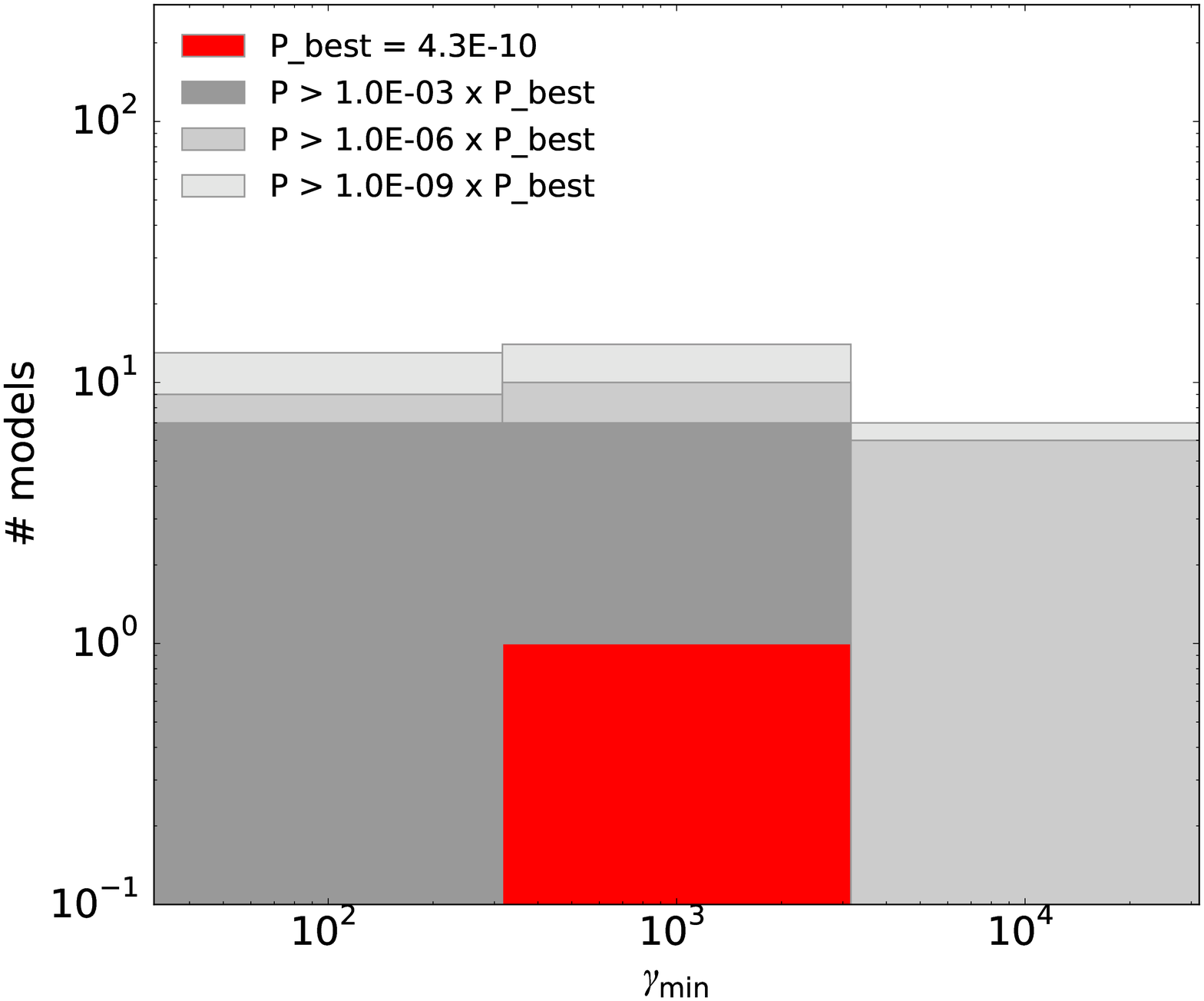}
\includegraphics[width=2.5in]{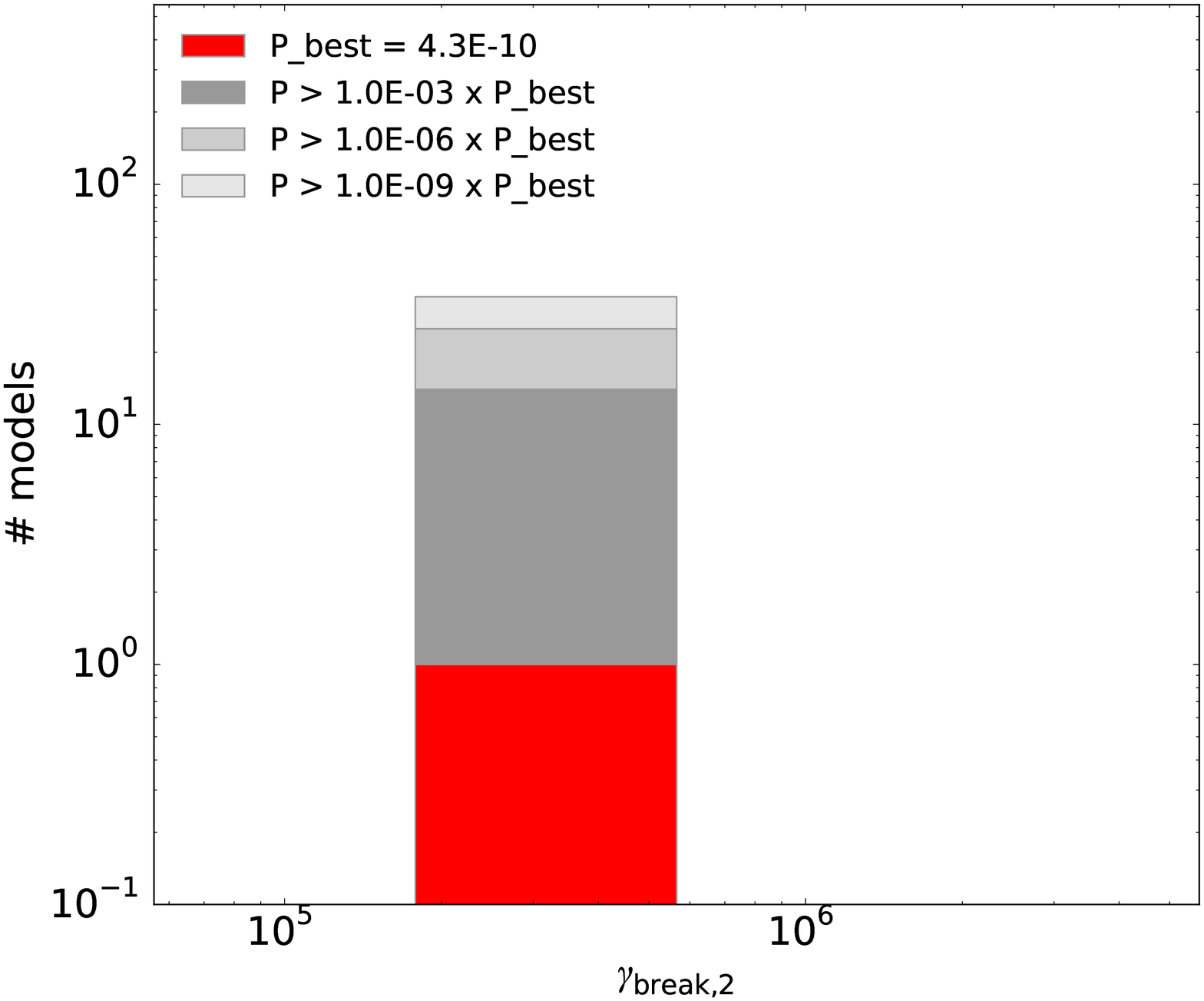}
\includegraphics[width=2.5in]{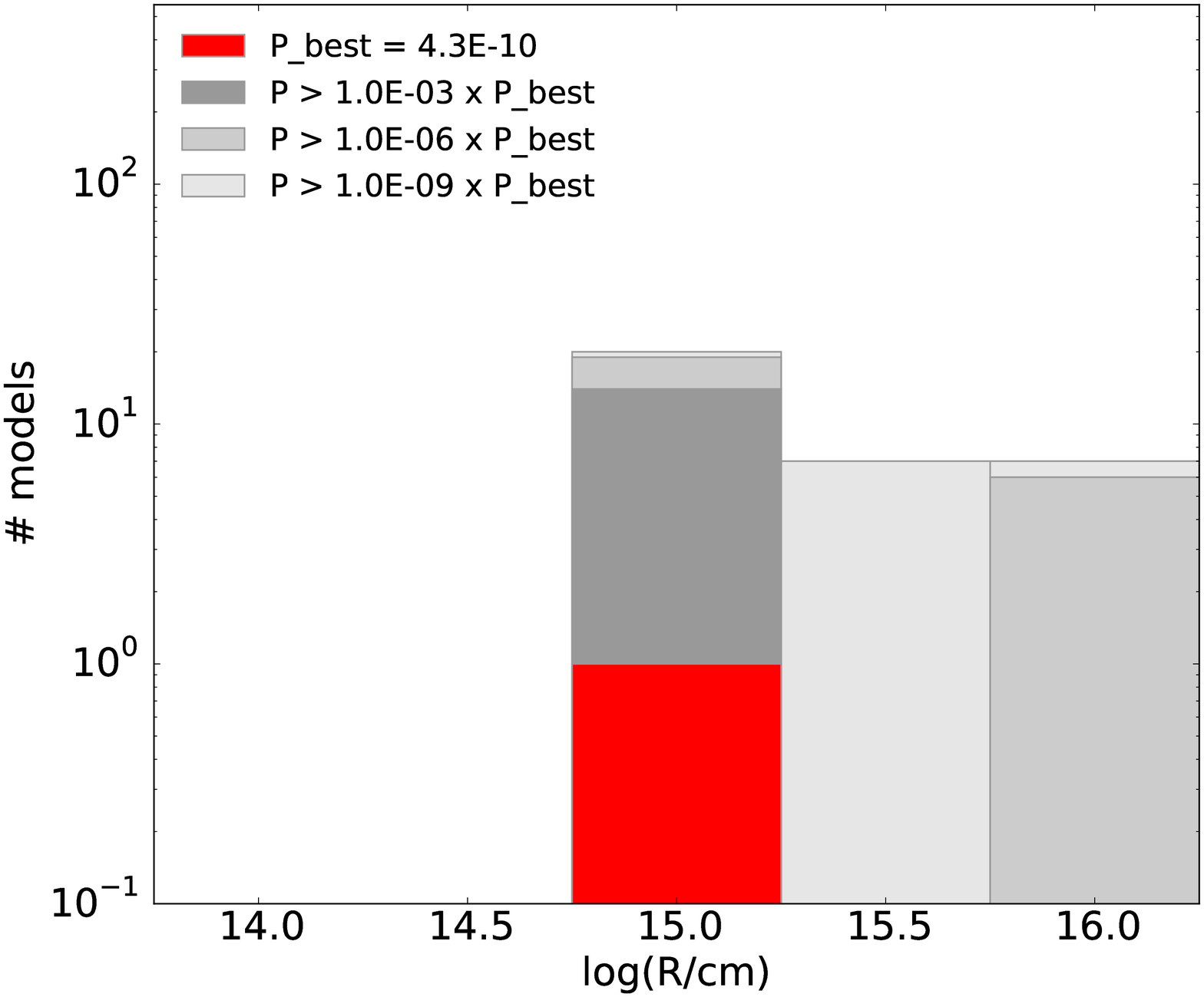}
	\end{center}
\end{minipage}
\hspace*{-5mm}
\begin{minipage}[t]{0.33\linewidth}
	\begin{center}
\includegraphics[width=2.5in]{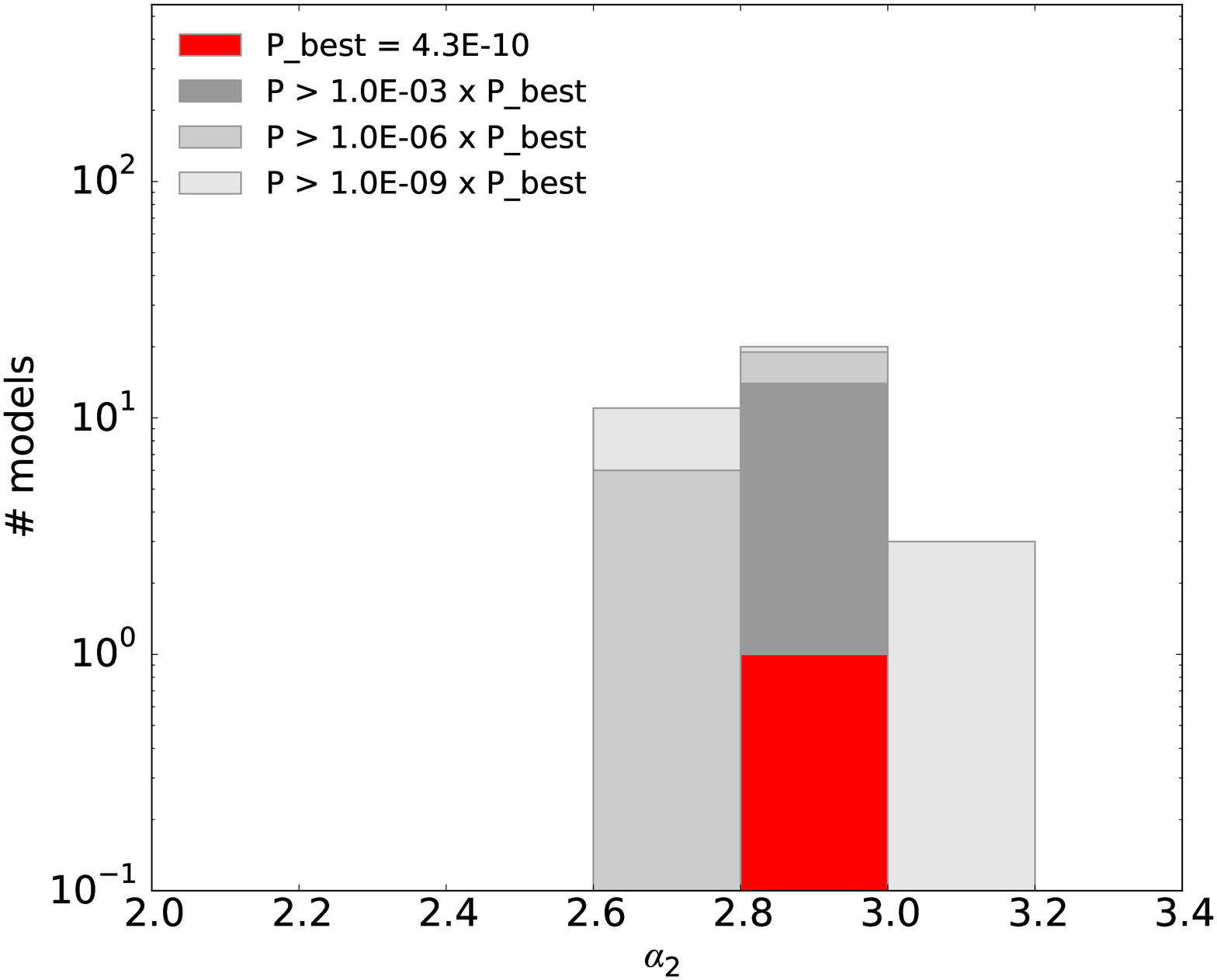}
\includegraphics[width=2.5in]{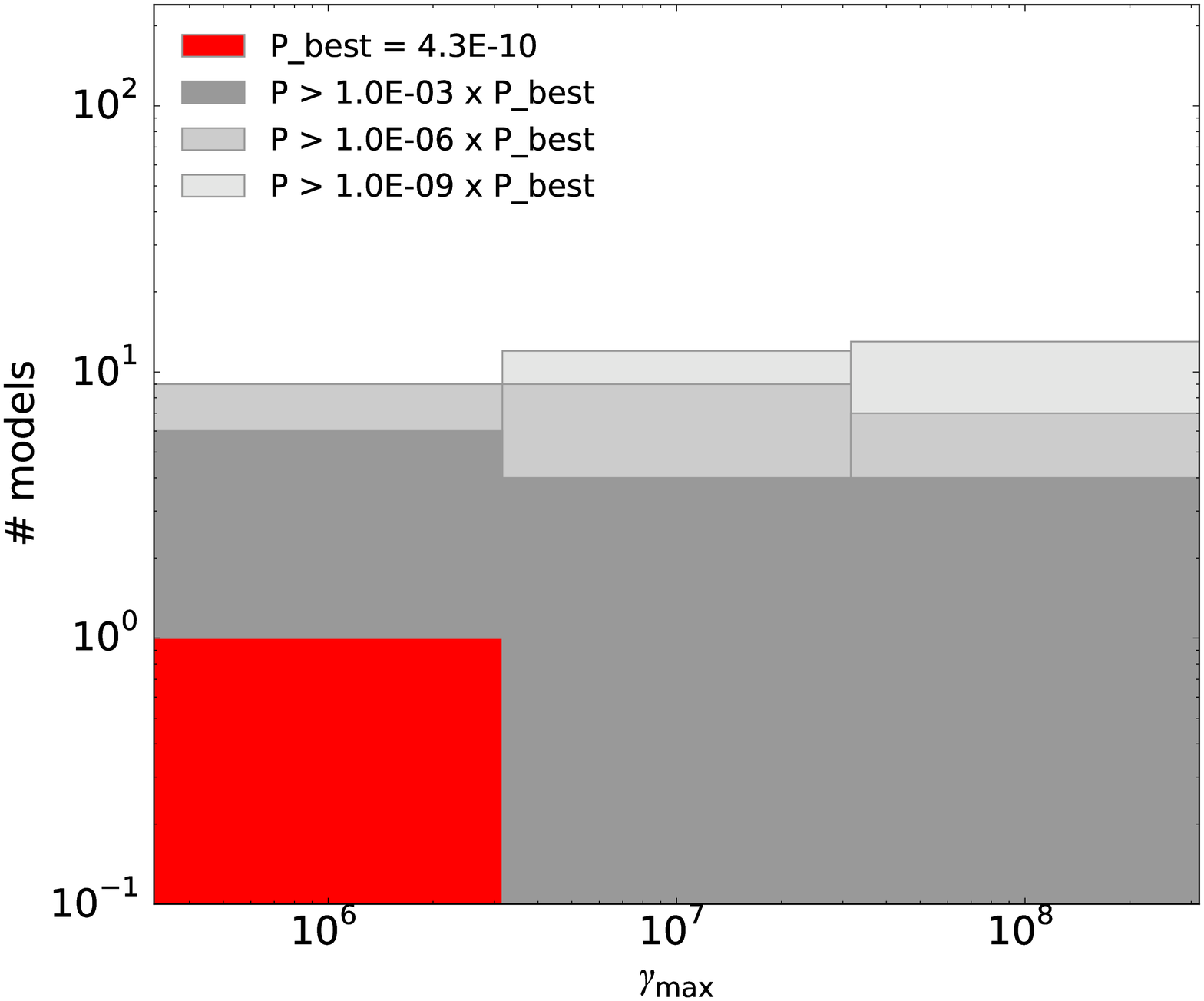}
\includegraphics[width=2.5in]{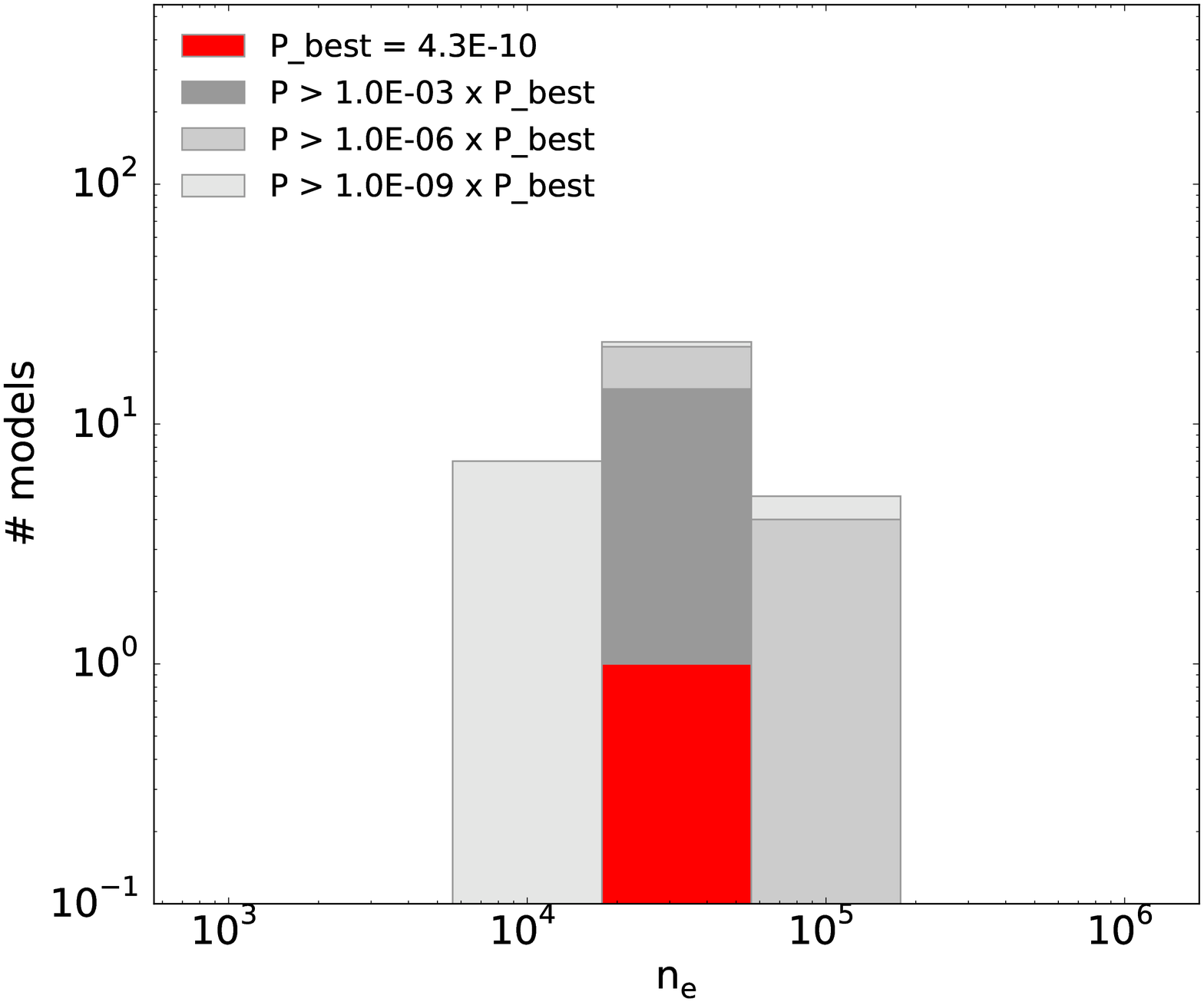}
\includegraphics[width=2.5in]{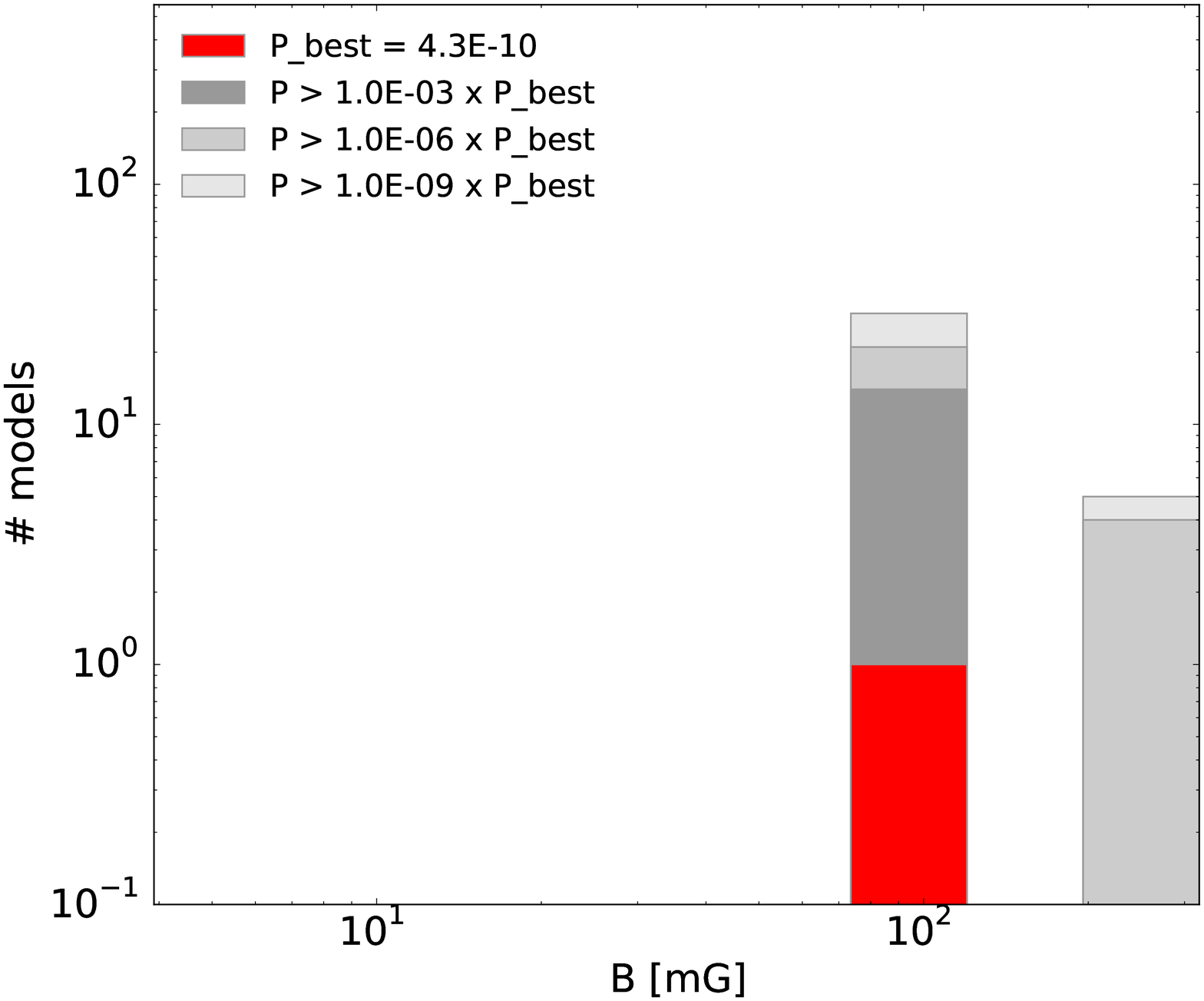}
	\end{center}
\end{minipage}
\hspace*{-5mm}
\begin{minipage}[t]{0.33\linewidth}
	\begin{center}
\includegraphics[width=2.5in]{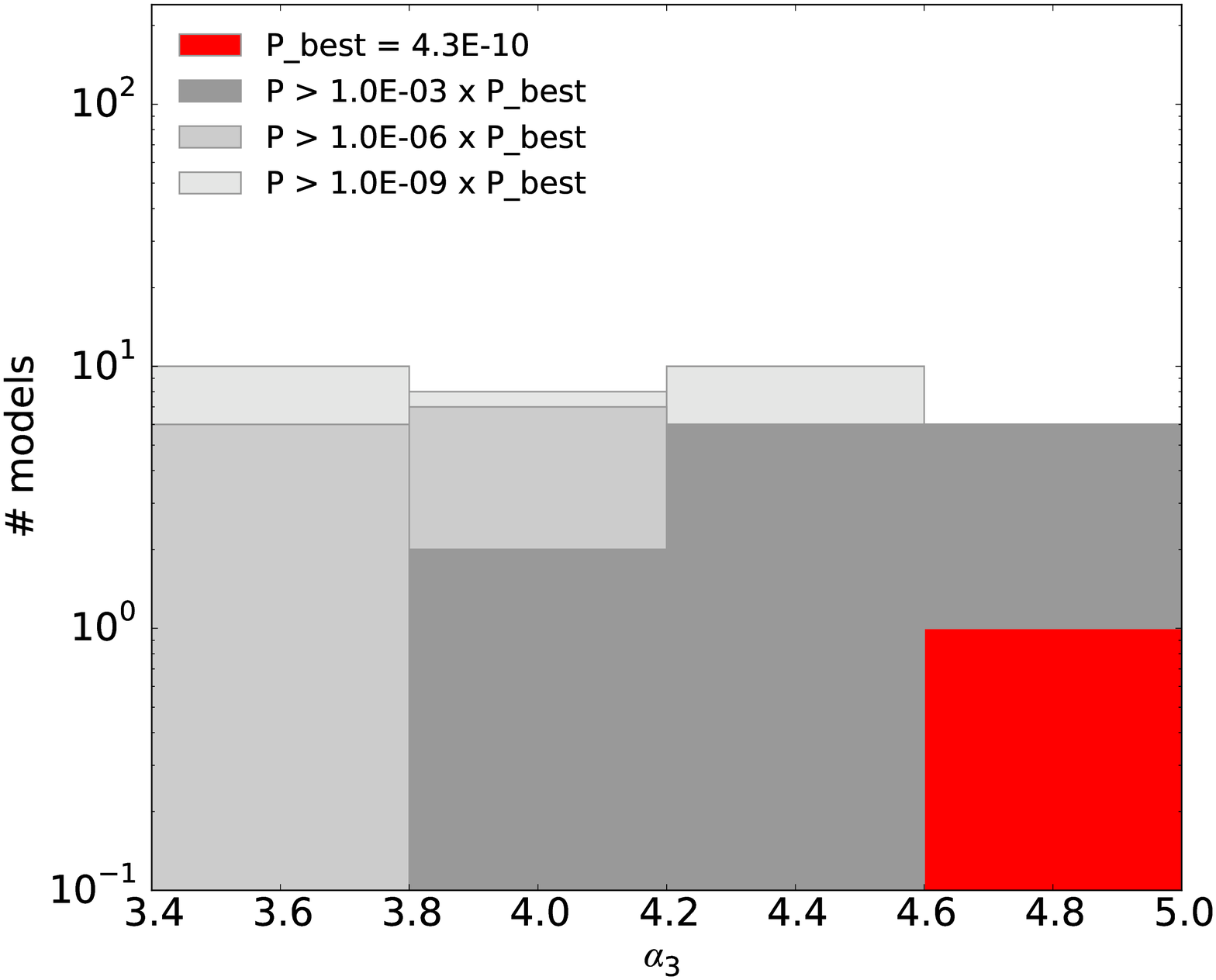}
\includegraphics[width=2.5in]{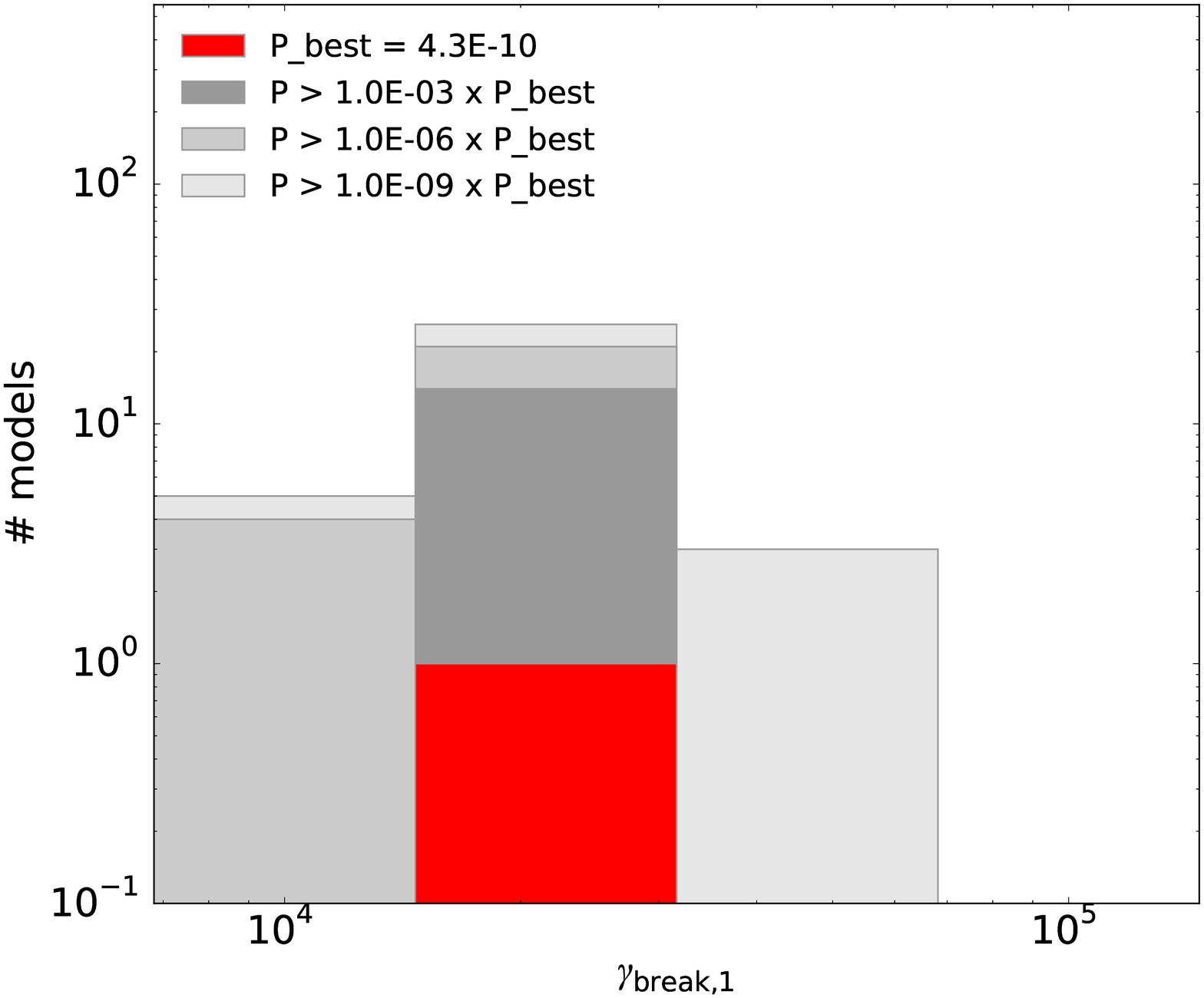}
\includegraphics[width=2.5in]{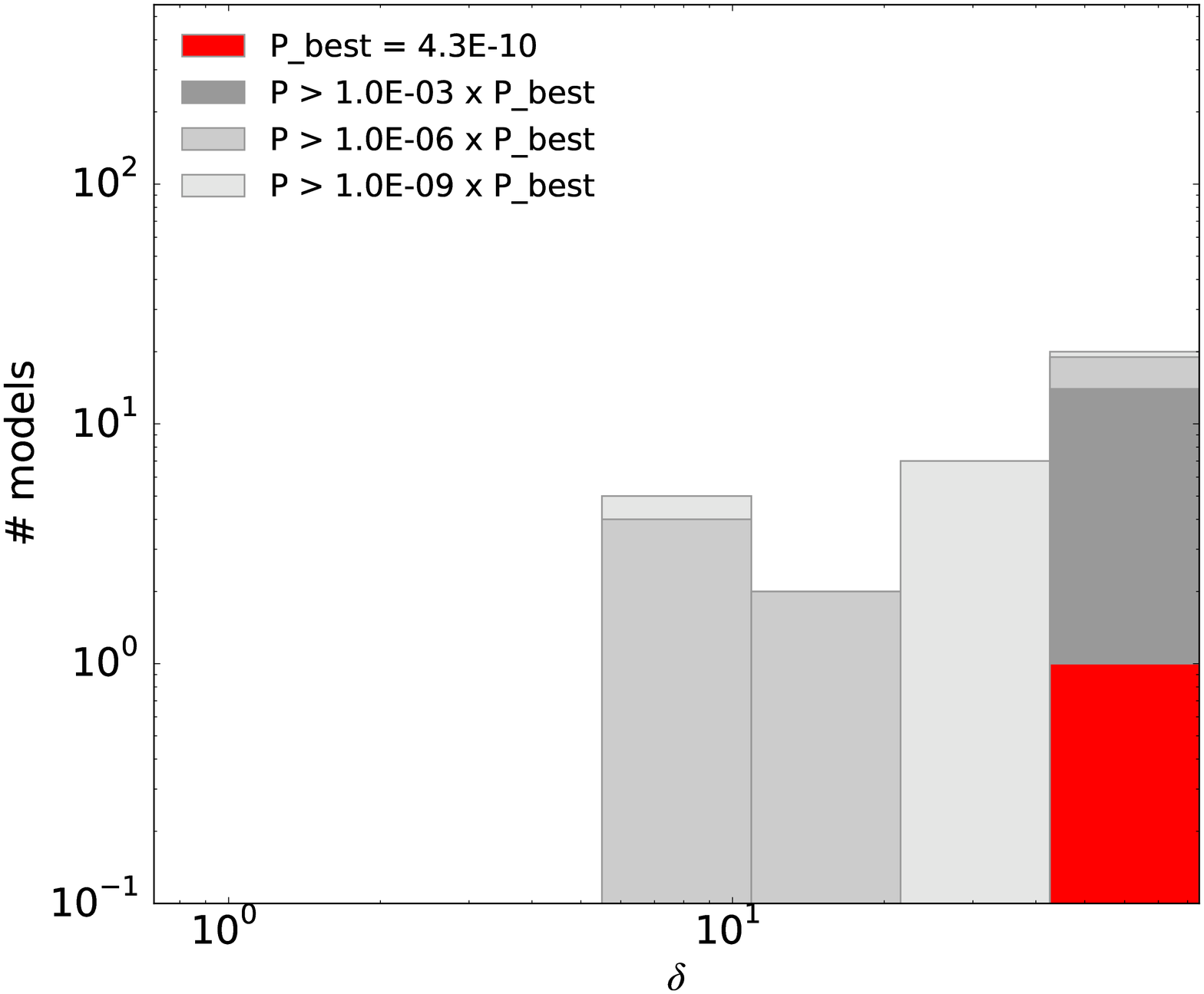}
	\end{center}
\end{minipage}
		\caption{Number of SSC model curves with a fit
                  probability above the given limits vs. each probed
                  value of each model parameter. Given are the results for the coarse parameter grid within a one-zone scenario for MJD~54973.  The X-axis of each plot spans over the probed range for each parameter. The figure shows the model with the highest probability of agreement to the data (red) and all models within several probability bands (different grey shades, see legend). }
	  \label{fig:paramrangesonezonecoarse}
\end{figure*}  

The results obtained for the one-zone scenario following the grid-scan from Tab.~\ref{tab:sscparameterspaceonezone} gave a best probability of agreement with the data points of $P_{\text{best}}  \approx4\times10^{-10}$ ($\chi^2/d.o.f. \approx 123/41$). We found that there are 14 additional SSC model curves with a model-to-data probability larger than 0.1\% 
of the best-matching model (i.e. $P > 10^{-3} \times P_{\text{best}}$), which we set as a generous probability threshold to consider the model-to-data agreement comparable. 
Given that  $P_{\text{best}}  \approx4\times10^{-10}$, even those models with comparatively best agreement $P > 10^{-3} \times P_{\text{best}}$ do not adequately describe the measured broadband SED. Yet this 
relatively bad model-to-data agreement is not worse than some of the agreements between (simple) models and SED data shown in some studies \citep[e.g.][]{2010ApJ...716...30A,2012A&A...541A.160G,2013ApJ...770...77D,2016MNRAS.459.3271A}. 
This occurs because, in most studies involving broadband SEDs, the
models are adjusted ``by eye'' to the data without any rigorous
mathematical procedure that quantifies the model-to-data agreement. Differences on the order of 20\%-30\% in a log-log plot spanning many orders of magnitude do not ``appear to be problematic'', despite these differences could be (statistically) significant due to the small errors from some of the data points (e.g. optical/UV and X-ray). If the differences between the data and model are not substantial (regardless of the statistical agreement), the models are considred to approximately describe the data and used to extract some physical properties of the source and its environment. 

\begin{figure*}[!th]
	\begin{center}
\includegraphics[width=5.1in]{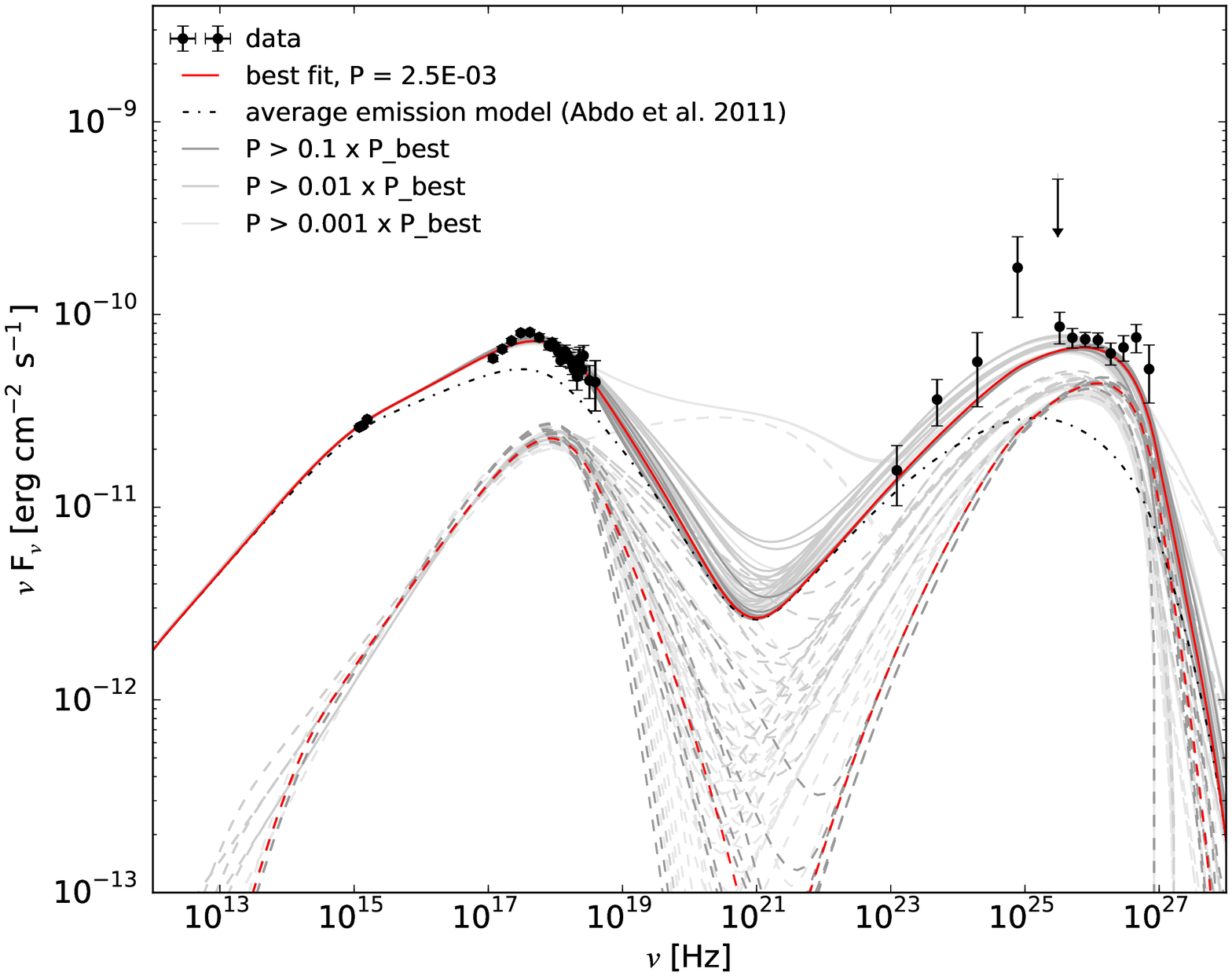}
\end{center}
\caption{SED grid-scan modelling results for the flaring episode around MJD 54973 in the scope of a two-zone SSC scenario. The total emission (solid lines) is assumed to stem from a first quiescent region (black dot-dashed lines) responsible for the average state \citep{2011ApJ...727..129A} plus a second emission region (dashed lines). The model with the highest probability of agreement with the data is highlighted in red. Model curves underlaid in grey show the bands spanned by models with a fit probability better than $0.1\times P_{\text{best}}$, $0.01\times P_{\text{best}}$ and $0.001\times P_{\text{best}}$, respectively. Data points have been corrected for EBL absorption according to the model by \cite{Franceschini:2008vt}.}
\label{fig:flare2twozonesedmodeling_coarse}
\end{figure*}
\begin{figure*}[!th]
\begin{minipage}[b]{0.33\linewidth}
	\begin{center}
%\vspace*{-5mm}
\includegraphics[width=2.5in]{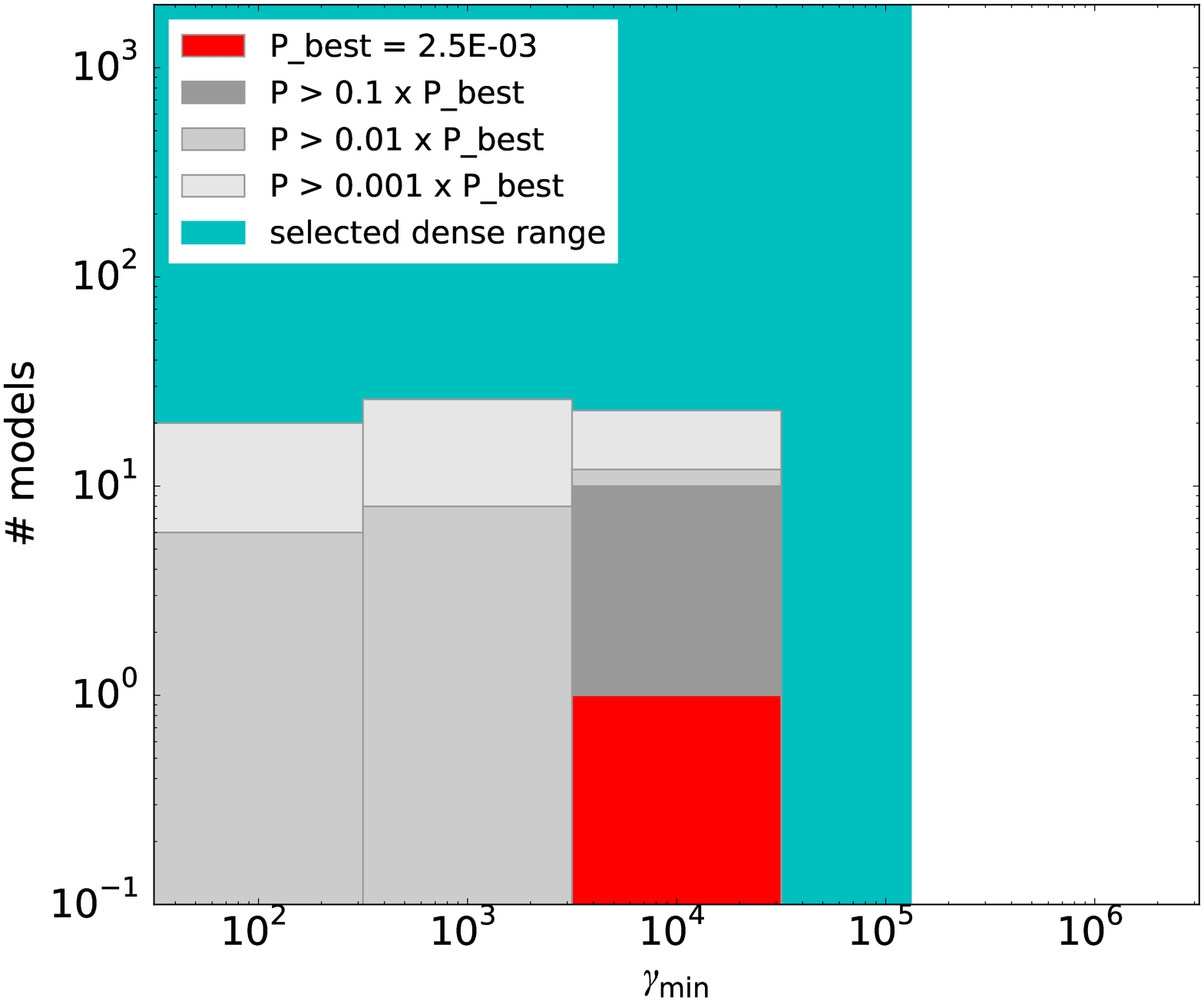}
\includegraphics[width=2.5in]{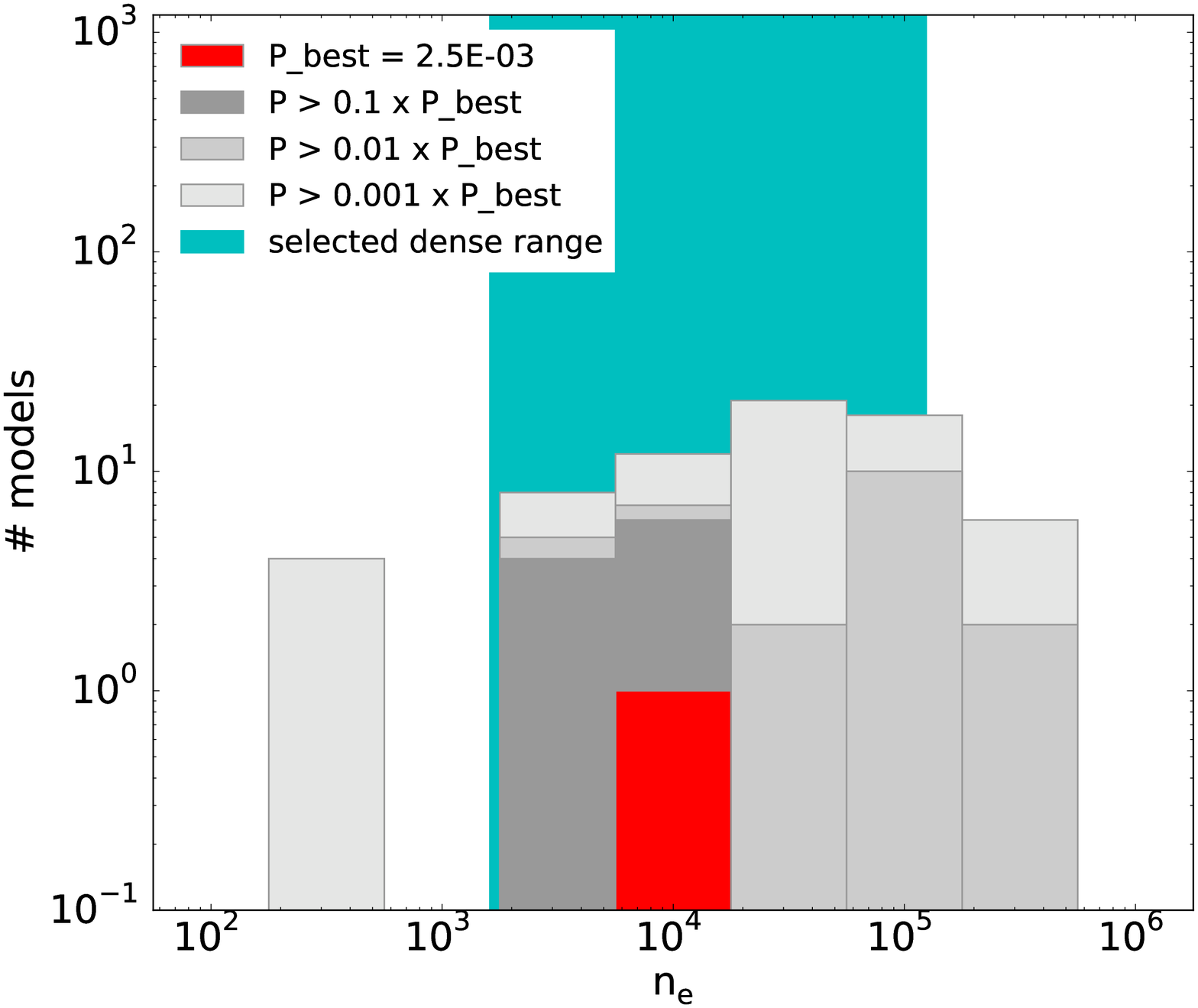}
\includegraphics[width=2.5in]{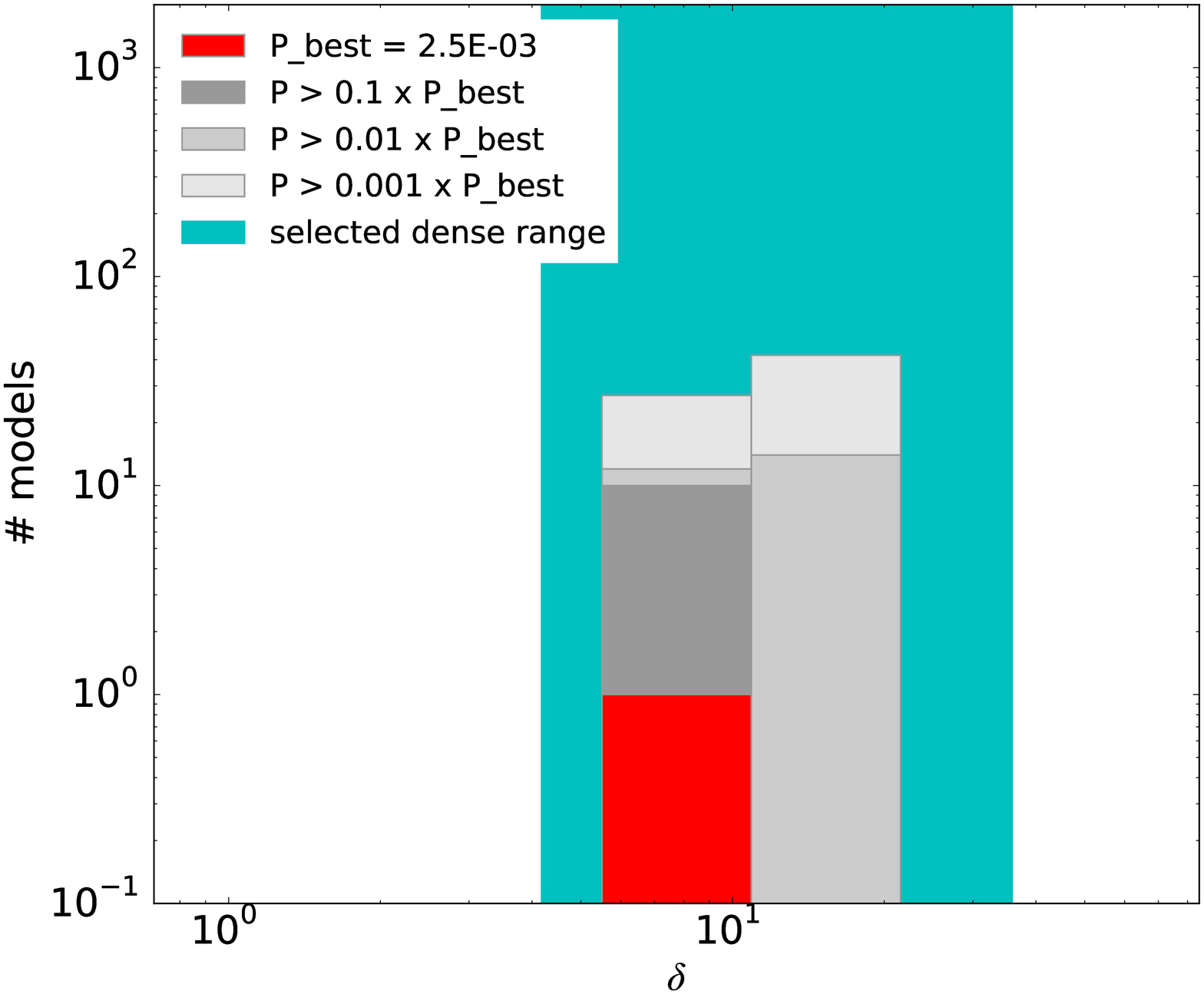}
	\end{center}
\end{minipage}
\hspace*{-5mm}
\begin{minipage}[b]{0.33\linewidth}
	\begin{center}
%\vspace*{-5mm}
\includegraphics[width=2.5in]{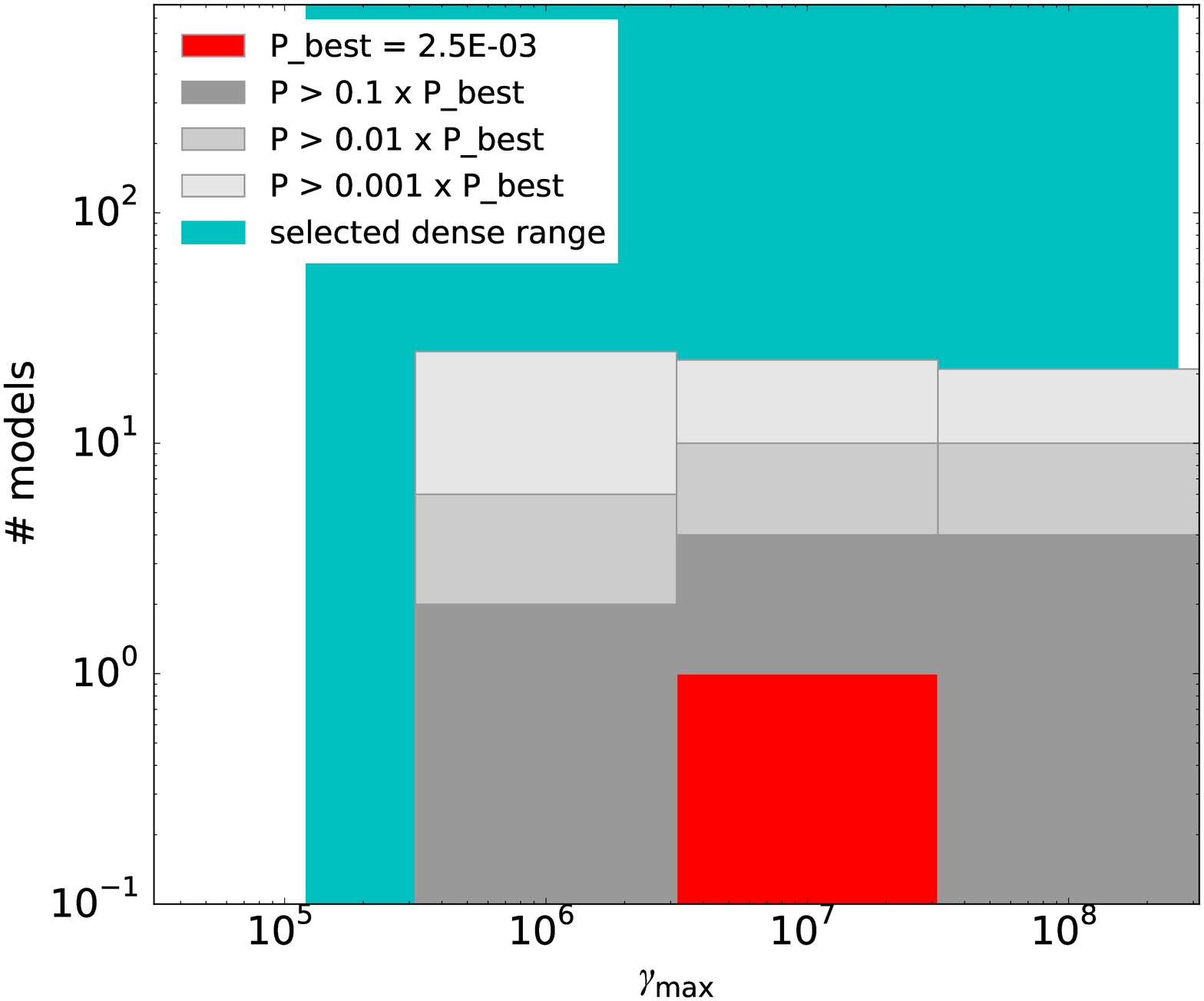}
\includegraphics[width=2.5in]{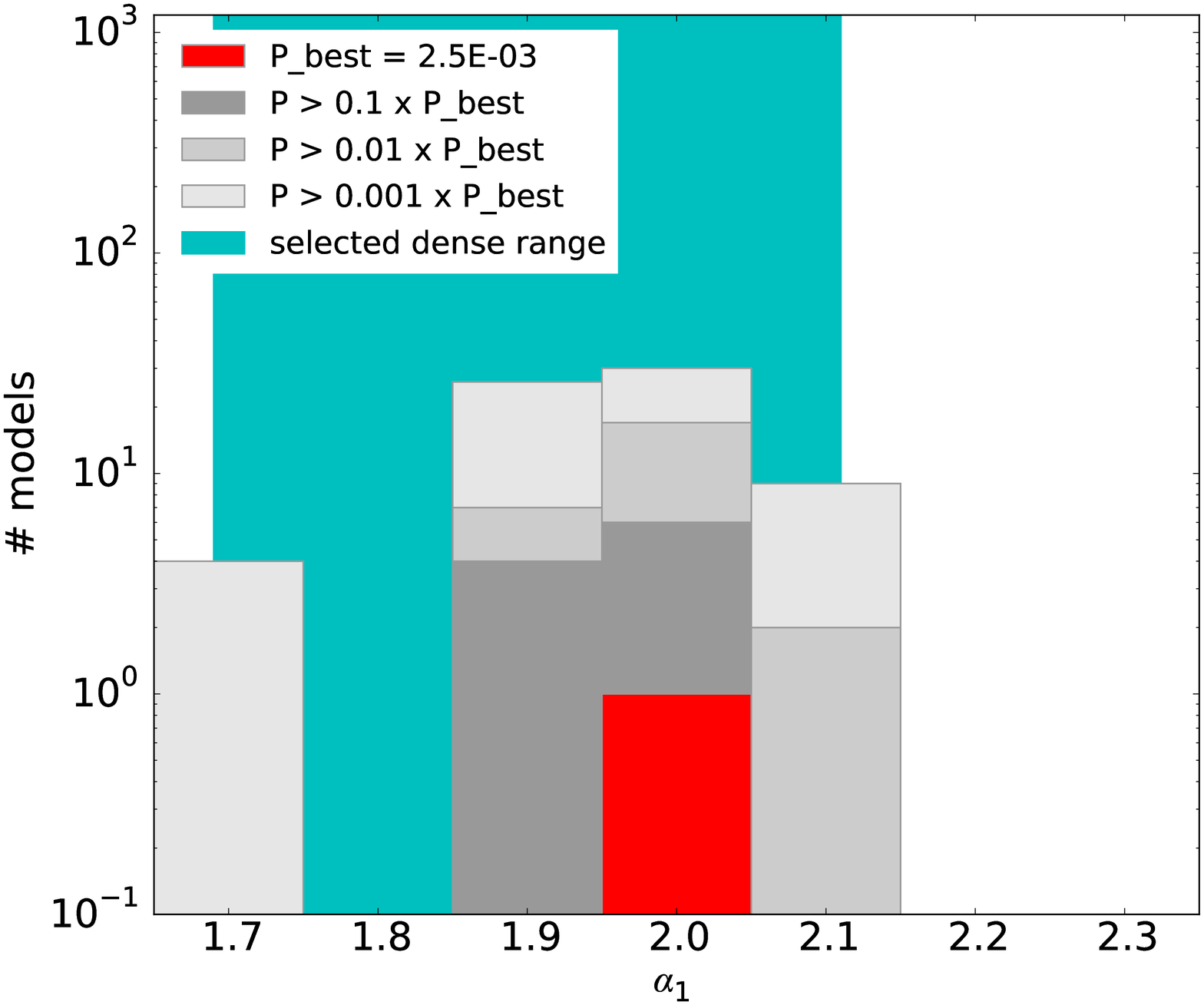}
\includegraphics[width=2.5in]{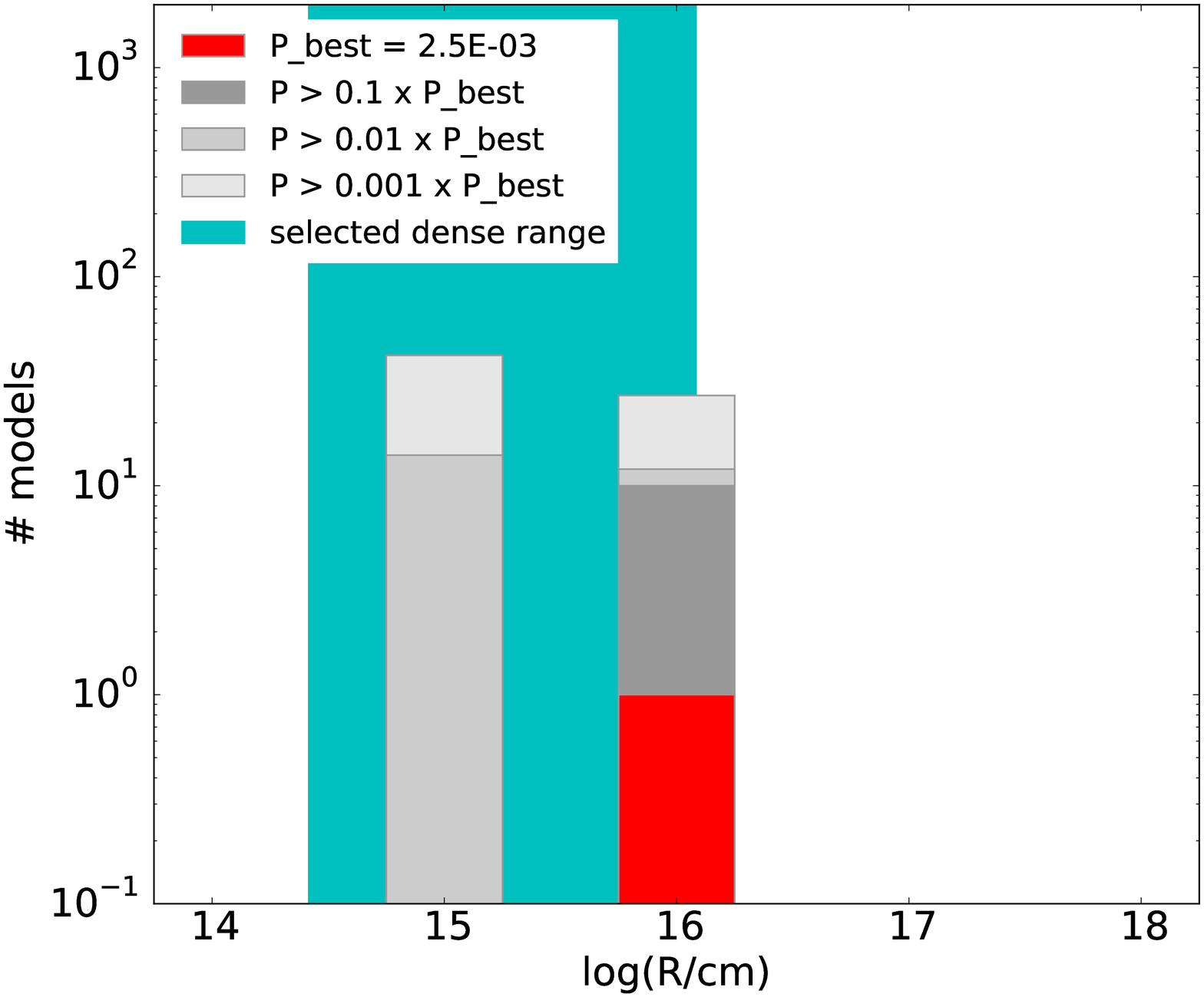}
	\end{center}
\end{minipage}
\hspace*{-5mm}
\begin{minipage}[b]{0.33\linewidth}
	\begin{center}
%\vspace*{-5mm}
\includegraphics[width=2.5in]{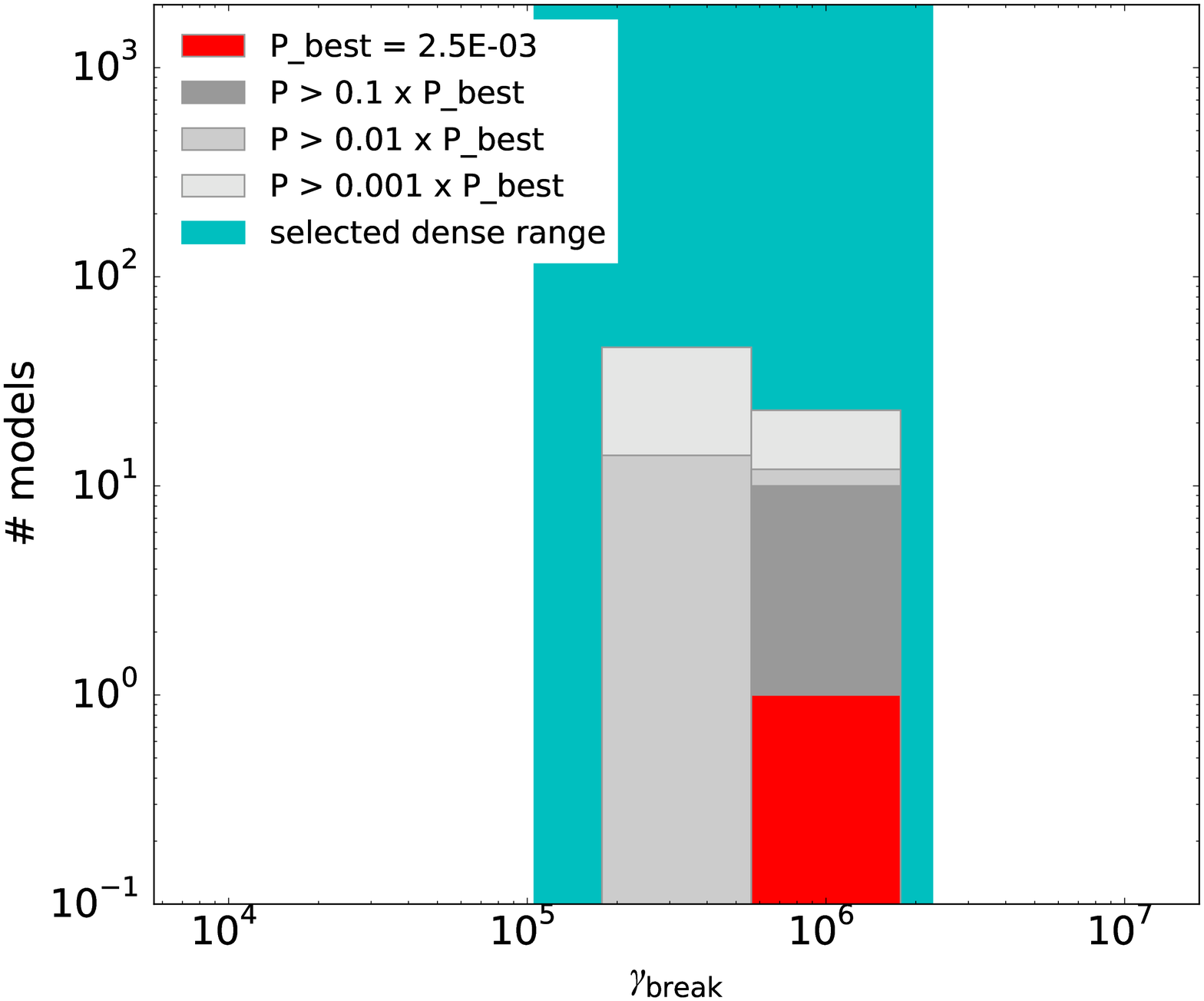}
\includegraphics[width=2.5in]{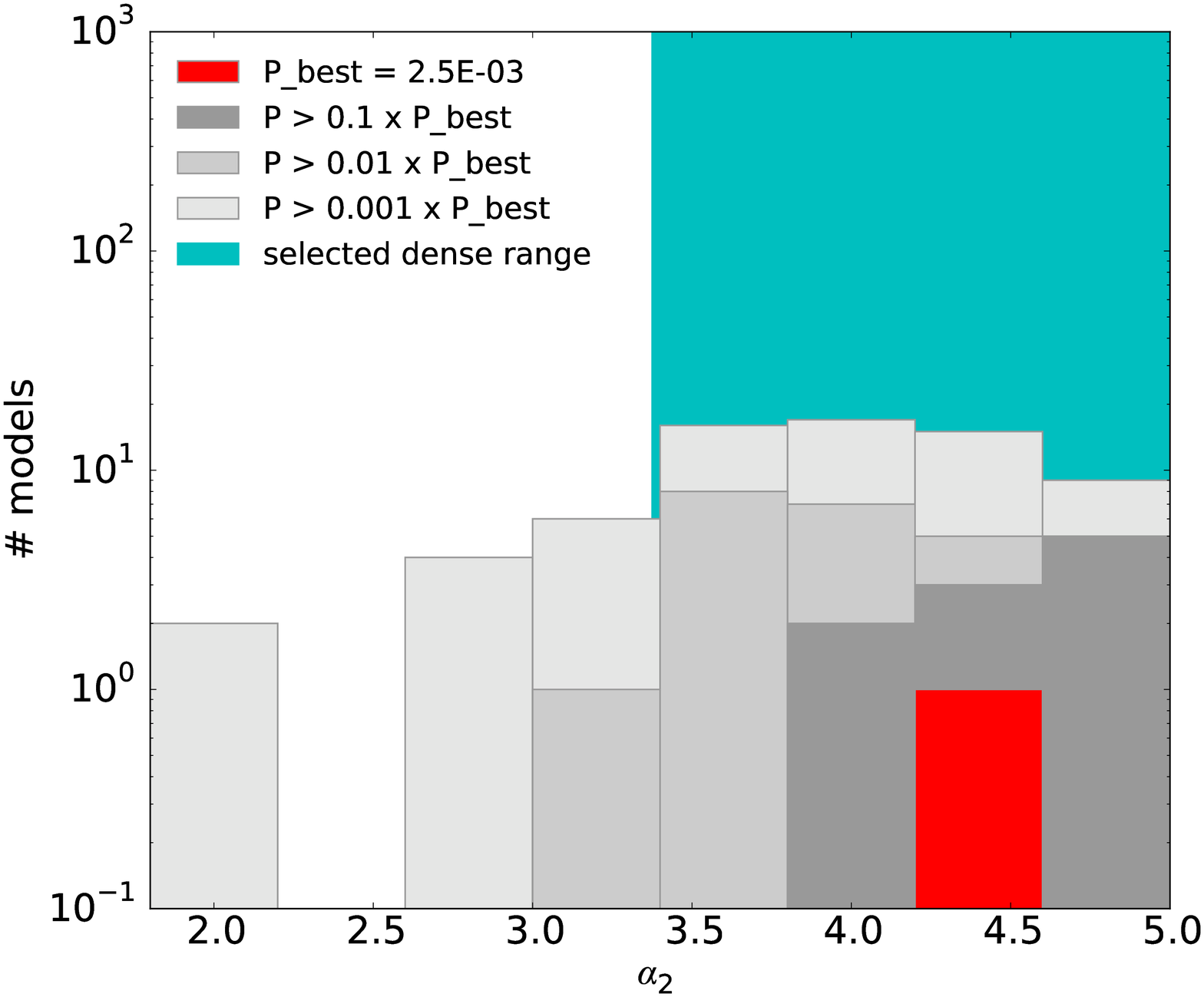}
\includegraphics[width=2.5in]{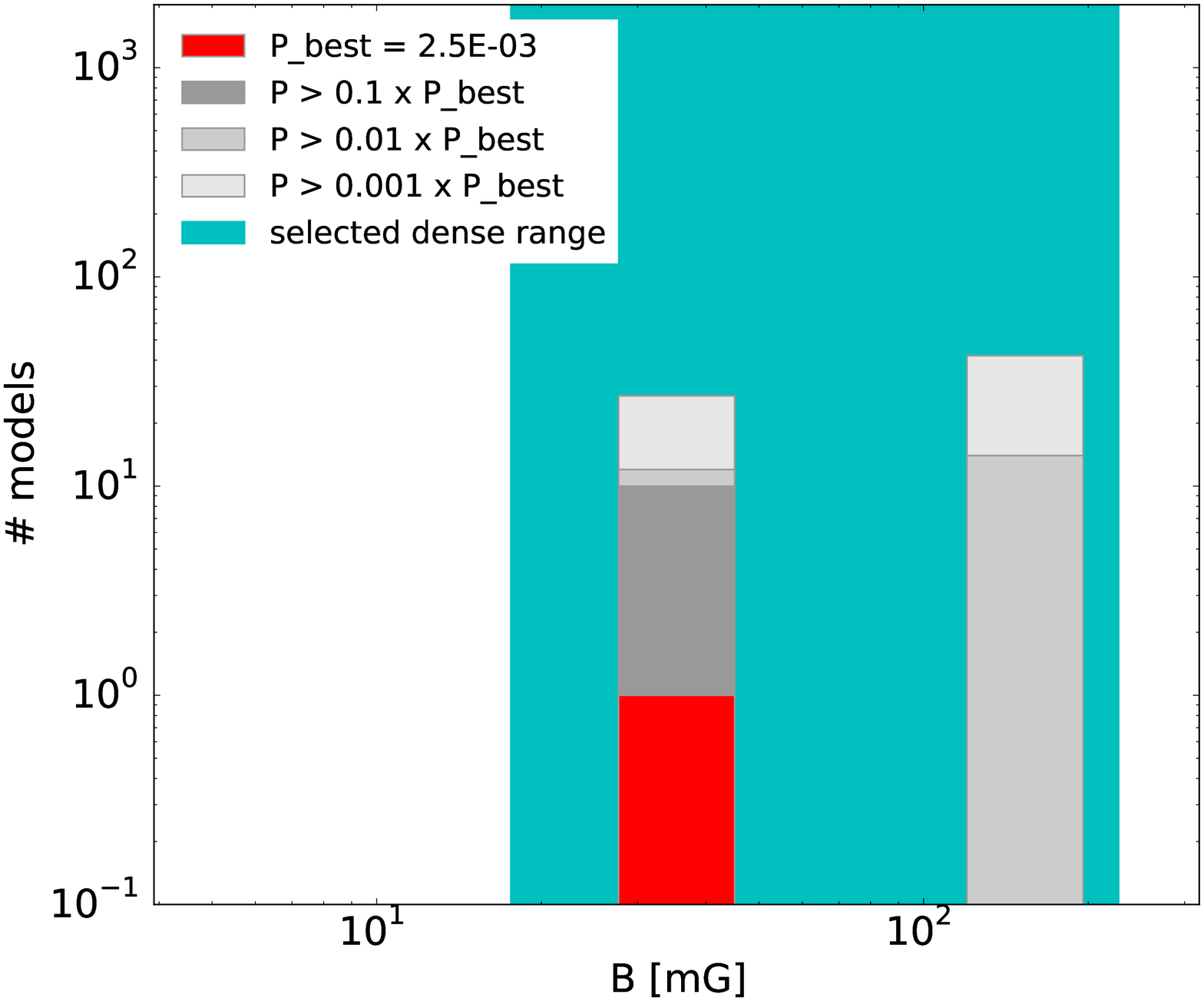}
	\end{center}
\end{minipage}
		\caption{Number of SSC model curves which fulfill the given limits for the fit probability vs. each probed value of each model parameter. Shown are results for the coarse parameter grid-scan within a two-zone scenario for MJD~54973. The X-axis of each plot spans over the probed range for each parameter. Given are the model with the highest probability of agreement with the data and all models within the given probability bands (see legend). The parameter ranges chosen for the dense scan are also shown in each plot. }
	  \label{fig:sedparamrangescoarse}
\end{figure*} 

Figure~\ref{fig:flare2onezonesedmodeling} depicts the best SSC model curves from the one-zone scenario, with the model featuring the best agreement to the data shown with a red curve, and the other 14 SSC models with comparable (down to 0.1\%) model-to-data agreement shown with dark-grey curves. Given the very low number of SSC model curves in this group, we decided to depict additionally those SSC models with model-to-data probability of agreement larger than $10^{-6} \times P_{\text{best}}$ and $10^{-9} \times P_{\text{best}}$ with lighter grey shades (see legend), which increased the number of SSC model curves depicted to 34. The thresholds used of $10^{-6} \times P_{\text{best}}$ and $10^{-9} \times P_{\text{best}}$ are somewhat arbitrary, and could be changed without any major qualitative impact in the reported results. The inclusion of these additional 20 models in the figure helps illustrate the behaviour of the SSC model curves that start being worse than the best-matching model. To guide the eye, the SSC model describing the average state is also shown \citep[from][dash-dotted black line]{2011ApJ...727..129A}. One can see that the most significant deviations of the model curves from the data points stem from the \textit{Swift} region. Therefore, while the hard X-ray and $\gamma$-ray bands can be satisfactorily modeled with a one-zone SSC scenario, this model realization fails at reconstructing both the soft X-ray data points and the UV emission at the same time. 
Figure~\ref{fig:paramrangesonezonecoarse} displays how many model curves produced for each point on the parameter grid yield a model-to-data agreement probability $P$ better than $10^{-3} \times P_{\text{best}}$, which are the models that are considered to be comparable. This is shown for each of the parameters separately. One can see that some parameters are more constrained than others: e.g.~$\gamma_{\text{break},1}$, $\gamma_{\text{break},2}$ and $\alpha_2$ show a narrower distribution than for instance $\gamma_{\text{max}}$ or $\alpha_3$, which lead to equally good models over essentially the entire range of values probed. Additionally, as done for Fig.~\ref{fig:flare2onezonesedmodeling}, with lighter grey shades we also report the parameter values for  $P>10^{-6} \times P_{\text{best}}$ and $P>10^{-9} \times P_{\text{best}}$.  One can see that the SSC models that are not comparable to the best-matching model (i.e. those with $P<10^{-3} \times P_{\text{best}}$), have a similar distribution for those parameters that are not constrained, like $\gamma_{\text{max}}$ or $\alpha_3$. On the other hand, on the parameters that can be constrained, like $\gamma_{\text{break},1}$, and $\alpha_2$, these additional models extend the range of parameter values with respect to the distributions for the models with  $P>10^{-3} \times P_{\text{best}}$.  The parameter $\gamma_{\text{break},2}$ seems to be quite well constrained, and even the models with $P<10^{-3} \times P_{\text{best}}$ converge to the same value of  3.2$\times 10^{5}$.  The implications of these distributions on the possibility to constrain the different model parameters will be further discussed in Sect.~\ref{sec:discussion}.

We also evaluated the model-to-data agreement for the one-zone scenario that uses the more simple grid-scan defined by Table~\ref{tab:sscparameterspacetwozone}, which is related to a grid of 9 parameters (instead of 11), but with a somewhat extended regions for some of these parameters. We found that this grid-scan did not provide any additional SSC model with $P>10^{-3} \times P_{\text{best}}$, and only five additional SSC models with $P>10^{-9} \times P_{\text{best}}$. Hence  this grid-scan did not bring any practical improvement with respect to that from Table~\ref{tab:sscparameterspaceonezone}, that led to 14 SSC models with $P>10^{-3} \times P_{\text{best}}$, and 34 SSC models with $P>10^{-9} \times P_{\text{best}}$.

When using the above-mentioned two-zone SSC scenario, with the quiescent emission characterized by the model parameters from the average SED reported in Table~\ref{tab:sscaveragestate}, and the spatially independent region responsible for the flaring activity modeled based on the coarse grid parameter values reported in Table~\ref{tab:sscparameterspacetwozone}, we find a substantial improvement, with respect to the one-zone models, in describing the measured broadband SED (including the UV emission), with a  best model-to-data probability of $P_{\text{best}}  \approx2.5\times10^{-3}$ ($\chi^2/d.o.f. \approx 71/41$). 
The two-zone model provides a better description of the SED than the one zone model, but it still does not reproduces the data perfectly.

Fig.~\ref{fig:flare2twozonesedmodeling_coarse} displays the 69 SSC model curves with model-to-data agreement probability $P$ better than $10^{-3} \times P_{\text{best}}$, which is the generous probability threshold that we adopted to consider the probability of agreement comparable. Because of the relatively large number of SSC model curves (in comparison to those surviving the same selection criteria in the one-zone scenario), we decided to split those models into three groups according to their model-to-data probability $P$ being better than $10^{-1} \times P_{\text{best}}$, $10^{-2} \times P_{\text{best}}$ and $10^{-3} \times P_{\text{best}}$. Since $P_{\text{best}} \approx$0.25\%,  these models start providing an acceptable representation of the data, with the different bands reporting slightly different levels of success in the model-to-data agreement.   The parameter values for those models are depicted in Fig.~\ref{fig:sedparamrangescoarse}.  As occurred for the one-zone scenario, some parameters are better constrained than others: e.g. $\gamma_{\text{break}}$ shows a narrow distribution, while $\gamma_{\text{min}}$  and $\gamma_{\text{max}}$ show a rather flat distribution. Despite the parameter $\gamma_{\text{min}}$ being probed up to $10^6$, the highest $\gamma_{\text{min}}$ values used in the SSC models that can adequately describe the broad-band SED go only up to $10^4$, which, for the highest B field values reported in Fig.~\ref{fig:sedparamrangescoarse} ($\sim$0.15~G), relate to a cooling time of  $3.5\times10^{6}$ seconds. This is one order of magnitude larger than the dynamical timescale set by the highest R values reported in Fig.~\ref{fig:sedparamrangescoarse} ($10^{16}$cm), hence ensuring the existence of a low-energy cutoff. See Sect.~\ref{sec:discussion} for further discussions of this topic.

In order to refine the adjustment of the different model parameters even further, a second iteration of the grid-scan modelling, referred to as a  dense grid-scan, is performed. The dense grid-scan focuses on the parameter ranges that provide the best model-to-data agreement in the coarse grid-scan, which are depicted in Fig.~\ref{fig:sedparamrangescoarse}.  Following this strategy, the chosen parameter ranges can be narrowed in favour of a smaller step size in the individual parameters, while keeping the computing time at a reasonable amount. The new dense grid ranges and number of steps for each of the parameters are given in Table~\ref{tab:sscparameterspacefine}.

The model with the highest probability of model-to-data agreement in
the dense grid-scan yields $P_{\text{best}} \approx6.6\times10^{-2}$ ($\chi^2/d.o.f. \approx 55.4/41$), which implies an order of magnitude improvement with respect to the best-matching model obtained with the coarse grid-scan. If this model curve had been obtained through a regular mathematical fit, and conservatively considering that the nine dimensions of the grid relate to nine independent and free parameters in the fitting procedure, we would have obtained a {\it p-value} $\approx 6.3\times10^{-3}$ ($\chi^2/d.o.f. \approx 55.4/32$).
The dense grid-scan focussed on relatively good regions of the parameter space, which yielded a large number of SSC curves with a good model-to-data agreement despite the fact that the parameter values of this dense grid-scan still vary largely (implying very different physical conditions in the source).  Because of the large number of model curves, we can be more demanding with the probability threshold for considering the probability of agreement comparable to that of the best-matching model: 
a probability threshold of $0.1 \times P_{\text{best}}$ still keeps 1684 SSC models, which is a large increase in statistics, in comparison to the results obtained with the coarse scan. Given that $P_{\text{best}} \approx$6.6\%, all the models above this probability threshold provide a decent representation of the data. We split these models into three groups according to their model-to-data agreement being $P>0.9 \times P_{\text{best}}$, $P>0.5 \times P_{\text{best}}$, and $P>0.1 \times P_{\text{best}}$, hence reporting somewhat different levels of success in describing the measured broadband SED.

\begin{figure*}[!th]
\begin{minipage}[b]{0.33\linewidth}
	\begin{center}
%\vspace*{-5mm}
\includegraphics[width=2.5in]{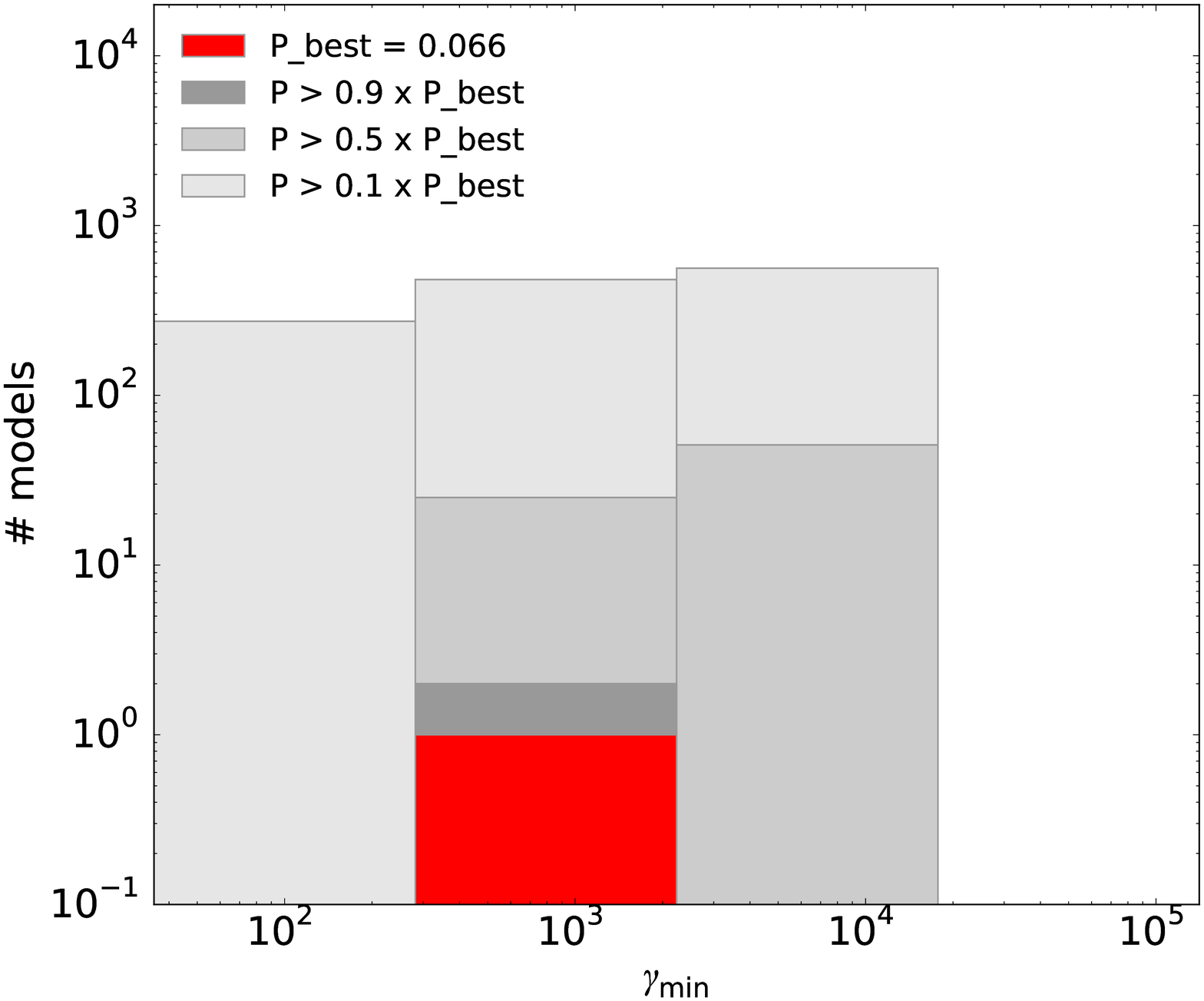}
\includegraphics[width=2.5in]{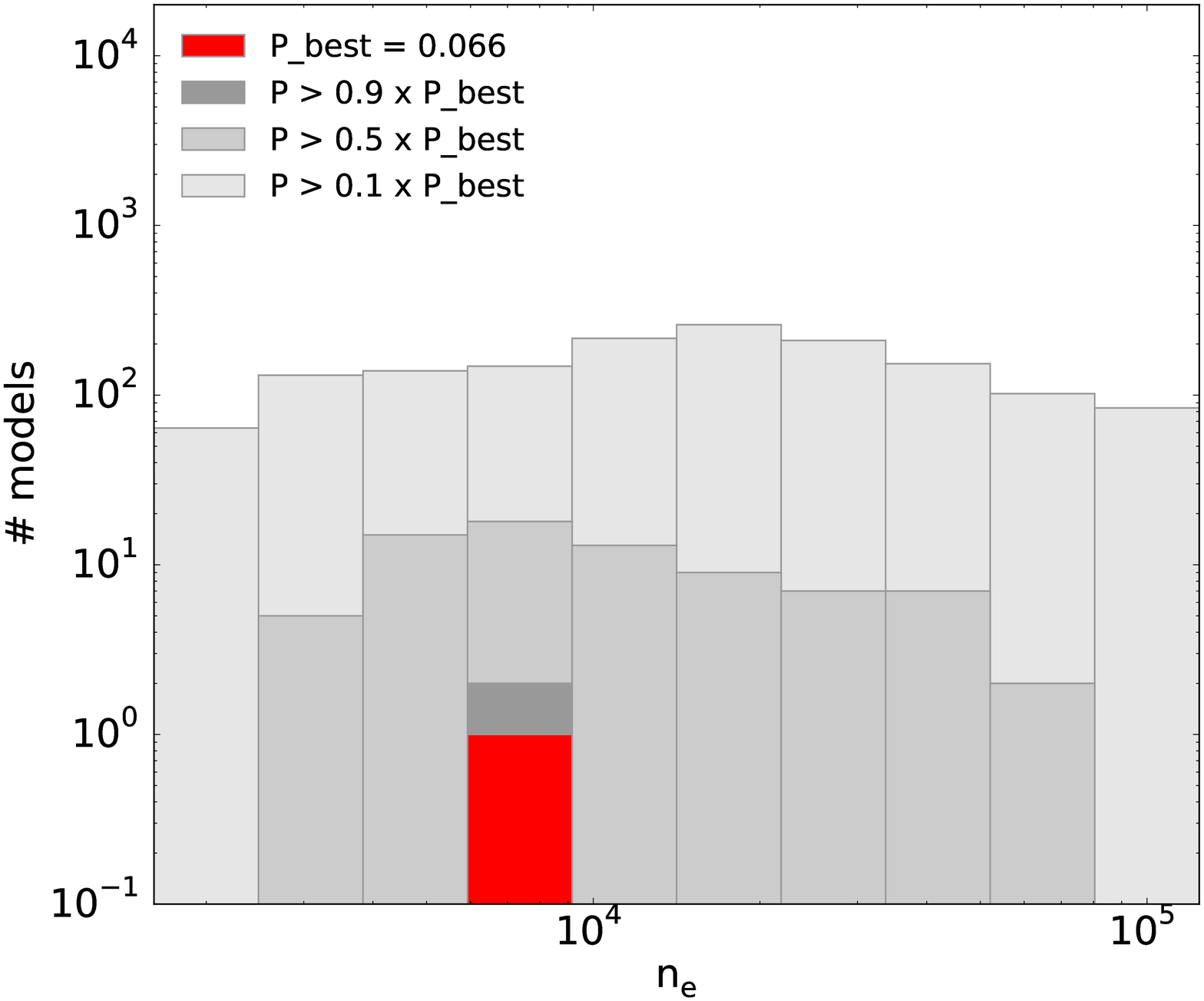}
\includegraphics[width=2.5in]{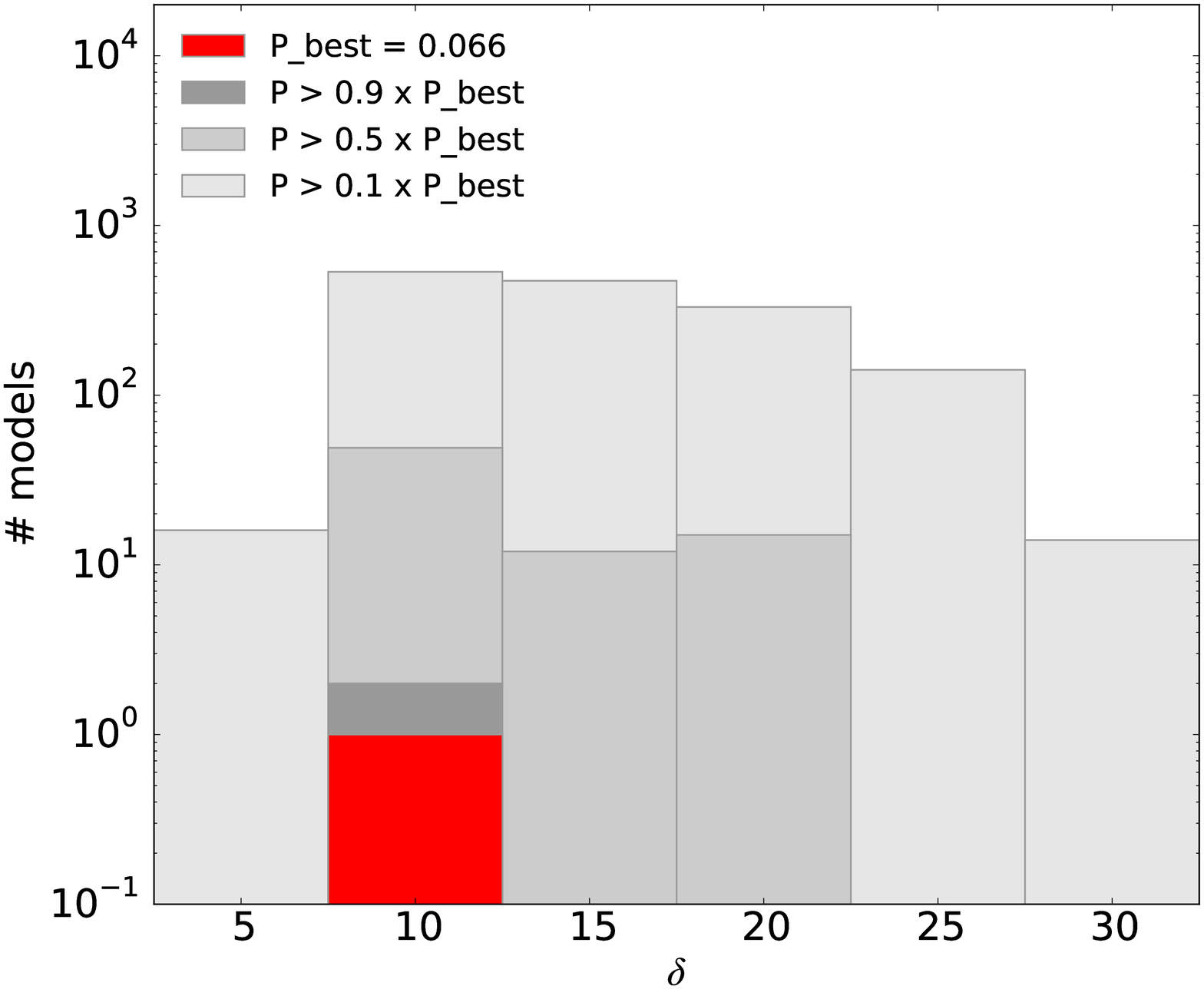}

	\end{center}
\end{minipage}
\hspace*{-5mm}
\begin{minipage}[b]{0.33\linewidth}
	\begin{center}
%\vspace*{-5mm}
\includegraphics[width=2.5in]{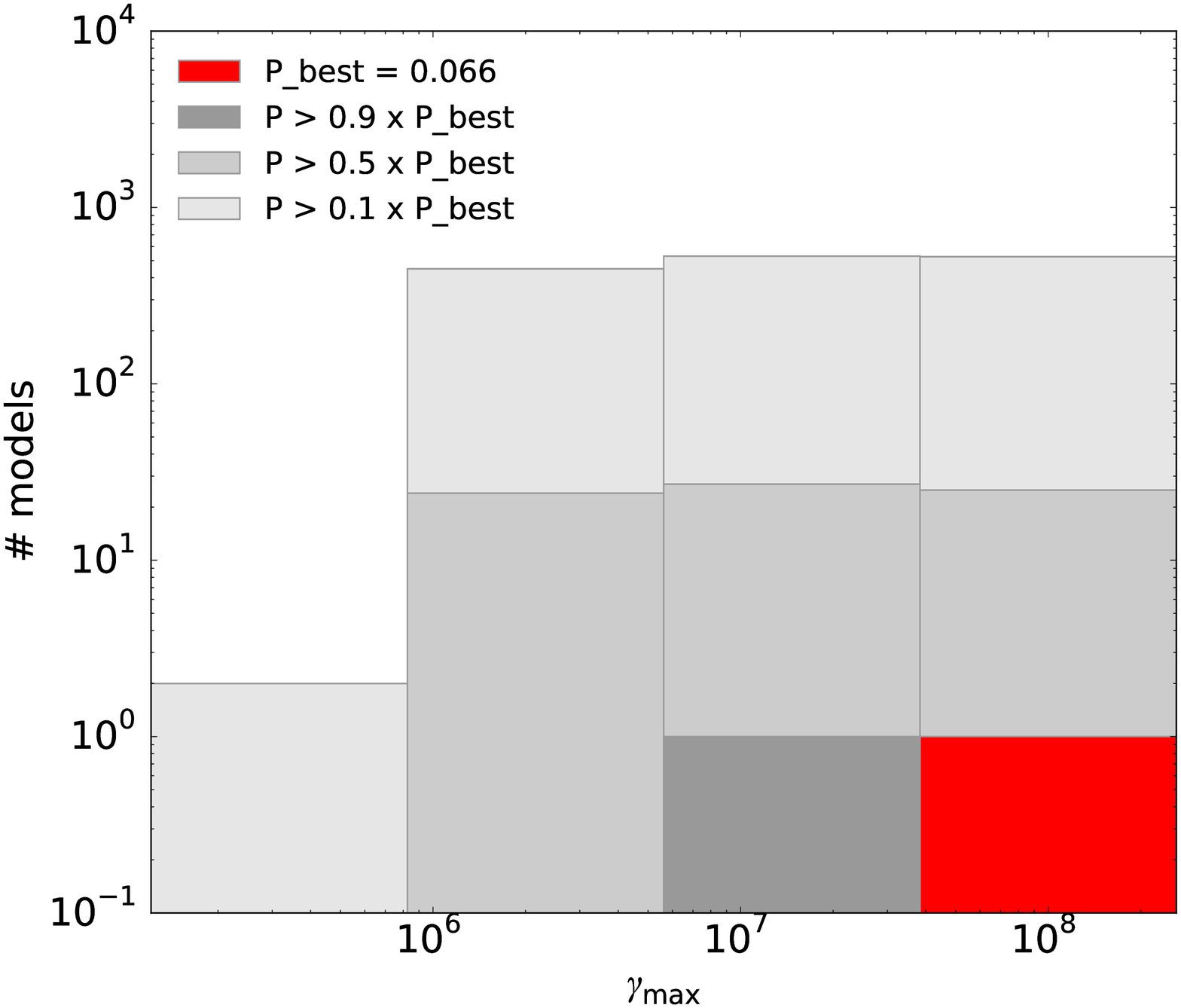}
\includegraphics[width=2.5in]{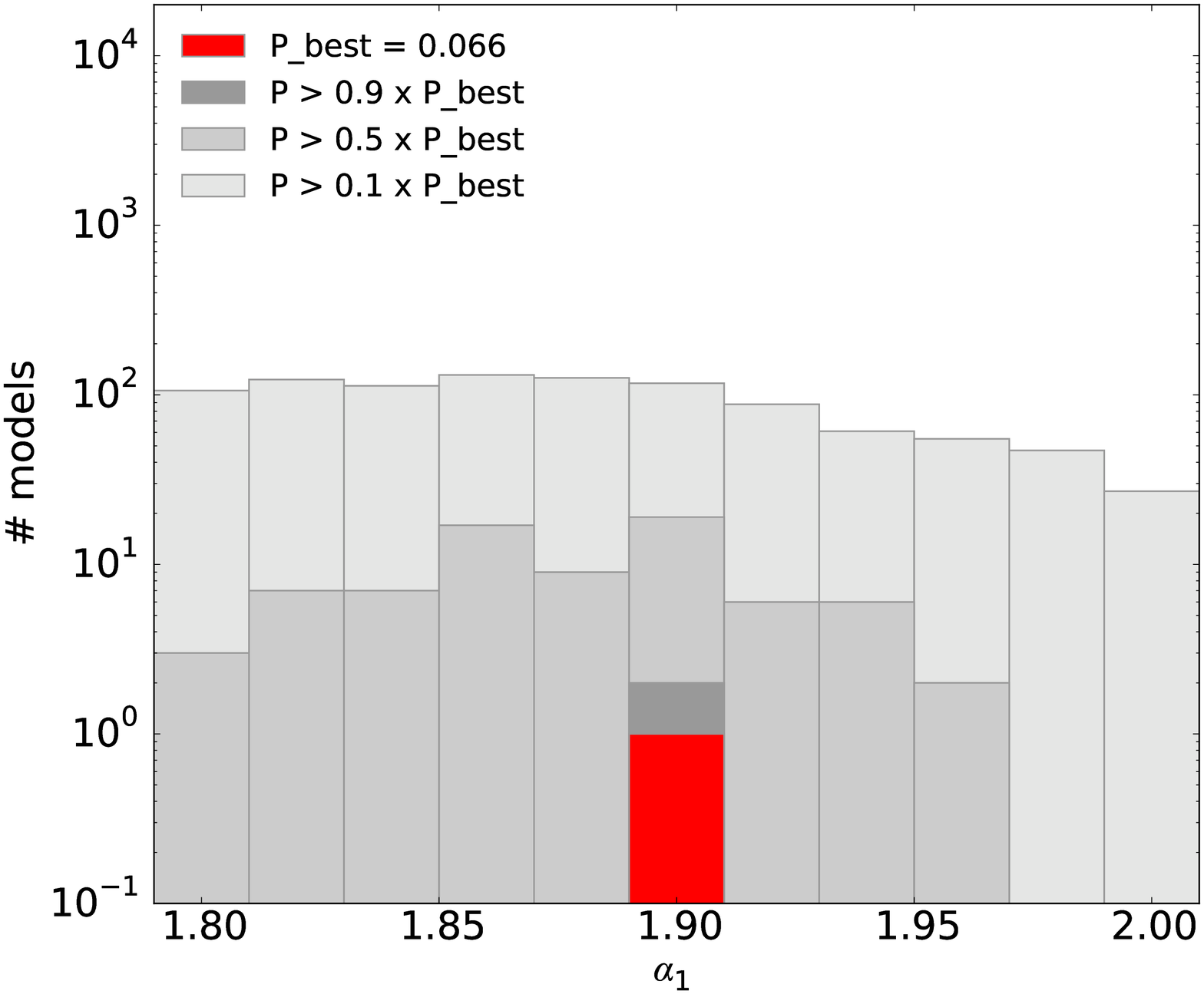}
\includegraphics[width=2.5in]{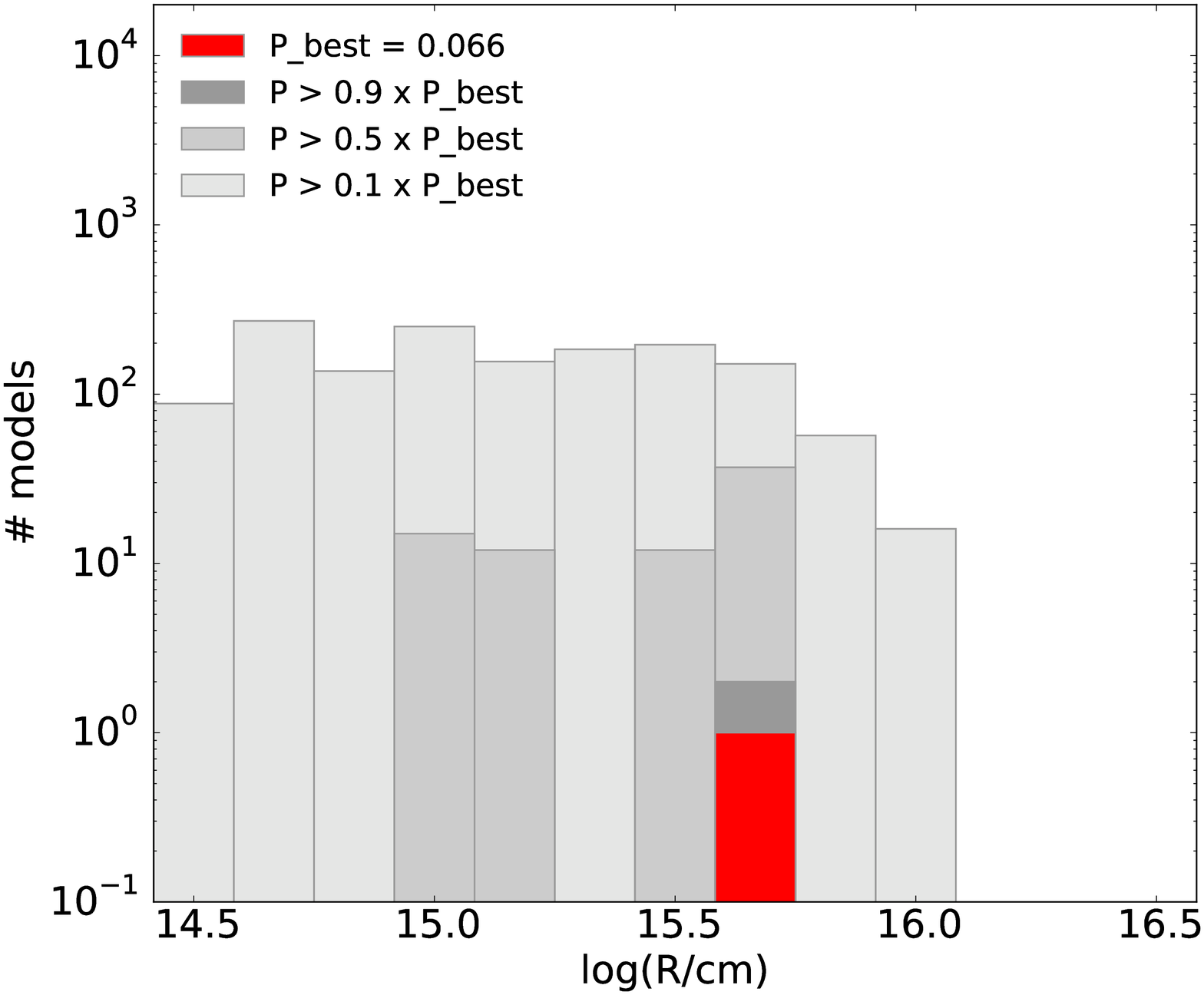}

	\end{center}
\end{minipage}
\hspace*{-5mm}
\begin{minipage}[b]{0.33\linewidth}
	\begin{center}
%\vspace*{-5mm}
\includegraphics[width=2.5in]{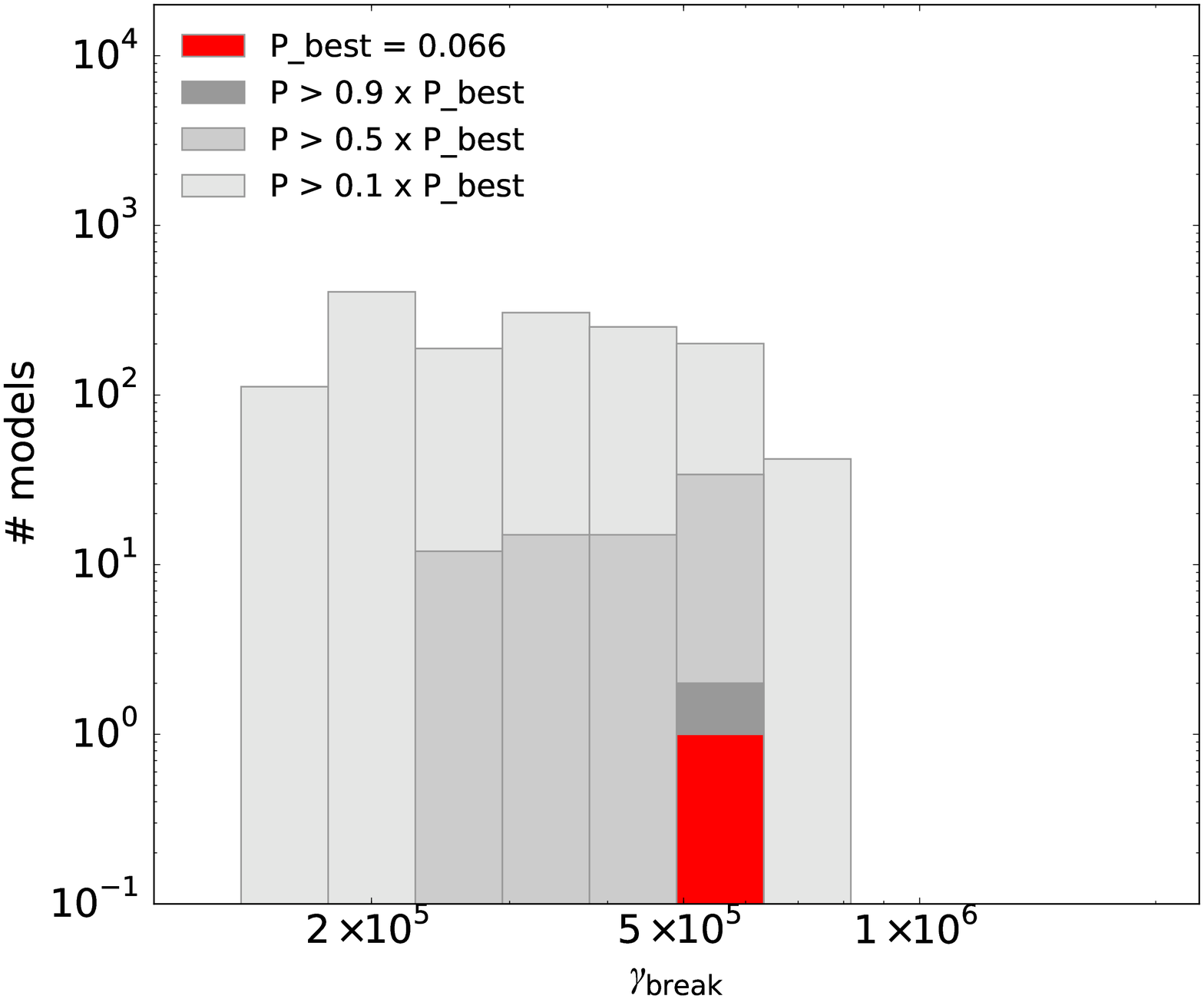}
\includegraphics[width=2.5in]{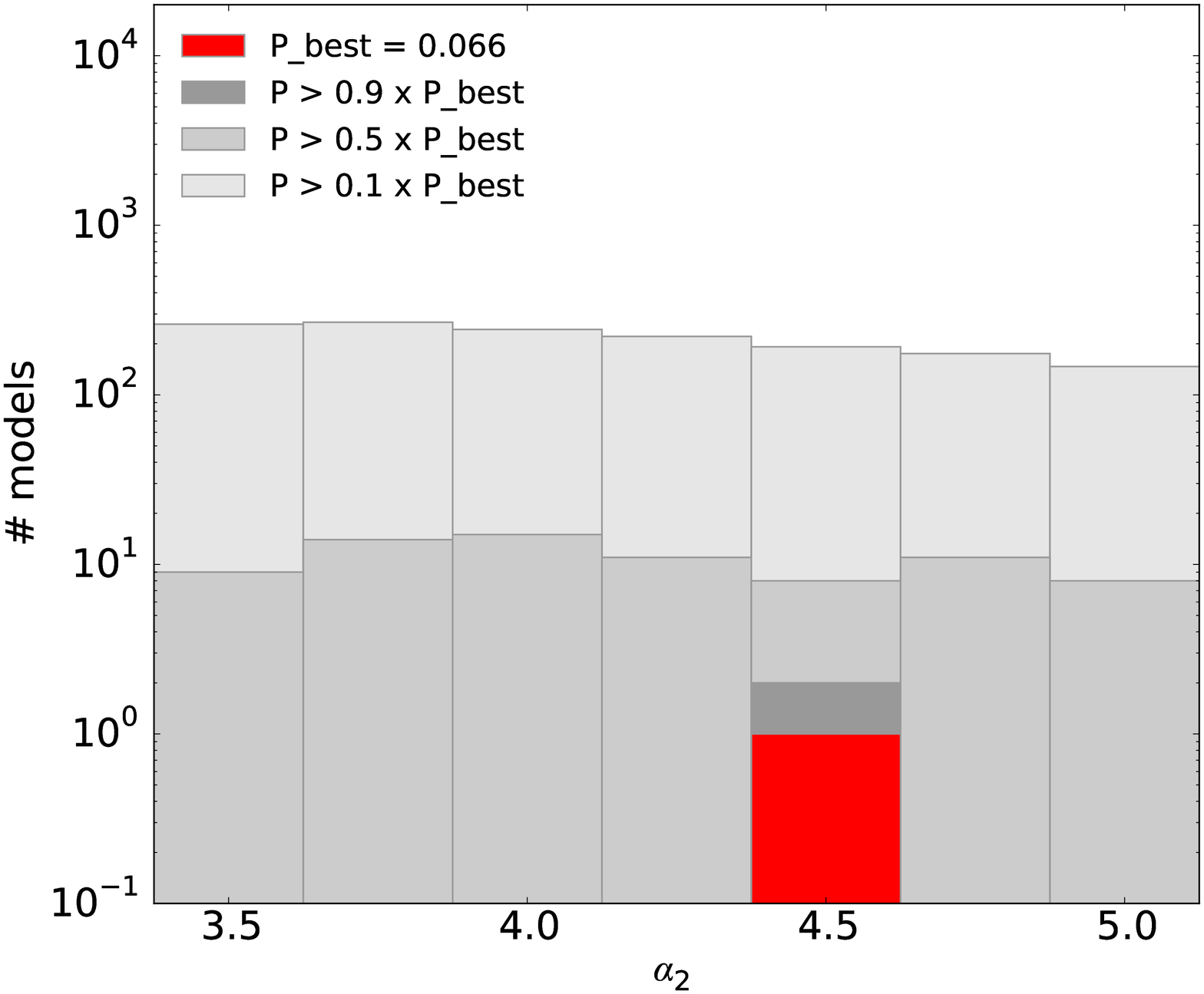}
\includegraphics[width=2.5in]{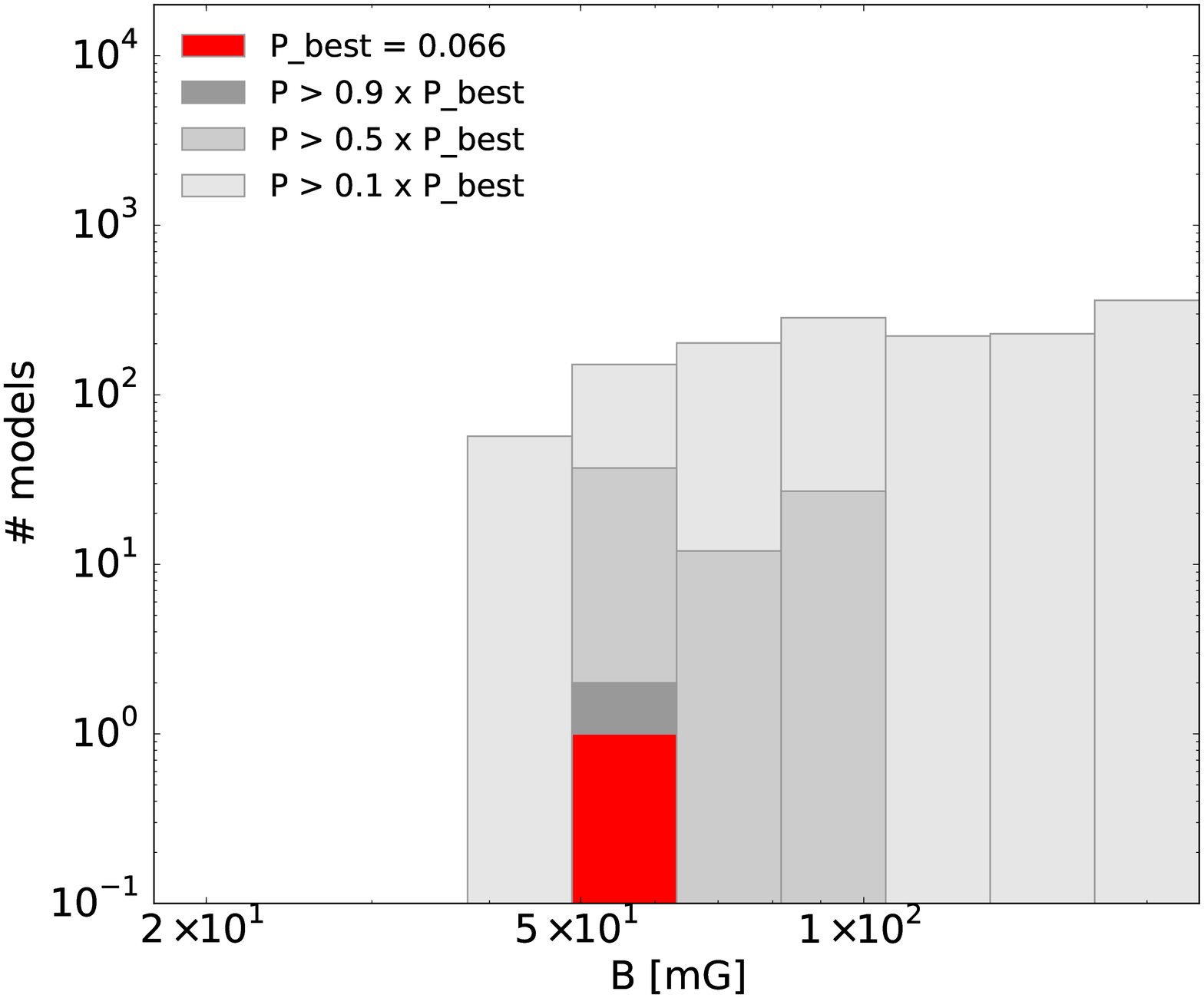}
	\end{center}
\end{minipage}
		\caption{Distributions of the investigated models in the individual model parameters for the dense parameter grid and a two-zone scenario for MJD~54973. The X-axis of each plot spans over the probed range for each parameter. Shown are the model with the highest probability of agreement with the data and all models which populate the given probability bands (see legend). 
}
	  \label{fig:sedparamranges}
\end{figure*}

Also here we investigate the spread - or degeneracy - of the different models within the dense grid space of model parameters. Fig.~\ref{fig:sedparamranges} shows again the distribution of the best model (red) and the models with $P>0.9\times P_{\text{best}}$, $P>0.5\times P_{\text{best}}$ and $P>0.1\times P_{\text{best}}$, respectively, over the entire dense grid parameter space. In comparison to Fig.~\ref{fig:sedparamrangescoarse}, one can notice an apparent larger degree of degeneracy, with distributions with entries in most of the probed parameter ranges depicted in the figure. The larger spread in the parameter values shown in Fig.~\ref{fig:sedparamranges} is caused by the selected parameter range, which is narrower and intentionally only covers regions with an already reasonable agreement between model and data, as derived from the coarse grid scan. Despite the large spread, one can see that there are regions with slightly better model-to-data agreement, like the region around  $\gamma_{\text{break}} \sim 5\times10^{5}$, or the region around $\alpha_1 \sim 1.9$.  The results will be further discussed in Sect.~\ref{sec:discussion}.

\begin{table*}[htbp]
\begin{center}
\caption{Grid parameter space probed for two-zone models within the dense grid-scan applied to the flare around MJD~54973.   The number of SSC models required to realize this grid-scan amounts to 212 million. See text for further details. 
}
\label{tab:sscparameterspacefine}
\vspace{5mm}
%\hline
	\begin{tabular}{r|ccccccccc}
dense grid &&&&&&&&& \\
two-zone		&$\gamma_{\text{min}}$ 	&$\gamma_{\text{max}}$ &	$\gamma_{\text{break}}$& 	$\alpha_{1}$ &	$\alpha_{2}$  	&n$_{e}$ &	$\frac{\mathrm{B}}{\mathrm{mG}}$ &	$\log(\frac{R}{cm}) $&$\delta$ 	 \\

\hline
min 	&	$2.1\times 10^{1}$ 	&	$3.2\times 10^{5}$ 	&	$1.2\times 10^{5}$				&	1.7 &	3.5 	&$2 \times 10^{3}$ &	20 	&	14.5&	5 	\\
max		&	$5\times 10^{4}$	&	$1\times 10^{8}$	&	$2 \times 10^{6}$			&		2.1 &	5.0	&		$1\times 10^{5}$	&200	&	16.0	&30	\\
$\#$ of steps	&	5		&	4	&		12			&			21	&	7		&10	&	10		&10	&6	\\
spacing	&	log		&	log		&	log			&		lin	&	lin	&	log		&log	&	lin	&	lin	\\
	%&&&&&&&&&& \\
\end{tabular}
\end{center}
\end{table*}

\begin{figure*}
	\begin{center}
\vspace{-5mm}
\includegraphics[width=6.1in]
{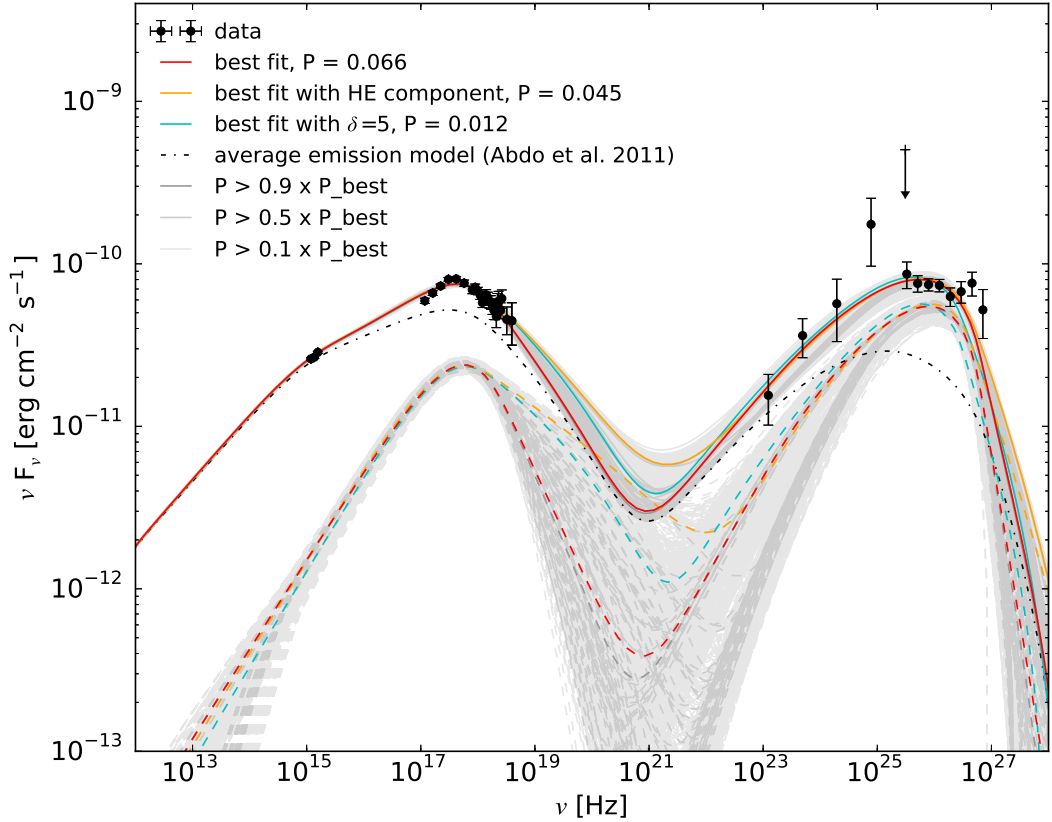}
\end{center}
\caption{Modelling of the SED of Mrk~501 compiled from measurements collected during the high state observed around MJD~54973. Two-zone SSC models have been inspected following the grid-scan strategy. The total emission (solid lines) is assumed to stem from a first quiescent region (black dot-dashed lines) responsible for the average state \citep{2011ApJ...727..129A} plus a second emission region (dashed lines). Highlighted are the model with the highest probability of agreement with the data (red), a model featuring a prominent  high-energy component in the EED (orange), and a model with low Doppler factor (cyan, $\delta=5$). 
Model curves underlaid in grey show the bands spanned by models with a fit probability better than $0.9\times P_{\text{best}}$, $0.5\times P_{\text{best}}$ and $0.1\times P_{\text{best}}$, respectively. The data points have been corrected for EBL absorption according to the model by \cite{Franceschini:2008vt}.
}
	  \label{fig:secondflaresedgridmodeled}
\end{figure*}

The SED models which are picked as a result of the dense grid-scan for two-zone SSC models are presented in Fig.~\ref{fig:secondflaresedgridmodeled}.
The figure highlights three SSC model curves: the model which gave the best agreement with the SED data points, a model featuring a prominent high-energy component in the EED, and a model that features 
a low Doppler factor of $\delta=5$.  The parameter values for these three specific SSC model curves are given in Table~\ref{tab:ssctwozonefineresults}, showing once more that three very distinct sets of SSC model parameters can provide comparable agreement with the experimental data.

\begin{table*} [htbp]
\begin{center}
\caption{Results of the dense grid-scan SED modelling of the flaring episode around MJD 54973 in the scope of a two-zone SSC scenario. Quoted here are the three models highlighted in Fig.~\ref{fig:secondflaresedgridmodeled}: the model with the best agreement to the data, a model with a prominent high-energy electron component, and a model with a remarkably low Doppler factor ($\delta = 5$). Besides the model parameters, the reduced $\chi^2$ values, the fit probability compared to the best achieved fit probability, the departure from equipartition and the implied minimum variability timescale are also reported. 
}
\label{tab:ssctwozonefineresults}
\vspace{5mm}
	\begin{tabular}{r|ccccccccccccc}
selected &\multirow{2}{*}{$\gamma_{\text{min}}$}&\multirow{2}{*}{$\gamma_{\text{max}}$}&\multirow{2}{*}{$\gamma_{\text{break}}$}&\multirow{2}{*}{$\alpha_{1}$}&\multirow{2}{*}{$\alpha_{2}$}&\multirow{2}{*}{n$_{e}$}&\multirow{2}{*}{$\frac{\mathrm{B}}{\mathrm{mG}}$}&\multirow{2}{*}{$\log(\frac{R}{cm}) $}&\multirow{2}{*}{$\delta$}&\multirow{2}{*}{$\frac{\chi^2}{\mathrm{d.o.f.}}$}&\multirow{2}{*}{$\frac{\mathrm{P}}{\mathrm{P}_{\mathrm{best}}}$}&\multirow{2}{*}{$ \frac{u_e}{u_B} $}&\multirow{2}{*}{$\frac{t_{\text{var}_{\text{min}}}}{\text{hour}}$}\\
models& 	& &	& 	 &	  	& &	 &		& 	&  &  &  & \\
\hline 
best $\chi^2$ & $1\cdot 10^{3}$& $1.0\cdot 10^{8}$& $5.6\cdot 10^{5}$&1.90 & 4.5& $7.4\cdot 10^{3}$& 56 & 15.7& 10 & 55.4/41
&1.00& 933 & 4\\
HE comp & $1.0\cdot 10^{3}$& $1.5\cdot 10^{7}$& $4.3\cdot 10^{5}$& 1.86& 3.5& $4.8\cdot 10^{3}$& 56&15.7 & 10 &57.5/41
&0.68& 919 & 4\\
Low $\delta$ & $1.5\cdot 10^{2}$& $1.5\cdot 10^{7}$& $5.6\cdot 10^{5}$& 1.82& 3.5& $2.0\cdot 10^{3}$& 72&16.0 & 5 &64.2/41
&0.18& 424 & 19\\
\end{tabular}
\end{center}
\end{table*}

%\input{discussion}
%\newpage

\section{Discussion}
\label{sec:discussion}

\subsection{Variability and correlations}

For Mrk~501, an increase of the fractional variability with energy has been reported in the past within the X-ray and VHE band \citep{Gliozzi:2006it,Albert:2007bt,PichelMrk501MW2009}. In the work presented here, we extend this trend throughout all wave-bands from radio to VHE $\gamma$-rays, showing that the source is relatively steady at radio/optical frequencies, but variable ($F_{\text{var}}\geq 0.2$) and very variable ($F_{\text{var}}\geq 0.4$) in the X-ray and VHE $\gamma$-ray bands, respectively, with a clear increase in the fractional variability with energy (observed in all the bands where we can measure). A similar variability pattern was reported in \citet{2015A&A...573A..50A} and, during the preparation of this study, also in \citet{Mrk501MW2012}, in relation to the extensive campaigns on Mrk~501 performed in 2008 and 2012, respectively. This suggests that this variability vs.~energy behaviour is an intrinsic characteristic in Mrk~501. On the other hand, \citet{2015arXiv150904936F} recently reported a different fractional variability vs.~energy pattern based on observations taken in 2013, where the observed variability at X-rays is similar to that at VHE.

The multiband variability pattern that has been observed in Mrk~501 is quite different from that observed in Mrk~421 during the multi-instrument campaigns from 2009, 2010 and 2013, as reported in \cite{2015A&A...576A.126A}, \cite{2015A&A...578A..22A} and  \cite{2016ApJ...819..156B}. In those works a double-bump structure in the fractional variability plot was found (instead of a continuous increase with energy) which relates to the two bumps in the broadband SED, and where the highest variability occurs at X-rays and VHE, at comparable levels.

A clear correlation of the X-ray and VHE $\gamma$-ray emission was
observed during the large and long $\gamma$-ray activity  from 1997,
as reported in e.g. \cite{1998ApJ...492L..17P,Gliozzi:2006it}, but
this correlation was only marginally detected during the $\gamma$-ray
flare observed in 2005 \citep{Albert:2007bt}. The low significance in the X-ray-to-VHE correlation during the flares in 2005 was ascribed to the lack of sensitive X-ray measurements during this observing campaign; only \textit{RXTE}/ASM data, which has limited sensitivity to detect Mrk~501, was available for this study. A positive correlation 
between X-ray and VHE $\gamma$-rays was reported, for the first time, also during very low X-ray and VHE activity, but only at 99\% confidence level \citep{2015A&A...573A..50A}. The marginally significant correlation observed during this low activity, using 
data from the multi-instrument campaign in 2008, was ascribed to the very low variability during that campaign, where the measured $F_{var}$ values were about 0.1 for X-rays and 0.2 for VHE. As reported above, during the multi-instrument campaign in 2009, Mrk~501 was mostly in its low/typical state, but we also measured two flaring activities in May 2009. 
The measured $F_{var}$ is about 0.3 for X-rays and 0.8 for VHE, while if we exclude the two flaring episodes, we obtain $F_{var}$ values of about 0.2 for X-rays and 0.5 for VHE. However, despite the larger variability observed in 2009 (with respect to 2008), we did not observe any significant correlation between the X-ray and the VHE emission (including and excluding the flaring episodes). This may appear to be a controversial result, but we would like to stress that a very significant correlation with past data was only observed during the very large and long flare in 1997. 
Recently, \citet{2015arXiv150904936F} reported a significant X-ray-to-VHE correlation, using data from the multi-instrument campaign in 2013. This correlation is dominated by the large X-ray and VHE activity observed during four consecutive days in July 2013: although it still remains at 2 $\sigma$ (for the 0.3-3 keV energy band) and 5 $\sigma$ (for the 3-7 keV band) when removing the flaring activity.   In conclusion,  some multi-instrument campaigns on Mrk 501 do not show a clear X-ray to VHE correlation when the source is not flaring strongly or persistently high. However, for the other archetypical TeV blazar, Mrk 421, the X-ray-to-VHE correlation is significantly detected during both low- \citep[e.g.][]{2015A&A...576A.126A,2016ApJ...819..156B} and high-activity states \citep[e.g.][]{2008ApJ...677..906F,2011ApJ...738...25A,2015A&A...578A..22A}.

The X-ray-to-VHE correlation and the fractional variability vs.~energy pattern observed in Mrk~421 suggests that the X-ray and
VHE emissions are produced by the same electrons, within the framework of SSC scenarios, and that the highest variability is produced by the highest-energy and most-variable electrons, which dominate the emission at the keV and the TeV bands, respectively. 
Instead, in Mrk~501 we observe a continuous increase in the variability with energy and  absence of persistent correlation between the keV and TeV emissions. This suggests that the highest-energy electrons, in the framework of SSC scenarios, are not responsible for the keV  emission, while they are responsible (at least partially) for the TeV emission. Alternatively,  there could be an additional (and very variable) component contributing to the $\gamma$-ray emission, in addition to that coming from the SSC scenario, like  inverse-Compton of the high-energy electrons off some  external low-energy photon fields \citep{1992A&A...256L..27D,1994ApJ...421..153S,2016arXiv160205965F}.

\subsection{The VHE flaring state SEDs}

The first flaring event (MJD~54952) is characterized by a fast and large outburst in the VHE band, which was apparently not accompanied by a substantial increase of the X-ray flux, and hence appeared to be like an ``orphan flare''  \citep[see e.g.][]{2004ApJ...601..151K}. In fact, based on these observations, this event was indeed tentatively categorized as an ``orphan flare'' event \citep{Pichel:2011we,2012A&A...541A..31N}, which would substantially challenge the currently favoured SSC emission models  (for HBLs). 
However, a detailed look at the SED of the flaring episode reveals a hardening of the X-ray spectrum measured by \textit{Swift}/XRT (see Sect.~\ref{sec:variability}), which more likely corresponds to a shift of the synchrotron bump towards higher energies. 
During the outstanding activity in 1997, the synchrotron peak was shifted to beyond 100\,keV, as accurately measured by BeppoSax \citep{1998ApJ...492L..17P}. Such a large increase in the location of the synchrotron peak position could have occurred in the MJD~54952 flare discussed here. Additionally, the peak of the high-energy $\gamma$-ray bump at the time of this flare also appears to shift towards higher energies, as occurred in 1997.  
This suggests a more general appearance of such phenomena, and that,
even though the measured keV and TeV flux are not correlated during
this flaring activity, the overall broad band X-ray and VHE emission
may still be correlated, which could have been measured if X-ray
observations at several 10s of keV  had been available during this
flaring episode. Such a shift of the entire SED has been interpreted
as a shift in the energy distribution of the radiating electron
population \citep[e.g.][]{1998ApJ...492L..17P,Albert:2007bt}. In this
context, the small change in the inverse Compton peak position
compared to that of the synchrotron peak location could be ascribed to
Klein-Nishina effects. High-energy electrons can efficiently produce
high-energy synchrotron photons; however their effectiveness to upscatter photons reduces with respect to the lower-energy electrons because the Klein-Nishina cross-section is smaller than the Thomson cross-section \citep{1998ApJ...509..608T,2011ApJ...729....2A}.

We tried to parameterize the broadband SED during this first flare (MJD~54952) using a wide range of SSC emission scenarios following the grid-scan strategy defined in Sect.~\ref{TheGridScan}, allowing for models with one or two (independent) emission zones and covering a wide range in the space of model parameters. We found that none of the tested models could satisfactorily reproduce the changes observed in the spectral distribution. This broad-band SED may be explained with more sophisticated theoretical models, like the inhomogeneous time-dependent models reported in \citet{2005A&A...432..401G,2008ApJ...689...68G,2011MNRAS.416.2368C,2015MNRAS.447..530C,2016MNRAS.458.3260C}, which provide a more elaborate physical scenario, at the expense of an increase in the number of degrees of freedom of the model. However, a caveat has to be taken into account when interpreting these results, which is the lack of strict simultaneity of the different data sets, in particular the Swift/XRT and the VHE $\gamma$-ray data. The individual exposure times are separated by seven hours, while we see flux changes by a factor of $\sim$4 on sub-hour timescales, as reported in \citet{Pichel:2011we} and  \citet{PichelMrk501MW2009}. Therefore, it is somewhat uncertain whether the measurements of the synchrotron peak and the high-energy peak probe the same source state.  In the recent study reported in \citet{PichelMrk501MW2009}, the broadband SED derived for the 3-day time interval MJD 54952–55 could be satisfactorily parameterized  with a one-zone SSC scenario that differs from the one used here and in \citet{2011ApJ...727..129A} in various aspects, including the template used to describe the host galaxy contribution.

Compared to the first flaring event, the second flare ($\approx$
MJD~54973) occurs during VHE flux changes of factors of $\sim$2 on
timescales of a few days, and hence the lack of strict simultaneity in the X-ray/VHE observations is a much smaller caveat than for the first flaring event.  In this case, again following the grid-scan modeling approach, one-zone SSC models were found unable to describe the measured SED, reaching best probabilities of agreement $\sim10^{-10}$). The two-zone SSC models were able to reproduce the experimental observations better, reaching best probabilities of agreement $\sim10^{-3}$). Therefore, the two-zone scenario appears to be favoured compared to the one-zone scenario considered here. Building on the range of two-zone model parameters providing decent data-model agreement, a fine grid-scan was performed, yielding hundreds of two-zone SSC models with probabilities of agreement $\sim10^{-2}$. The obtained set of two-zone SSC models providing the best agreement comprises several set-ups with quite different implications for the parameters defining the EED and the surrounding region of the second emission region (see Sect.~\ref{sec:sed2}). 

Comparing the configurations obtained for the emission region responsible for the second flare with the parameter values describing the emission region assumed to create the quiescent emission, some general trends can be stated:  while the parameters describing the EED and the Doppler factor are found to populate roughly the same ranges of values, the electron density n$_e$ is increased by 1-2 orders of magnitude, the magnetic field is larger by $\approx1$ order of magnitude and the size of the emission region $\mathrm{R}$ is found to be smaller by 1-2 orders of magnitude. The latter result is affected by the requirement of a minimum  variability timescale of  a day, in order to account for the variability seen in the data.

Besides the general observations made above, some interesting model configurations stand out from the set of adequate scenarios: models which feature a prominent high-energy component in the EED, and models with Doppler factors as low as $\delta=5$ can be used to adequately model the flaring SED. In the paragraphs below we discuss the benefits of these two families of models. 

Synchrotron self-Compton models with a strong high-energy component
are interesting not only to explain the SED collected during the
presented campaign, but also in the context of other observations of
Mrk~501. During the extreme flare seen in 1997, a strong increase in
the regime of hard X-rays, around 100 keV, was observed
\citep{1997A&A...320L...5B}. This increase can be interpreted as the
emergence of a strong high-energy component adding to the overall SED,
which only becomes visible sometimes during extreme flaring states. Moreover, Cherenkov telescope observations often give hints of an additional hard component in the EED during flaring times: in \cite{Albert:2007bt} a significant spectral hardening during flaring states was observed and reported for the first time, and in the course of several more observational campaigns this ``harder when brighter'' behaviour has been established as typical for Mrk~501. Ultimately, a tendency for this behaviour was also seen during the campaign presented in this paper. In this light, SSC models with such a high-energy contribution to the EED could be favoured, as they can also explain such mentioned observations. Naturally, with the data set at hand and the lack of hard X-ray observations above $\approx 20$\,keV, they can neither be confirmed nor discarded.

The finding that models with $\delta=5$ can also adequately
reconstruct the data is particularly interesting. Quite high values
(above $\delta=10$, up to $\delta=50$ or more) are usually required to model the SEDs of blazars \citep{1998ApJ...509..608T,2001ApJ...559..187K,2004ApJ...616..136S}. These high Doppler factors are in tension with regard to expectations from the small (typically less than $2c$) apparent velocities observed in the 43\,GHz radio emission of various high-peaked BL Lac objects, and particularly with that measured for Mrk~501 \citep{2002ApJ...579L..67E,2004ApJ...600..115P,2010ApJ...723.1150P}. This has posed a common problem for TeV sources, which has been dubbed  the ``bulk Lorentz factor crisis'' \citep[][]{2006ApJ...640..185H}, and requires the radio and TeV emission to be produced in regions with different bulk Lorentz factors. Debates on this problem  \citep[see for example][]{2003ApJ...594L..27G,2007ApJ...671L..29L,2008MNRAS.383.1695S} have led to a series of sophisticated models, e.g.~the "spine-sheath" model from \cite{2005A&A...432..401G}, in which the jet is structured transverse to its axis into a fast ``spine'' that produces the high-energy emission, and a slower ``layer'' which dominates the radio emission. 
The modeling results presented in this paper show that it is actually possible to model the SED of a flaring activity of Mrk~501 using a relatively small Doppler factor to describe the flaring emission, hence alleviating the tension with the radio interferometric observations. Naturally, such low Doppler factors cannot be used for broad-band SEDs related to periods with fast (sub-hour) variability, such as the one from MJD~54952; but they can be be used for broad-band SEDs related to flaring activities with day timescales, such as the one from MJD~54973, which are more commonly observed in high-peaked BL Lac objects.

In addition to individual models which give a good reproduction of the data, the degree of degeneracy of well-fitting models in the individual parameters has also been studied, revealing a wide range of equally good models in the SSC parameter space, and showing that some model parameters can be constrained much better than others. While this can be seen already for the one-zone SSC models, where e.g.~$\gamma_{break,1}$ and $\gamma_{break,2}$ show a narrower distribution than for instance $\gamma_{max}$ or $\alpha_3$, it is particularly interesting to study this for the more applicable two-zone scenario, which is the one that describes suitably the data. We find that for both the coarse and the dense grid-scan, the distribution of parameter values giving a good agreement with the data is quite well constrained for some parameters, such as the Lorentz factor at the break energy of the electrons $\gamma_{\text{break}}$, while other parameters show a rather broad distribution, like $\gamma_{\text{max}}$ or the index of the EED after the break $\alpha_2$, which points to a real degeneracy in these parameters. To some extent this degeneracy can be explained by the unequal sampling of the SED: the density and accuracy of measurements at or around the positions of the synchrotron and the IC peak is rather dense, which leads to a good definition of the spectral break in the EED. However, moving from the peak positions up to higher energies, the uncertainties of the measurements increase (especially for the synchrotron peak) and parameters such as the spectral index after the break $\alpha_2$ or the Lorentz factor where the EED is cut off $\gamma_{\text{max}}$ cannot be constrained equally well.

This result has several implications for the modeling of SEDs in
general. On one hand, it shows that an actual fitting procedure, which
moves along the direction of the steepest gradient in the parameter
space towards a minimum in e.g.~the $\chi^2$ of the model-to-data
agreement, does not necessarily reveal the entire picture of possible
descriptions of the data in the context of the applied model. Usually 
one ``best'' solution is quoted as the result, while most of the time a wide range of models explain the data equally well. 
We also see that, in order to be able to put stronger constraints on the parameters defining SED models, we need data sets which are characterized by a better coverage in energy and by smaller uncertainties in flux. We see in Fig.~\ref{fig:secondflaresedgridmodeled} that especially the hard X-ray regime, but also the HE and VHE $\gamma$-ray regime, are allowing for a wide range of possible model curves. 

Unfortunately, the here-presented exercise indicates that the SED modeling results which are performed for less constrained data sets, e.g.~for ``weak sources''  which are sampled with much less coverage in energy and with less statistics, should be taken with caution because are likely to have substantial degeneracies in the model parameters. Such modeling exercises can demonstrate that a particular scenario (e.g. one-zone or two-zone SSC) is capable of reproducing the measured data, but they certainly cannot claim the exclusiveness or even the prominence of the particular set of parameter values that has been chosen or found to be ``best''.

\subsection{Change in optical polarization during VHE $\gamma$-ray flare}
The first VHE flare (MJD~54952) was found to coincide with an observed change in the optical polarization.  While simulations of turbulent processes in blazar jets show that a rotation of this dimension can be ascribed to random behaviour \citep{2014ApJ...780...87M}, the coinciding occurrence of the change in rotation and a flare of the VHE $\gamma$-ray flux suggests a common origin of these events.  Such combined events have already been seen in low-frequency peaked BL Lac type sources (LBL) and flat spectrum radio quasars (FSRQ), but it has been observed for the first time for an HBL in the course of the 2009 campaign, and already reported in \citet{Pichel:2011we,PichelMrk501MW2009}.

These observations show similarities to double or multiple flaring events seen in the LBL BL~Lacertae in 2005 and in the FSRQ PKS~1510-089 in 2009, which were discussed by \cite{Marscher:2008ii} and \cite{Marscher:2010hu}, respectively.

Exhibiting different peak frequencies for the synchrotron and the IC bump, the optical variability seen in BL~Lac could be seen as corresponding to the X-ray variability in Mrk~501. While a strong flare in the VHE band has been observed during the first flare of Mrk~501, BL Lac gives hints for activity in that band during the first optical outburst \citep{2007ApJ...666L..17A}. A coincidence of a flaring event and a change in the optical polarization is seen in all three data sets. 
The observed degree in optical polarization in Mrk~501 of $\approx 5\%$ appears to be small in comparison to that in BL~Lacertae (up to $18\%$). Still, the optical flux in Mrk~501 is strongly dominated by the host galaxy, so that the jet contribution amounts to only $\sim$1/3. Therefore, the measured degree of polarized light in Mrk~501 corresponds to  a fraction of $\approx 15-20\%$ of polarized emission from the jet, which is comparable to BL~Lacertae. The second episode of high activity in both Mrk~501 and BL~Lac was characterized by an increased flux at the synchrotron bump over a longer time span.

In the case of BL~Lac, \cite{Marscher:2008ii} suggested that the ``first'' flare and the change in polarization may have occurred when a blob of highly energetic particles travels along the last spiral arm of a helical path within the acceleration and collimation zone of the jet, and finally leaves this zone to enter a more turbulent region. The ``second'' flare seen in BL~Lac has been identified with the passage of the feature through the shocked region of the radio core. The observed behaviour of Mrk~501 suggests that the discussed scenario could be applicable here. Despite the lack of simultaneous interferometric radio observations during both flares, an enhancement of the activity in the VLBA 43 GHz core emission in May 2009 (with respect to the previous months) was observed, supporting the interpretation of the blob traversing a standing shock region during the second flaring episode.

The polarization data collected during this campaign could also hint
at a different physical scenario. After the VHE flare on MJD~54952,
not only did the EVPA rotation stop, but it also started rotating in the reverse direction during the following 2 days, and the polarization degree did not drop to the "typical low-state values" of about 1--2\%, but only decreased from 5.4\% to 4.5\% (see the two bottom panels from Fig.~\ref{fig:mwllcs}). The characteristics of the polarization data after the VHE flare may be better explained as resulting from  light-travel-time effects in a straight shock-in-jet model with helical magnetic fields, as proposed by \citet{2014ApJ...789...66Z,2015ApJ...804..142Z}. This shock-in-jet model, which uses a full three-dimensional radiation transfer code and takes into account all light–travel–time and other geometric effects (for some assumed geometries), may be more successful in explaining the broad-band SED (and variability patters) observed during the VHE flare from MJD 54952, which could not be explained with the "relatively simple" one-zone and two-zone SSC scenarios described in Sect~\ref{sec:sed}. However, the lack of strictly simultaneous X-ray/VHE data during the MJD~54952 VHE flare, and the relatively scarce polarization observations after the VHE flare would be an important limitation in the full application of this theoretical scenario to the here-presented multi-instrument data set.

\section{Summary and concluding remarks}
\label{sec:summary}

\noindent
We presented a detailed study of the MWL variability of the HBL Mrk~501, based on a multi-instrument campaign that was conducted over 4.5 months in 2009, with the participation of MAGIC, VERITAS, the Whipple 10\,m, {\em Fermi}-LAT, {\em RXTE}, {\em Swift}, GASP-WEBT, and several optical and radio telescopes. Mrk~501 shows an increase in the fractional variability with energy, from a steady flux at radio and optical frequencies to fast and prominent flux changes in the VHE $\gamma$-ray band. 
Overall, no significant correlation was found between any of the
measured energy bands, particularly no correlation was seen between
X-rays and VHE $\gamma$-rays, despite the relatively large variability
measured in these two energy bands. This suggests that the
highest-energy (and most variable) electrons that are responsible for
the VHE $\gamma$-rays measured by MAGIC, VERITAS and Whipple 10\,m, do
not have a dominant contribution to the $\sim$1~keV emission measured
by {\em Swift}/XRT. These high-energy electrons may have a dominant
contribution to the hard X-ray emission above 10--50 keV, where the
instrumentation used in this campaign did not provide sensitive
data. Alternatively, there could be a component contributing to the
VHE $\gamma$-ray emission in addition the component coming from the
SSC scenario (e.g. external Compton), which is highly variable and
further increases the variability of Mrk\,501 at VHE $\gamma$-rays with respect to that expected from the pure SSC scenario.

This paper discusses two prominent flaring events at VHE $\gamma$-rays
with different characteristics that were seen during the campaign. The first flare is dominated by a fast outburst in the VHE range, which does not appear to be accompanied by a large flux increase in the X-ray band, but shows a hardening in the X-ray spectrum that can be associated with a shift of the synchrotron bump to higher energies. On the other hand, the second flare is characterised by a flux increase in both the VHE and the  X-ray band. For the parameterisation of the broadband SEDs from these two VHE $\gamma$-ray flares, we applied a novel variation of the {\em grid-scan} approach in the space of model parameters. For the two theoretical scenarios investigated, the one-zone and two-zone SSC models, we probed multi-dimensional grids with the various model parameters, evaluating the model-to-data agreement for tens of millions of SSC models. This strategy allowed us to identify disjointed regions of equally good model configurations, and provided a quantification of the degeneracy in the model parameters that describe the measured broadband SEDs. The presented methodology provides a less biased  interpretation than the  commonly used ``single-curve model adjustment procedure'' typically reported in the literature.

A lack of strict simultaneity in the X-ray/VHE observations of the
first flare, which is characterized by large VHE flux changes in
sub-hour timescales,  does not permit us to draw final conclusions on
the underlying mechanism; but the SED modeling with the grid-scan
suggests that a simple one-zone or two-independent-zone SSC model is
not sufficient to explain the measured broadband emission. The
broadband SED derived for the second flare also lacks strictly
simultaneous observations, but the flux changes here are smaller and
on longer timescales, and hence substantially less problematic than
for the first flare.  The overall SED from the second flare cannot be
properly described by a one-zone SSC model (with an EED with two
spectral breaks), while it can be reproduced satisfactorily within a
two-independent-zone SSC scenario. In the two-zone models applied
here, one zone is responsible for the quiescent emission from the
averaged 4.5-month observing period, while the other one, which is
spatially separated from the first, dominates the flaring emission
occurring at X-rays and VHE $\gamma$-rays. The grid-scan shows that
there is a large number of SSC model realizations that describe the
broadband SED data equally well, and hence that there is substantial
degeneracy in the model parameters despite the relatively
well-measured broadband SEDs. For instance, regarding the features of
the EED, the position of the break(s) appear to be well constrained,
while the highest Lorentz factor and the high-energy spectral index vary more strongly within the best-fitting model realizations. While the few models with the best relative agreement to the data feature Doppler factors $\delta$ in the range 10--20, the data can also be reproduced using substantially lower Doppler factors of $\delta=5$ while still reaching fit probabilities higher than $10\%\,  P_{\text{best}}$. This shows that it is possible to reproduce the observed SED from Mrk~501 assuming boost factors well below the usually required values of $\delta \approx 10 - 50$, which 
may loosen a bit the tension posed between large values of $\delta$ required for modeling and low values imposed from radio velocity measurements, which has been dubbed the ``bulk Lorentz factor crisis''.

A change in the rotation of the EVPA was measured in temporal coincidence with the first VHE flare, at MJD~54952, as reported in \citet{Pichel:2011we,PichelMrk501MW2009}. Here we also show that during the first VHE flare, the degree of polarization increased by a factor of $\sim$3 with respect to the polarisation measured before and after this flaring activity. This is the first time that such behaviour is observed in Mrk~501, or in any other HBL object, and suggests a common origin of the VHE flare and the optical polarized emission.
With the coincidence seen of a VHE flare and changes in the trend of the optical polarization, this two-flare event resembles prior events observed in the LBL BL~Lacertae and the FSRQ PKS~1510-089, which were discussed in \cite{Marscher:2008ii} and \cite{Marscher:2010hu}, respectively. The common features suggest a similar interpretation of the flaring event of Mrk~501 as an emission region which is traveling upstream of the radio core along a spiral path in a helical magnetic field, entering a region of turbulent plasma and crossing the standing shock region of the radio core.  After the VHE flare at MJD~54952, the polarization degree decreased from 5.4\% to only 4.5\% (instead of the typical low-state value of 1--2\%), and there is also a small EVPA rotation in the reverse direction during the following 2 days, which may be difficult to explain with the typical helical pattern motions mentioned above; they may be better explained as light-travel-time effects in a shock-in-jet model in a straight, axisymmetric jet embedded in a helical magnetic field, as reported in \citet{2014ApJ...789...66Z,2015ApJ...804..142Z}. Beyond the interpretation of the flaring event itself, the observational results obtained in the course of this MWL campaign reveal phenomena that have not been seen for any HBL before, but have already been studied for LBLs and FSRQs. This gives a strong indication of the intrinsic similarity of these blazar subclasses, despite showing different jet characteristics in general, such as apparent jet speed and the overall power output. Observations of rapid variability in the LBL BL~Lacertae \citep{2013ApJ...762...92A} support this further, as such fast flux changes had only been seen in HBL observations.

Additional multi-instrument observations of Mrk~501 will be crucial to
confirm and extend several of the observations presented here. First
of all, the large degeneracy in the model parameter values providing
an acceptable description of the broadband SED is largely dominated by
the poor coverage at hard X-rays, as well as the somewhat limited
resolution at VHE $\gamma$-rays.  Since mid-2012,
NuSTAR\footnote{http://www.nustar.caltech.edu}  provides excellent
sensitivity above 10 keV, which narrows down the large gap in the SED
between soft X-rays and the \textit{Fermi}-LAT regime for campaigns
from 2012 onwards. In the coming years, observations with
ASTROSAT\footnote{http://astrosat.iucaa.in} will also be possible,
hence facilitating a much more accurate characterization of the
evolution of the synchrotron bump of Mrk~501, including the
determination of hard components in the EED whose synchrotron emission
may peak at 50 or 100 keV. Moreover, in the regime of VHE
$\gamma$-rays, data sets of much higher quality are already being
collected, yielding a better resolution and an extended energy
coverage. This is achieved on the one hand by the operation of the MAGIC
telescopes as a stereo system, which gave a remarkable improvement in
the overall performance compared to the mono mode which was still
operational during the campaign reported in this paper \citep{2012APh....35..435A}. On the other hand, both MAGIC and VERITAS underwent major upgrades in the years 2011 and 2012, which gave a further substantial push to the performance \citep{2016APh....72...61A,2016APh....72...76A,2013arXiv1307.8360Z}. In the future, the Cherenkov Telescope Array (CTA) promises to deliver further, substantial improvement both in terms of the energy coverage and the resolution of the flux measurement  \citep{Actis:2011cg}.  Additionally, the temporal coverage extending over many years will permit variability studies (including many flares) over large portions of the electromagnetic spectrum and with good sensitivity, which will permit one to evaluate whether the association of EVPA rotations and polarization degree changes with VHE $\gamma$-ray flares are rare or regular events, whether these events occur together with a second flaring activity with contemporaneous enhancement of the VLBA radio core emission, and whether the measured multiband variability and lack of 1~keV--1~TeV correlation is a typical characteristic in Mrk~501 that gets repeated over time.

\begin{acknowledgements}
%\footnotesize{{\bf Acknowledgment:}{\\

The authors thank the anonymous referee for providing a very
  detailed and constructive list of remarks  that helped us to clarify and improve some of the results reported in the
  manuscript.  \\

The MAGIC Collaboration would like to thank 
the Instituto de Astrof\'{\i}sica de Canarias
for the excellent working conditions
at the Observatorio del Roque de los Muchachos in La Palma.
The financial support of the German BMBF and MPG,
the Italian INFN and INAF,
the Swiss National Fund SNF,
the ERDF under the Spanish MINECO
(FPA2015-69818-P, FPA2012-36668, FPA2015-68278-P,
FPA2015-69210-C6-2-R, FPA2015-69210-C6-4-R,
FPA2015-69210-C6-6-R, AYA2013-47447-C3-1-P,
AYA2015-71042-P, ESP2015-71662-C2-2-P, CSD2009-00064),
and the Japanese JSPS and MEXT
is gratefully acknowledged.
This work was also supported
by the Spanish Centro de Excelencia ``Severo Ochoa''
SEV-2012-0234 and SEV-2015-0548,
and Unidad de Excelencia ``Mar\'{\i}a de Maeztu'' MDM-2014-0369,
by grant 268740 of the Academy of Finland,
by the Croatian Science Foundation (HrZZ) Project 09/176
and the University of Rijeka Project 13.12.1.3.02,
by the DFG Collaborative Research Centers SFB823/C4 and SFB876/C3,
and by the Polish MNiSzW grant 745/N-HESS-MAGIC/2010/0. \\

VERITAS is supported by grants from the US Department of Energy Office of Science, the U.S. National
Science Foundation and the Smithsonian Institution, by
NSERC in Canada, by Science Foundation Ireland (SFI
10/RFP/AST2748) and by STFC in the U.K. The VERITAS collaboration acknowledges
the excellent work of the technical support staff at the Fred
Lawrence Whipple Observatory and at the collaborating institutions
in the construction and operation of the instrument.\\

The $Fermi$ LAT Collaboration acknowledges support from a number of agencies and institutes for both development and the operation of the LAT as well as scientific data analysis. These include NASA and DOE in the United States, CEA/Irfu and IN2P3/CNRS in France, ASI and INFN in Italy, MEXT, KEK, and JAXA in Japan, and the K.~A.~Wallenberg Foundation, the Swedish Research Council and the National Space Board in Sweden. Additional support from INAF in Italy and CNES in France for science analysis during the operations phase is also gratefully acknowledged.\\

We acknowledge the use of public data from the {\it Swift} and {\it
  RXTE} data archive. The Mets\"ahovi team acknowledges the support from the Academy of Finland
to our observing projects (numbers 212656, 210338, 121148, and others).This research has made use of data obtained from the National Radio
Astronomy Observatory’s Very Long Baseline Array (VLBA), projects
BK150, BP143 and MOJAVE. St.Petersburg University team acknowledges support from Russian RFBR
grant 15-02-00949 and St.Petersburg University research
grant 6.38.335.2015. AZT-24 observations are made within an agreement between  Pulkovo,
Rome and Teramo observatories.
The Abastumani Observatory team acknowledges financial support by the
Shota Rustaveli National Science Foundation under contract
FR/577/6-320/13.
This research is partly based on observations with the 100-m telescope of the MPIfR (Max-Planck-Institut f\"ur Radioastronomie) at Effelsberg, as well as with the Medicina and Noto telescopes operated by INAF–Istituto di Radioastronomia. The Submillimeter Array is a joint project between the Smithsonian Astrophysical Observatory and the Academia Sinica Institute of Astronomy and Astrophysics and is funded by the Smithsonian Institution and the Academia Sinica. 
The OVRO 40 m program was funded in part by NASA Fermi Guest
Investigator grant NNX08AW31G,  and the
NSF grant AST-0808050, and the Steward observatory by the NASA Fermi Guest Investigator grant NNX08AW56G

\end{acknowledgements}

%\newpage

%\cleardoublepage

\bibliographystyle{aa} % style aa.bst

\newpage

\end{document}